%% file: main.tex
\documentclass[acmsmall]{acmart}
\AtBeginDocument{%
  \providecommand\BibTeX{{%
    \normalfont B\kern-0.5em{\scshape i\kern-0.25em b}\kern-0.8em\TeX}}}

\usepackage[tikz]{bclogo}
\usepackage{transparent}
\usepackage[most]{tcolorbox}
\usepackage{dirtytalk}
\usepackage{comment}
\usepackage{booktabs}
\usepackage{array}

\usepackage{amssymb,amsfonts}
\usepackage{amsmath}
\usepackage{graphicx}
\usepackage{adjustbox}
\usepackage{textcomp}
\usepackage{xcolor}
\usepackage{float}
\usepackage{xcolor,colortbl}
\usepackage{multirow}
\usepackage{subcaption} 
\usepackage{tcolorbox}
\usepackage{csquotes}
\usepackage{hyperref}
\usepackage{threeparttable}
\usepackage{standalone}
\usepackage{balance}

\usepackage[linesnumbered,ruled,vlined]{algorithm2e}

\usepackage{bm}
\usepackage[utf8]{inputenc}
\usepackage{pgfplots}
\DeclareUnicodeCharacter{2212}{−}
\usepgfplotslibrary{groupplots,dateplot}
\usetikzlibrary{patterns,shapes.arrows}
\pgfplotsset{compat=newest}

\newcommand{\keystate}[1]{
\begin{tcolorbox}[leftrule=1mm,toprule=0mm,bottomrule=0mm,left=1pt,right=2pt,top=2pt,bottom=2pt]
\em #1
\end{tcolorbox}
}

{\noindent\begin{minipage}[c]{\linewidth}%
\begin{bclogo}[couleur=gray!30,%
                arrondi=0.1,%
                logo=\bclampe,%
                ombre=true]{\normalsize Key Characteristic} {#1}}%
{\end{bclogo}\end{minipage}\vspace{2mm}}

\def\BibTeX{{\rm B\kern-.05em{\sc i\kern-.025em b}\kern-.08em
T\kern-.1667em\lower.7ex\hbox{E}\kern-.125emX}}

\usepgfplotslibrary{fillbetween}

\pgfplotsset{
 /pgfplots/ybar legend/.style={
        /pgfplots/legend image code/.code={
            \draw [##1,/tikz/.cd] (0cm,-0.3em) rectangle (0.2em,0.4em);
        },
    },
}

\makeatletter
\tcbset{
    myhbox/.style 2 args={%
        enhanced, 
        breakable,
        colback=white,
        colframe=teal!80!black,
        attach boxed title to top left={yshift*=-\tcboxedtitleheight}, 
        title={#2},
        boxed title size=title,
        boxed title style={%
            sharp corners, 
            rounded corners=northwest, 
            colback=tcbcolframe, 
            boxrule=0pt,
        },
        underlay boxed title={%
            \path[fill=tcbcolframe] (title.south west)--(title.south east) 
                to[out=0, in=180] ([xshift=5mm]title.east)--
                (title.center-|frame.east)
                [rounded corners=\kvtcb@arc] |- 
                (frame.north) -| cycle; 
        },
        #1
    }
}   
\makeatother

\newtcolorbox{myhbox}[2][]{%
    myhbox={#1}{#2}
}

\definecolor{mycolor}{rgb}{0.122, 0.435, 0.698}

\newtcbox{\mytag}{nobeforeafter,colframe=mycolor,colback=mycolor!30!white,boxrule=0.7pt,arc=0pt,
 boxsep=-3pt,left=6pt,right=6pt,top=4pt,bottom=5pt,tcbox raise base}

\newcommand{\approach}{\texttt{CausalFT}}

\pgfplotsset{compat=1.11,
    /pgfplots/xbar legend/.style={
    /pgfplots/legend image code/.code={%
       \draw[##1,/tikz/.cd,yshift=-0.25em]
        (0cm,0cm) rectangle (3pt,0.8em);},
   },
}

\setcopyright{rightsretained}
\acmJournal{TOSEM}
\acmYear{2025} \acmVolume{1} \acmNumber{1} \acmArticle{1} \acmMonth{1}\acmDOI{}

\begin{document}
\title{Causally Perturbed Fairness Testing}
\author{Chengwen Du}
\email{cxd394@student.bham.ac.uk}
\affiliation{
  \institution{IDEAS Lab, University of Birmingham}
  \city{Birmingham}
  \country{UK}
 }

\author{Tao Chen}
\authornote{Corresponding author: Tao Chen, t.chen@bham.ac.uk.}
\email{t.chen@bham.ac.uk}
\affiliation{
  \institution{IDEAS Lab, University of Birmingham}
  \city{Birmingham}
  \country{UK}
}

\input{abstract}

\ccsdesc[500]{Software and its engineering~Search-based software engineering}
\ccsdesc[500]{Software and its engineering~Software creation and management}
\ccsdesc[500]{Computing methodologies~Machine learning}

\keywords{Fairness Testing, AI/DNN Testing, Software Engineering for AI}

\maketitle

\input{introduction}

\input{background}
\input{methodology}
\input{study}

\input{results}
\input{discussion}

\input{threats_to_validity}

\input{related_work}

\input{conclusion}

\begin{acks}
This work was supported by an NSFC Grant (62372084) and a UKRI Grant (10054084).
\end{acks}

\balance
\bibliographystyle{ACM-Reference-Format}
\bibliography{references}

\end{document}

%% file: abstract.tex
\begin{abstract}

To mitigate unfair and unethical discrimination over sensitive features (e.g., gender, age, or race), fairness testing plays an integral role in engineering systems that leverage AI models to handle tabular data. A key challenge therein is how to effectively reveal fairness bugs under an intractable sample size using perturbation. Much current work has been focusing on designing the test sample generators, ignoring the valuable knowledge about data characteristics that can help guide the perturbation and hence limiting their full potential. In this paper, we seek to bridge such a gap by proposing a generic framework of causally perturbed fairness testing, dubbed \approach. Through causal inference, the key idea of \approach~is to extract the most directly and causally relevant non-sensitive feature to its sensitive counterpart, which can jointly influence the prediction of the label. Such a causal relationship is then seamlessly injected into the perturbation to guide a test sample generator. Unlike existing generator-level work, \approach~serves as a higher-level framework that can be paired with diverse base generators. Extensive experiments on $1296$ cases confirm that \approach~can considerably improve arbitrary base generators in revealing fairness bugs over $93\%$ of the cases with acceptable extra runtime overhead. Compared with a state-of-the-art approach that ranks the non-sensitive features solely based on correlation, \approach~performs significantly better on $64\%$ cases while being much more efficient. Further, \approach~can better improve bias resilience in nearly all cases.

\end{abstract}

%% file: introduction.tex
\section{Introduction}

In 2019, it was reported that the learned model, used by US hospitals to predict which patients needed extra medical care, always decided that black patients are more likely to pay for active interventions like emergency hospital visits---despite showing signs of uncontrolled illnesses \cite{doi:10.1126/science.aax2342}. Similarly, it has been revealed that 75\% of the current sentiment analysis models show clear bias against women, exacerbating social inequity~\cite{DBLP:journals/expert/DuYZH21,DBLP:conf/starsem/KiritchenkoM18}. 
Indeed, many software systems are increasingly leveraging the power of 
AI techniques, including machine/deep learning models, for making data-driven predictions and decisions in various domains with a large amount of tabular data, such as health care, criminal justice, civil service~\cite{tramer2017fairtest,chan2018hiring,berk2021fairness}, and software engineering~\cite{DBLP:conf/icse/XiangChen26,DBLP:journals/pacmse/Gong024,DBLP:conf/sigsoft/Gong023,gong2024dividable}. However, as shown in the above examples, an undesired property of AI systems is that they can make discriminatory, biased, and unfair predictions, leading to severe societal impacts~\cite{DBLP:journals/tosem/ChenZHHS24}.


Discrimination is often defined with respect to some sensitive features, such as age, race, gender, etc, against other non-sensitive counterparts~\cite{DBLP:conf/aies/Fleisher21,DBLP:conf/sigsoft/GalhotraBM17,DBLP:conf/sigsoft/AggarwalLNDS19}. What features are sensitive is domain-dependent and it is known in advance~\cite{DBLP:conf/icse/ZhangW0D0WDD20}. Intuitively, discrimination (or fairness bug) happens when an AI system/model tends to make different decisions/predictions for distinct individuals
(individual fairness~\cite{DBLP:conf/icse/ZhengCD0CJW0C22}) or subgroups (group fairness~\cite{DBLP:conf/icse/ZhangH21}), which only differ on the values of the sensitive feature(s). 

The fairness bugs related to unwanted discrimination are deeply hidden in AI systems. The reasons for that can be vast, for example, the data samples used to train the AI model might be unfair and biased in the first place, or the model's hyperparameters have not been well-tuned~\cite{DBLP:conf/icse/ChenChen26,DBLP:conf/icse/XiongChen25,DBLP:journals/corr/abs-2112-07303,DBLP:conf/sigsoft/0001L21,DBLP:conf/sigsoft/0001L24} to mitigate fairness bugs~\cite{DBLP:conf/esem/DuChen24}. All of these factors can unintentionally introduce fairness bugs in the trained AI system. However, finding those fairness bugs is complex, because (1) the full explainability of the AI model remains an open problem~\cite{DBLP:conf/icse/ZhangW0D0WDD20}; and (2) the possible input samples that the AI model can take are often intractable. For instance, the \textsc{Kdd} dataset has 19 features with $4.13\times 10^{15}$ possible samples~\cite{misc_census-income_(kdd)_117}.

To that end, fairness testing becomes a crucial step for engineering AI systems that process tabular data. The goal is to find as many individual discriminatory instances\footnote{In this work, we use instance and sample interchangeably.} as possible~\cite{DBLP:conf/icse/ZhangW0D0WDD20,DBLP:conf/issta/ZhangZZ21,DBLP:conf/sigsoft/AggarwalLNDS19}, each of which represents a fairness bug that demonstrates the existence of discrimination. Indeed, over the past few years, several test sample generators, including white-box~\cite{DBLP:conf/icse/ZhangW0D0WDD20,DBLP:conf/issta/ZhangZZ21,DBLP:conf/icse/ZhengCD0CJW0C22} and black-box~\cite{DBLP:conf/sigsoft/AggarwalLNDS19,DBLP:conf/icse/0002WJY022,DBLP:conf/icics/JiangSLWSG23} ones, have been proposed for automatically perturbing the testing data, generating new, unforeseen test samples that can reveal hidden fairness bugs in AI systems. Yet, despite the recent advancements in fairness testing, existing generators have mostly relied on purely randomized perturbation around (or guided by) the sensitive feature, i.e., randomly searching and exploring the sample space without additional guidance and information other than the known sensitive feature. As such, those generators have not considered the interrelations between sensitive features and the other non-sensitive counterparts, and how they can help the perturbation in fairness testing. Indeed, prior works, such as PC fairness \cite{DBLP:conf/nips/Wu0WT19}, have explored the interrelations between sensitive and non-sensitive features in fairness analysis. However, they mostly focus on fairness measurement rather than test case generation for fairness testing of AI systems.
Given the large search space to be perturbed in the fairness testing and the confirmed fact that some non-sensitive features can interact with the sensitive one to jointly influence the fairness of the AI model~\cite{DBLP:journals/tosem/ChenZHHS24,DBLP:journals/corr/abs-2104-14537}, missing such valuable information can inevitably limit the generator in finding fairness bugs, leaving its full potential untapped.

In this paper, we propose \approach, a generic framework for more effectively finding fairness bugs under the concept of causally perturbed fairness testing for AI systems that handle tabular data. The key novelty/idea is to extract the causal relationships between sensitive features and their non-sensitive counterparts, which are then injected into the perturbation to guide the test sample generation. What makes \approach~unique is that, unlike most existing work that focuses on the test sample generator~\cite{DBLP:conf/icse/ZhangW0D0WDD20,DBLP:conf/issta/ZhangZZ21,DBLP:conf/icse/ZhengCD0CJW0C22,DBLP:conf/sigsoft/AggarwalLNDS19,DBLP:conf/icse/0002WJY022,DBLP:conf/icics/JiangSLWSG23}, \approach~serves as a higher-level, generator-agnostic framework that can be seamlessly paired with different base generators to enable causal perturbation, including white- and black-box ones. Notably, our contributions are:


\begin{itemize}
    \item By leveraging the notion of causal inference, we build a causal graph from the training data and extract all directly and causally relevant non-sensitive features to the concerned sensitive feature that affects the class label.

    \item We then compute the causal effect between all non-sensitive features identified and the concerned sensitive feature, from which we select the non-sensitive feature with the highest causal effects, i.e., the one that is the most causally relevant to the sensitive feature and jointly influences the class label.

    \item Agnostic to the test sample generator, we inject the extracted causal relationship into a given generator for guiding how to perturb, forming a transformed causal perturbation, in two ways:
        \begin{itemize}
            \item The definition of individual discriminatory instances is relaxed to consider both the concerned sensitive feature and its most causally relevant non-sensitive counterpart as the source of discrimination rather than the concerned sensitive feature only.
            \item The perturbation takes place on the features other than the identified sensitive feature and the non-sensitive counterpart.
        \end{itemize}

    \item \approach~is experimentally evaluated on eight datasets with $2$-$3$ sensitive features, across four AI models, six generators, and three fairness metrics, leading to 1296 cases of investigation.
\end{itemize}

The evaluation reveals encouraging results: with acceptable extra overhead ($\approx 270$ seconds in the worst case), \approach~considerably improves a given generator in $1209$ out of the $1296$ cases ($93\%$), finding up to $420$ more fairness bugs; it also performs much better than correlation-ranked non-sensitive features over $831$ out of the $1296$ cases ($64\%$), together with $34\%$ cases of similar results, while making the AI system/model more robust to bias in nearly all cases.

To promote open science, all artifacts can be found at our repository: \textcolor{blue}{\texttt{\url{https://github.com/ideas-labo/causalft}}}.

 


The remainder of the paper is organized as below: Section~\ref{sec:pre} introduces the preliminaries and the observations that motivate our work. Section~\ref{sec:methodology} elaborates on the designs of \approach. Section~\ref{sec:exp} presents the experiment setup and Section~\ref{sec:results} analyzes the results, followed by a discussion in Section~\ref{sec:dis}. Sections~\ref{sec:tov},~\ref{sec:related}, and~\ref{sec:conclusion} present the threats to validity, related work, and conclusion, respectively.

%% file: background.tex
\section{Preliminaries}
\label{sec:pre}

\subsection{Model Fairness}


A fairness bug refers to any imperfection in an AI system that causes a discordance between the existing and required fairness conditions~\cite{DBLP:journals/tosem/ChenZHHS24}, which is relevant to two concepts:

\begin{itemize}
    \item \textbf{Sensitive feature:} The feature in the dataset that is known to be legally or ethically protected as it could influence outcomes in a way that leads to discrimination~\cite{DBLP:conf/ijcai/XuWYZW19}. Common examples are \textit{Gender}, \textit{Age}, and \textit{Race}.
    \item \textbf{Non-sensitive feature:} All remaining unprotected features in the dataset would be non-sensitive features~\cite{DBLP:conf/icse/ZhangH21}.
\end{itemize}

Fundamentally, an AI model could have a fairness issue if its prediction outcome could be solely influenced by different values of the sensitive feature. For example, assuming that an AI model is trained to predict whether a person has income above \$50K per year using features, such as $\{Age, Relationship,Workclass,Occupation\}$. Suppose that, in the discretized representation, a sample can be represented as $\{23,1,3,7\}$ for which the model predicts the label as false. Now, if the model predicts true for a new input sample, i.e., $\{33,1,3,7\}$, created from the original one by only changing the \textit{Age} from $23$ to $33$, then it means there is an unfair discrimination. Here, we say the sample $\{23,1,3,7\}$ is an \textbf{individual discriminatory instance}, which reflects a fairness bug.

\subsection{Fairness Testing and Problem Formulation}


Given a trained model, the problem of fairness testing is to generate new, unforeseen individual discriminatory instances for a concerned sensitive feature\footnote{In fairness testing, there is often only one concerned sensitive feature each time even if multiple sensitive features exist in the dataset~\cite{DBLP:conf/icse/ZhangW0D0WDD20,DBLP:conf/issta/ZhangZZ21,DBLP:conf/sigsoft/AggarwalLNDS19}.} that can reveal fairness bugs in the AI system under test, deriving from some testing data. This is challenging because there are too many possible samples to explore, e.g., for the datasets considered in this work, the size of unique samples ranges from $9.2\times 10^6$ to $4.13\times10^{15}$.

As such, an automatic test generator is often a search algorithm as part of the field of Search-based Software Engineering~\cite{DBLP:journals/tosem/ChenL23,DBLP:journals/tse/LiCY22}, wherein the key is to pair two samples, from some testing data, that only differ on the concerned sensitive feature and randomly perturb all non-sensitive features (or sometimes any features) on both samples as guided by some fitness, hoping that such a perturbation can find more real individual discriminatory instances that would actually cause the trained model to predict the different outcome when changing solely on the sensitive feature. For example, suppose that the feature at index 0 is the concerned sensitive feature. Then, the initial sample pair $\mathbf{x}_a$ and $\mathbf{x}_b$ on the left might lead to the pair on the right after the perturbation: 
\begin{equation}
   \begin{split}
       &\mathbf{x}_a = \{0,7,4,5,1\} \Longrightarrow \mathbf{x'}_a = \{0,\textcolor{blue}{2},\textcolor{blue}{8},5,\textcolor{blue}{3}\}\\
       &\mathbf{x}_b = \{1,7,4,5,1\} \Longrightarrow \mathbf{x'}_b = \{1,\textcolor{blue}{2},\textcolor{blue}{8},5,\textcolor{blue}{3}\}
   \end{split}  
\end{equation}
If both pairs can cause the trained model to generate different outcomes, then we have four individual discriminatory instances. Our goal of fairness testing here is to find as many (unique) individual discriminatory instances as possible.

Two types of metrics exist to evaluate the capability of test generators in finding the fairness bugs in a trained model: individual fairness~\cite{DBLP:journals/corr/abs-2310-19391,DBLP:conf/icse/ZhengCD0CJW0C22} and group fairness~\cite{DBLP:conf/icse/ZhangH21,DBLP:journals/access/LiHGY21,DBLP:conf/fat/MashiatGRFD22,DBLP:journals/corr/abs-2402-02663,DBLP:conf/iclr/JiangHFYMH22,DBLP:journals/corr/abs-2303-08040,DBLP:journals/corr/abs-2209-14670,DBLP:journals/corr/abs-2304-09779}. The former is mainly relevant to the number of individual discriminatory instances in the generated samples; while the latter means a fair outcome demands the existence of parity between different feature groups by, e.g., \textit{gender} or \textit{race}. It has been shown that individual fairness can be correlated with the group fairness counterpart~\cite{DBLP:conf/fat/Binns20,DBLP:conf/icde/LahotiGW19}.

\begin{figure}[t!]
    \centering
    \begin{subfigure}[t!]{0.32\columnwidth}
        \centering
        \includegraphics[width=\columnwidth]{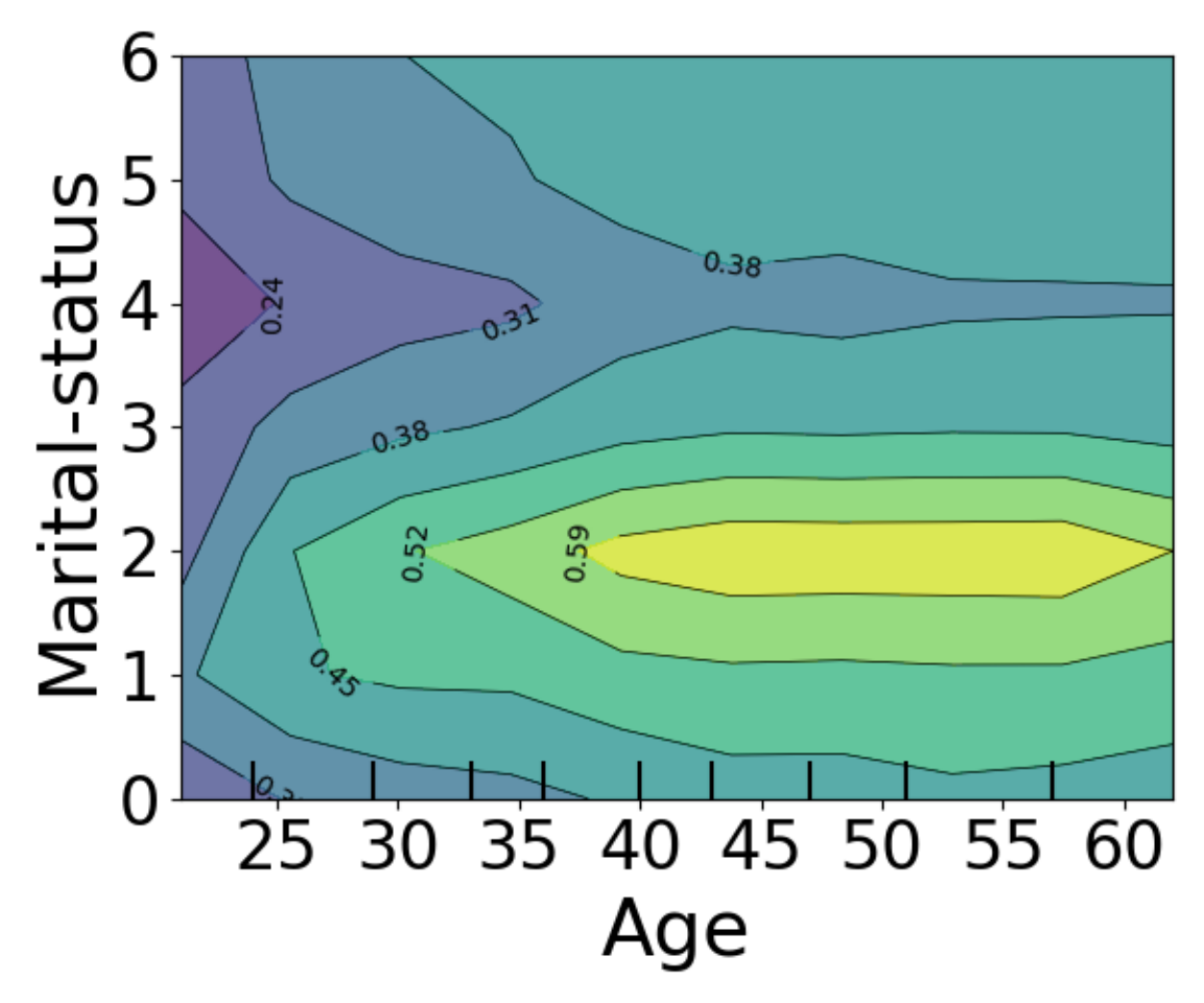}
\subcaption{\textit{Age} and \textit{Marital-status}}
    \end{subfigure}
      \hfill
    \begin{subfigure}[t!]{0.32\columnwidth}
        \centering        
        \includegraphics[width=\columnwidth]{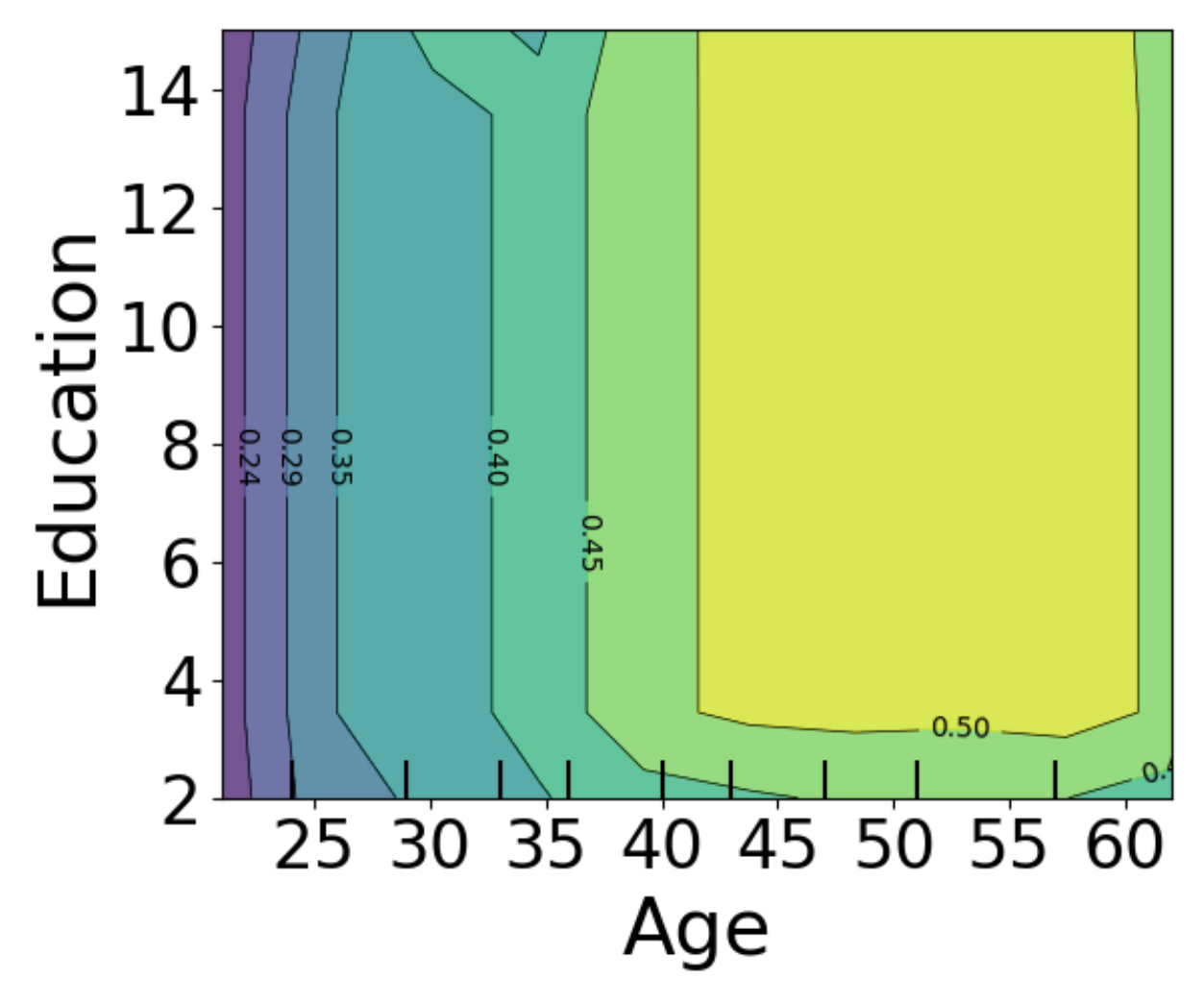}   
        \subcaption{\textit{Age} and \textit{Education}}
    \end{subfigure}
    \hfill
     \begin{subfigure}[t!]{0.32\columnwidth}
        \centering        
        \includegraphics[width=\columnwidth]{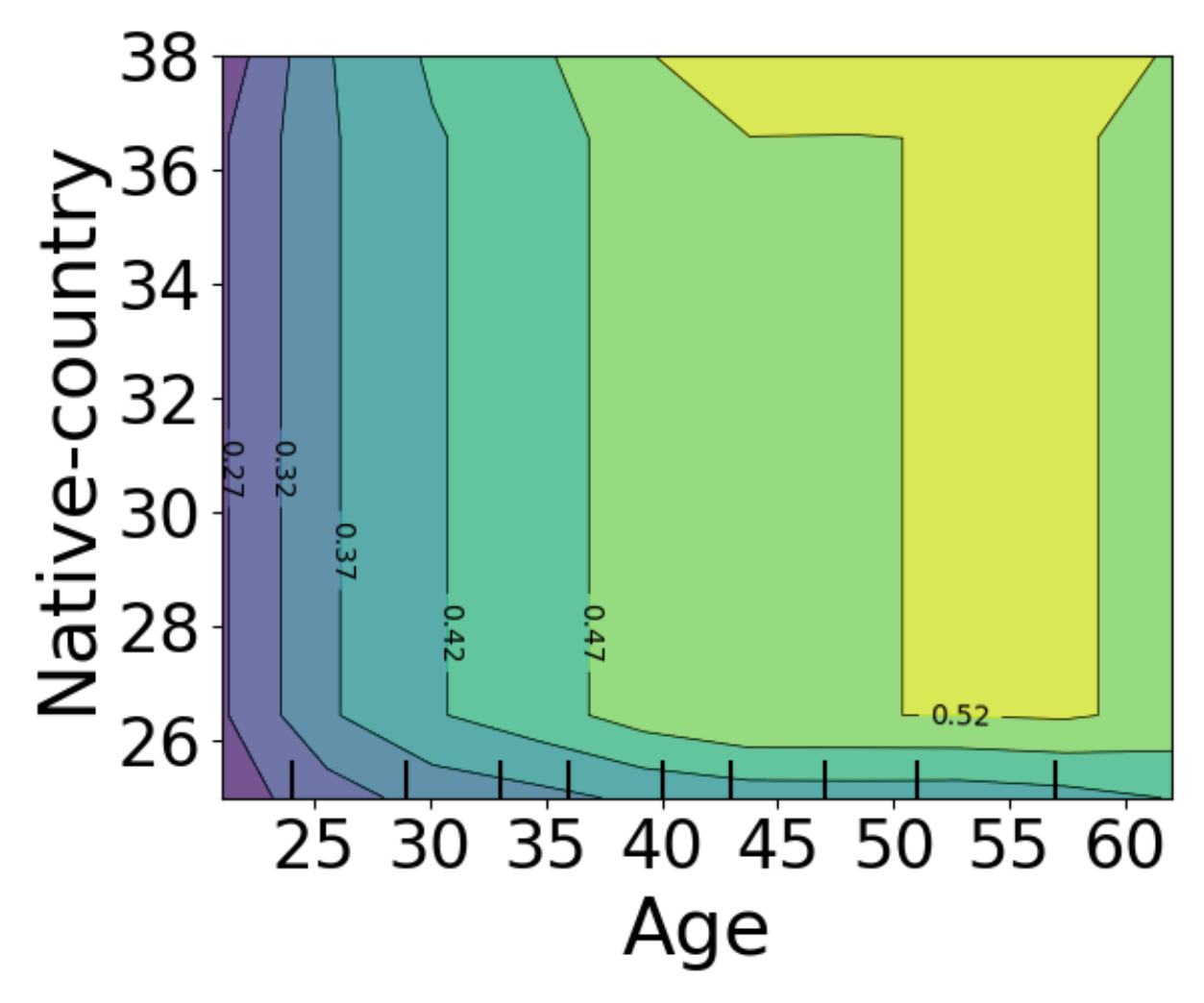} 
        \subcaption{\textit{Age} and \textit{Native-country}}
    \end{subfigure}
  
     \begin{subfigure}[t!]{0.32\columnwidth}
        \centering        
        \includegraphics[width=\columnwidth]{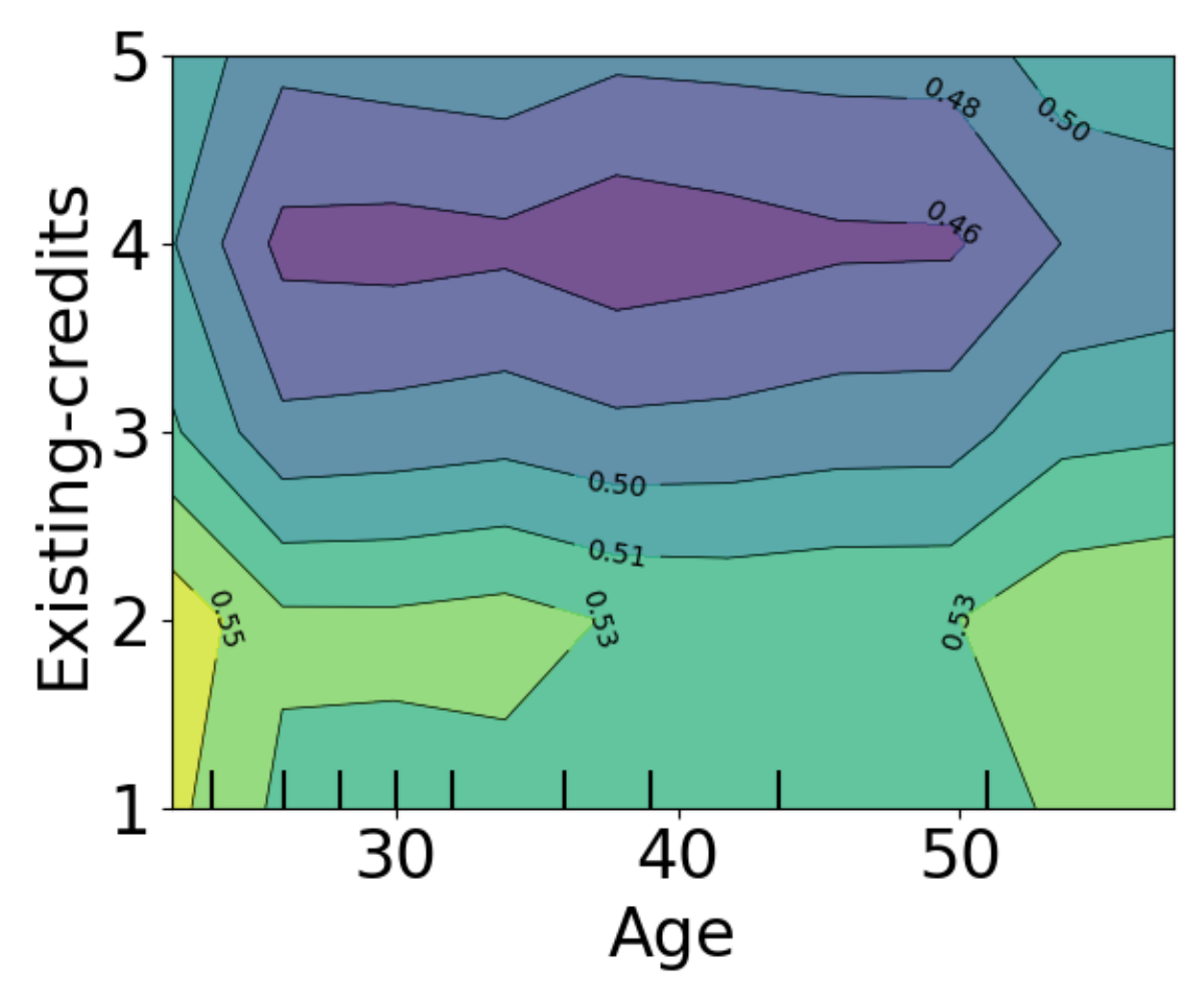} 
        \subcaption{\textit{Age} and \textit{Existing-credits}}
    \end{subfigure}
 \hfill
   \begin{subfigure}[t!]{0.32\columnwidth}
        \centering        
        \includegraphics[width=\columnwidth]{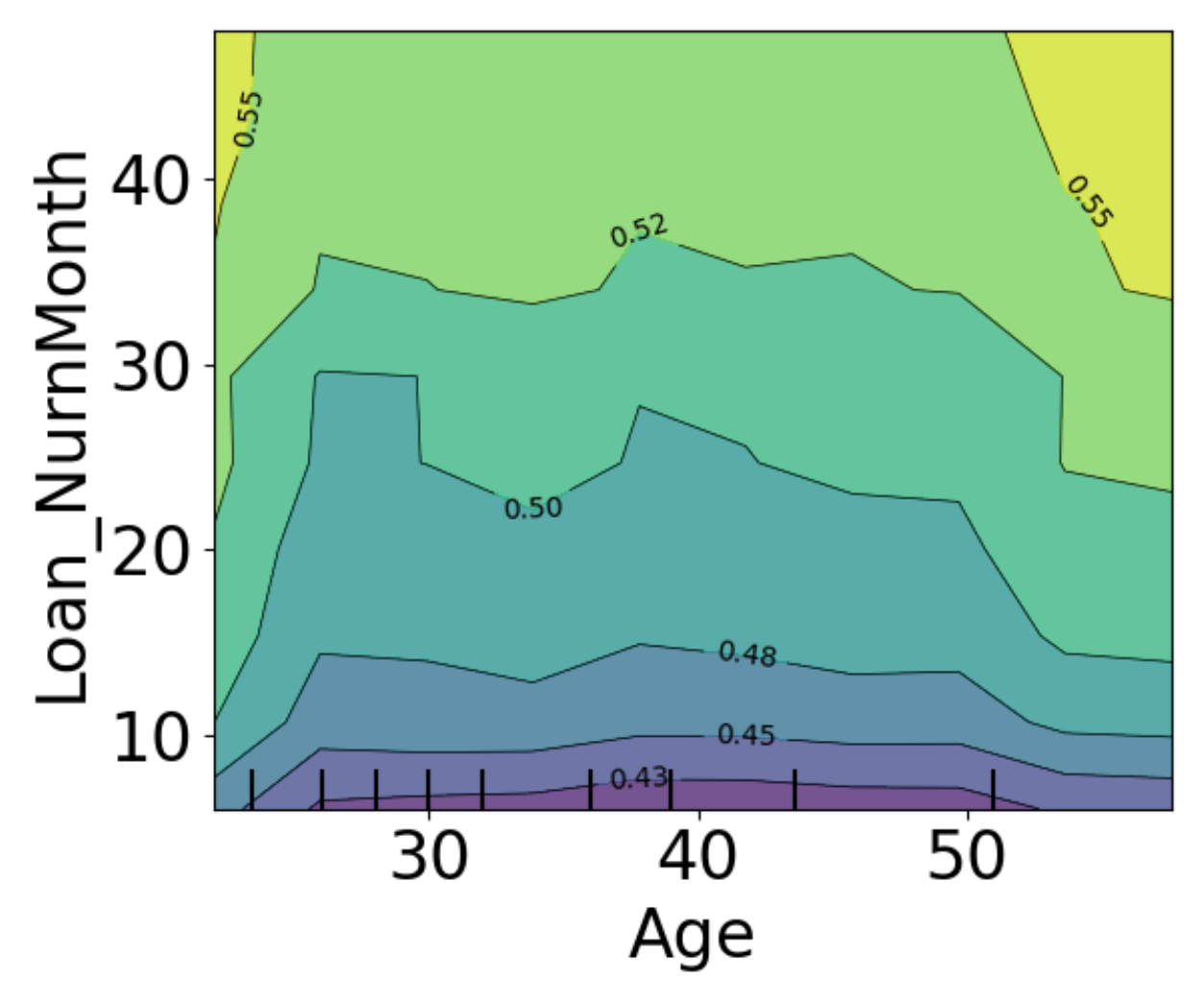} 
        \subcaption{\textit{Age} and \textit{Loan\_NumMonth}}
    \end{subfigure}
 \hfill
       \begin{subfigure}[t!]{0.32\columnwidth}
        \centering        
        \includegraphics[width=\columnwidth]{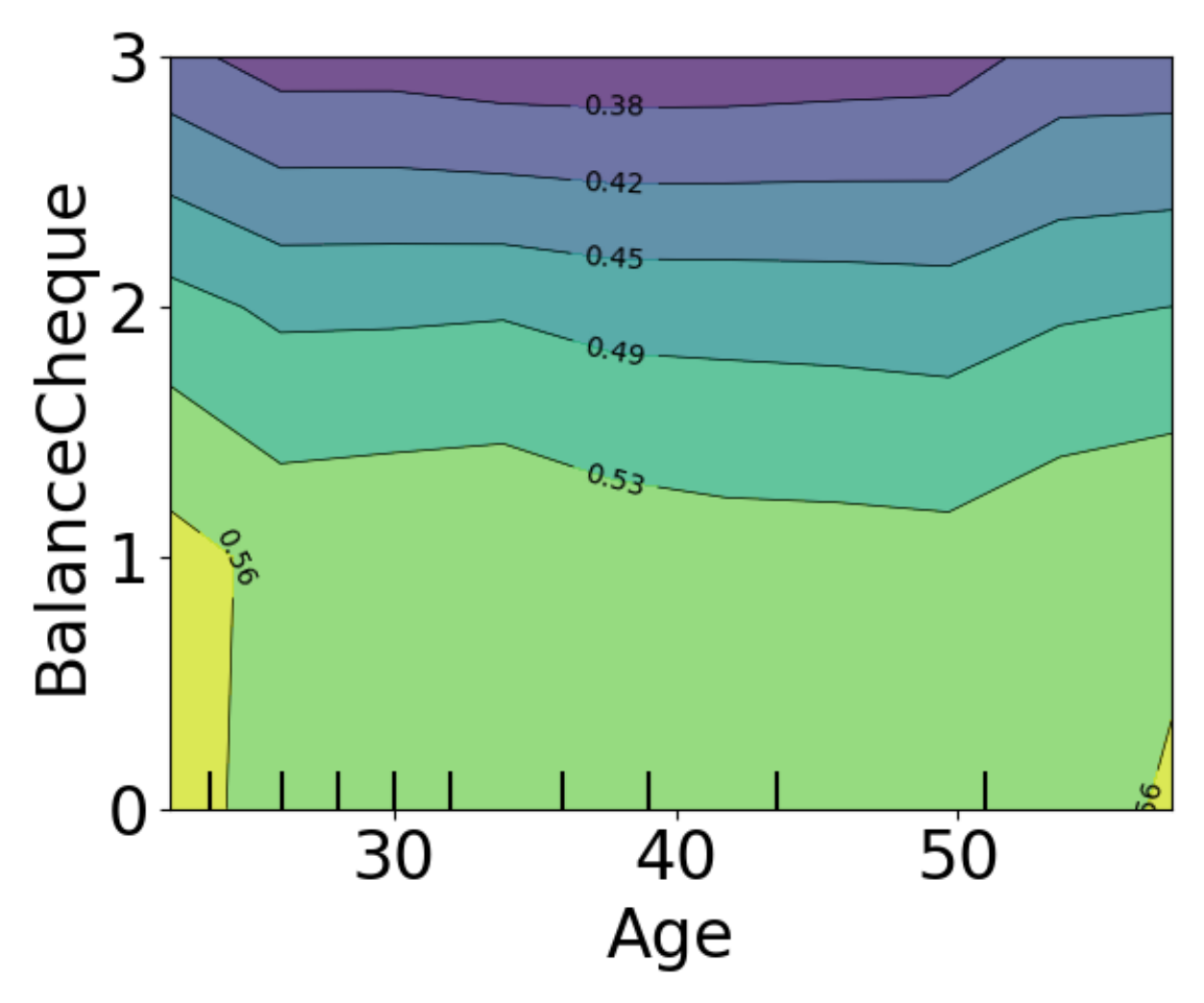} 
        \subcaption{\textit{Age} and \textit{BalanceCheque}}
    \end{subfigure}

         \begin{subfigure}[t!]{0.32\columnwidth}
        \centering        
        \includegraphics[width=\columnwidth]{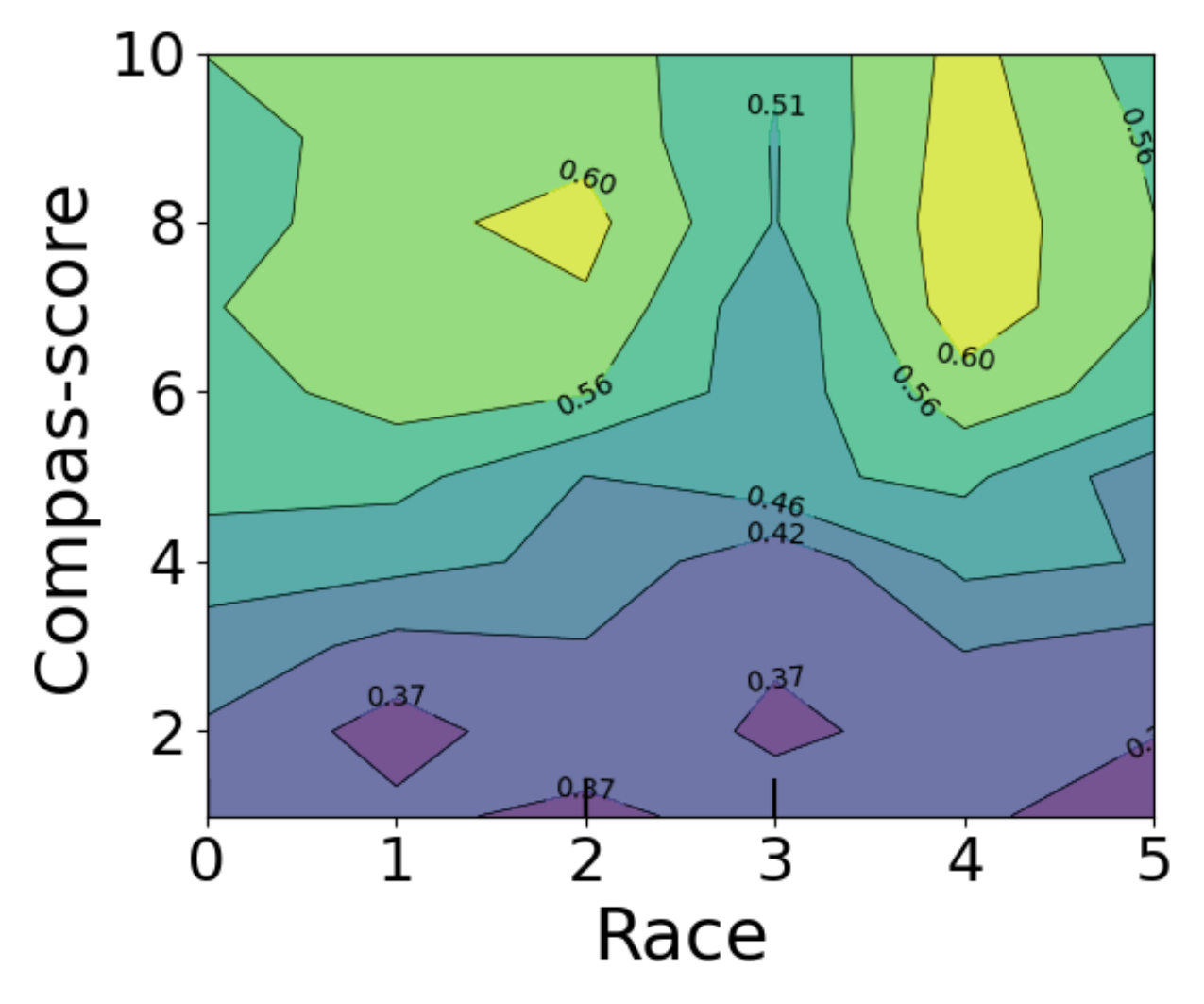} 
        \subcaption{\textit{Race} and \textit{Compas-score}}
    \end{subfigure}
 \hfill
   \begin{subfigure}[t!]{0.32\columnwidth}
        \centering        
        \includegraphics[width=\columnwidth]{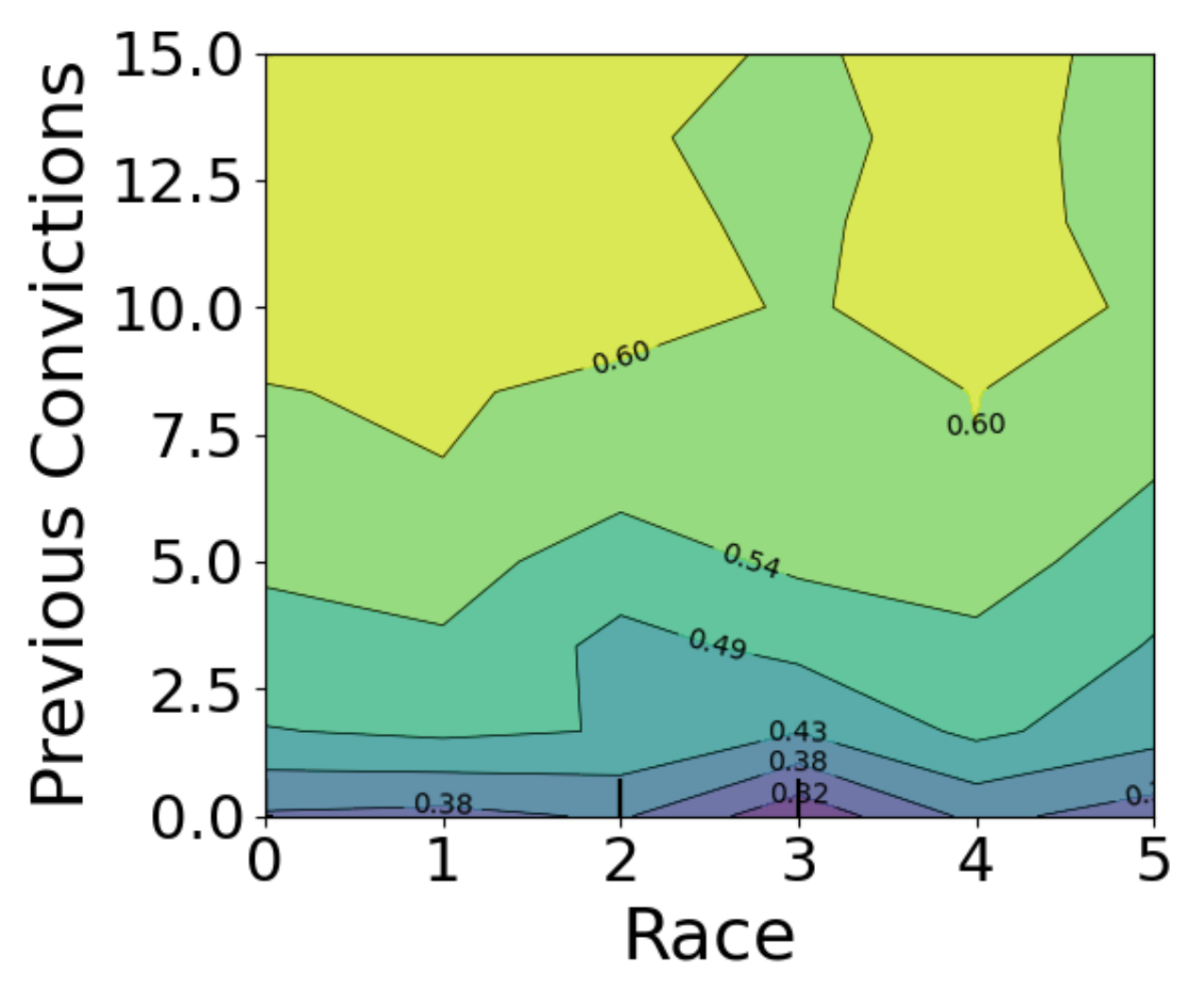} 
        \subcaption{\textit{Race} and \textit{Pre. Convictions} }
    \end{subfigure}
 \hfill
       \begin{subfigure}[t!]{0.32\columnwidth}
        \centering        
        \includegraphics[width=\columnwidth]{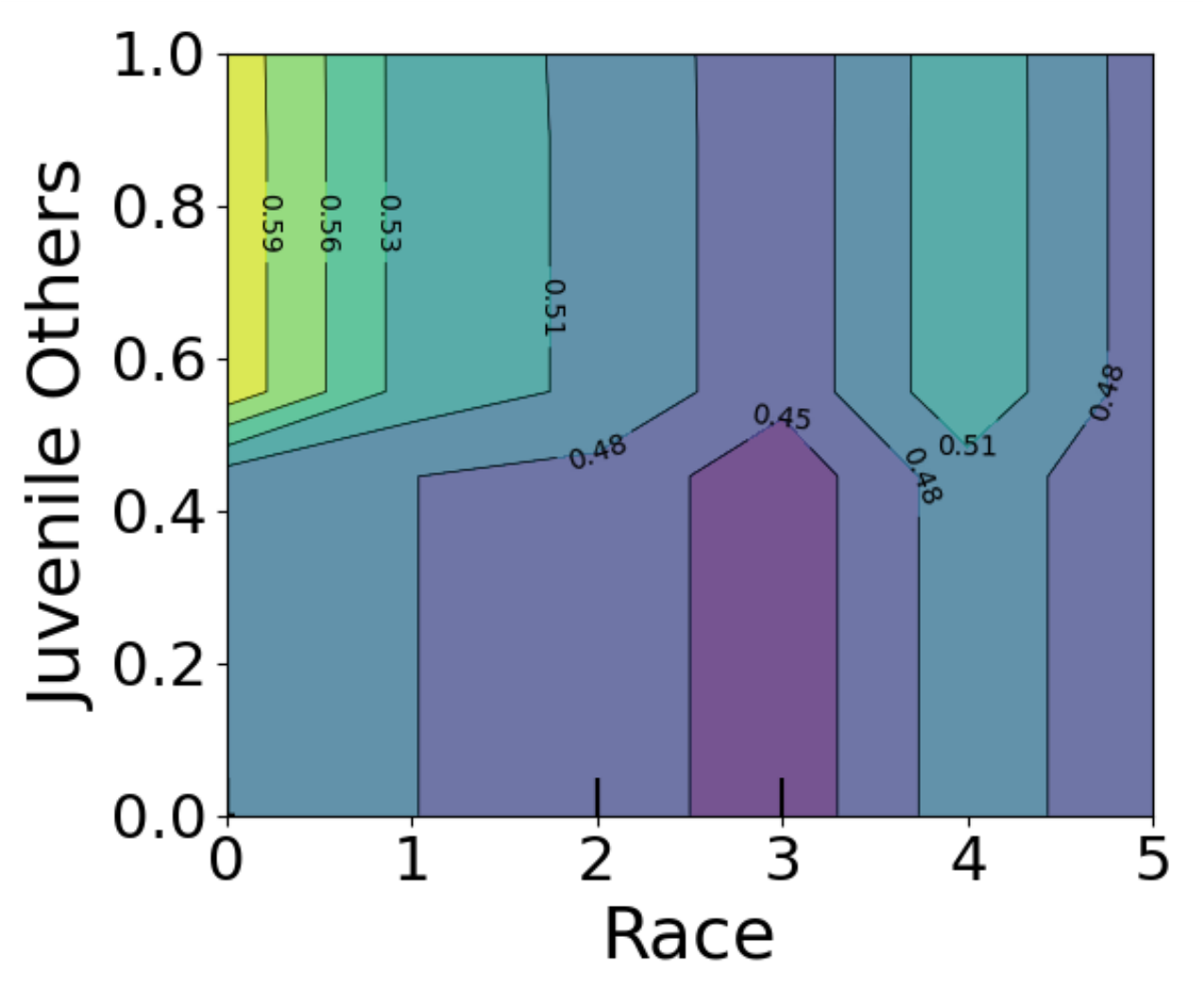} 
        \subcaption{\textit{Race} and \textit{Juvenile Others}}
    \end{subfigure}

    \caption{Exampled partial dependence plots for the most significant interactions between the sensitive and non-sensitive features in DNN prediction. (a), (b), and (c) are the \textsc{Adult} dataset when \textit{Age} is the sensitive feature;  (d), (e), and (f) are the \textsc{German} dataset when \textit{Age} is the sensitive feature;  (g), (h), and (i) are the \textsc{Compas} dataset when \textit{Race} is the sensitive feature. The numbers in the figures denote the inferred probability belongs to the positive class.}
    \label{fig:feature}
\end{figure}

\subsection{Observations and Motivation}


To understand how discrimination instances can easily cause AI systems to produce unfair outcomes, we examine the trained AI model and make predictions over several real-world tabular datasets. We randomly generate samples to test an AI system that contains a Deep Neural Network (DNN), and identify those that can reveal fairness bugs. As shown in Figure~\ref{fig:feature}, we observe the following patterns:

\begin{itemize}
    \item \textbf{Observation 1:} For a given sensitive feature, there exists a non-sensitive feature that is far more commonly interacting with it to create discrimination instances. For example, when the \textit{Age} is the sensitive feature for the \textsc{Adult} dataset, we see that there are more complex interactions between \textit{Age} and \textit{Marital-status}, which affect the decisions made by the model. In contrast, other non-sensitive features like \textit{Education} are more independent of the changes of \textit{Age} when influencing the AI model's prediction, as the probability boundary largely varies according to the values of \textit{Education} but not \textit{Age}.
    \item \textbf{Observation 2:} The non-sensitive feature that causes significant interactions with the sensitive counterpart to influence the prediction is uncertain. For example, even for the same sensitive feature \textit{Age}, it can create considerable interactions with varying non-sensitive features on \textsc{Adult} and \textsc{German} datasets, even though they share many common features.
\end{itemize}

The above makes sense in the real world, because, for example, in the \textsc{Adult} dataset, \textit{Age} can more considerably interact with the \textit{Marital-status} to determine the AI model's prediction: younger individuals are more likely to be single or in non-married relationships while older individuals are more likely to have other marital statuses \cite{DBLP:conf/fat/Binns18}.

Those observations lead to the following key insight when testing the fairness of AI systems:

\begin{tcbitemize}[%
    raster columns=1, 
    raster rows=1
    ]
  \tcbitem[myhbox={}{Key Insight}]   \textit{The strong interactions between a sensitive feature and its non-sensitive counterpart should be explicitly handled when testing the fairness of AI systems, as they serve as the key to determining model prediction, leading to discriminatory instances.}
\end{tcbitemize}

However, existing generators for fairness testing have failed to handle the above characteristics, since they typically perturb features in a random or heuristic manner without considering the interactions between a sensitive feature and its non-sensitive counterparts \cite{DBLP:conf/icse/ZhangW0D0WDD20,DBLP:conf/issta/ZhangZZ21,DBLP:conf/icse/ZhengCD0CJW0C22,DBLP:conf/sigsoft/AggarwalLNDS19,DBLP:conf/icse/0002WJY022,DBLP:conf/icics/JiangSLWSG23,DBLP:conf/sigsoft/GalhotraBM17,DBLP:conf/kbse/UdeshiAC18}. As such, existing generators generally treat all non-sensitive features as equally important in the perturbation. For example, when \textit{Age} is the sensitive feature for the \textsc{Adult} dataset, perturbing both \textit{Marital-status} and \textit{Education} with equal chance is clearly not ideal, as the former can more significantly interact with \textit{Age} to change the prediction outcome. This can cause several devastating consequences:


\begin{itemize}
    \item ineffective exploration of the most probable bias-induced regions in the input space;
    \item weak exploitation of the found fairness bugs, as the samples in the generated pairs are never re-paired with the other samples;
    \item wasting computational resources since those less influential non-sensitive features with limited interaction with the sensitive one can consume the majority of the computation.
\end{itemize}

All of the above motivate our work: we seek to overcome those limitations of AI system fairness testing by explicitly taking the newly discovered observations into account.

%% file: methodology.tex
\section{Fairness Testing with Causal Perturbation}
\label{sec:methodology}

\begin{figure}
    \centering
    \includegraphics[width=0.9\columnwidth]{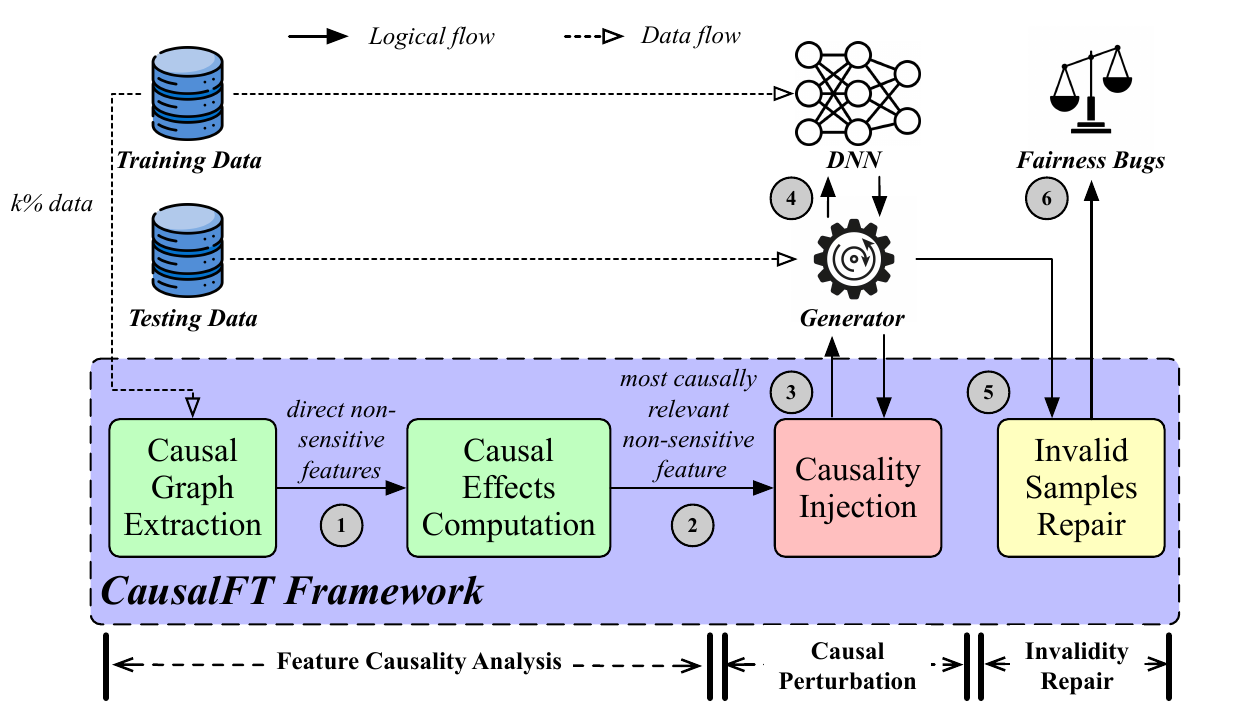}
    \caption{Workflow of \approach~for fairness testing.}
    \label{fig:overview}
\end{figure}

Since the perturbation is crucial for test sample generation in fairness testing, the ultimate goal of \approach~is to improve the effectiveness of perturbation by exploiting the learned causal relationships between sensitive and non-sensitive features. The key idea is that diversifying those non-sensitive features, which are more causally correlated with the concerned sensitive ones, as the starting point of the perturbation while keeping such diversity would be more likely to help find individual discriminatory instances that reveal fairness bugs. This is because those non-sensitive features and the sensitive counterpart have a larger chance to jointly influence the prediction of an AI system/model, and hence it is important to keep them diverse without being affected too much by the perturbation. \approach~aims to identify those non-sensitive features and explicitly handle them during perturbation.

As with existing work~\cite{DBLP:conf/icse/ZhangW0D0WDD20,DBLP:conf/issta/ZhangZZ21,DBLP:conf/sigsoft/AggarwalLNDS19}, we assume that there is only one concerned sensitive feature each time, known in advance. The benchmark dataset is split into training and testing data. The former trains an AI model and is used by the causal analysis in \approach~while the latter is used in the perturbation/generation process. We follow the ``rule-of-thumb'' that 70\% of a dataset is used for training while the remaining 30\% is used to seed the generator~\cite{DBLP:conf/kdd/PedreschiRT08}. From Figure~\ref{fig:overview} and Algorithm~\ref{alg:general_framework}, \approach~has three key phases:




\begin{itemize}
    \item \textbf{Feature Causality Analysis} (lines 1-3): Here, we seek to analyze the causal relationships between the concerned sensitive and other non-sensitive features, from which we pick the most causally relevant one to serve as part of the ``synthetic sensitive set'' along with the sensitive one.
    
    
    \item \textbf{Causal Perturbation} (lines 4-13): We inject the most causally relevant non-sensitive feature selected into the existing perturbation of a base generator. This transforms the generator in a way that its perturbation is guided by causal knowledge, diversifying the identified non-sensitive feature and its sensitive counterpart while reducing the space of perturbation.

    \item \textbf{Invalidity Repair} (lines 14-16): When possible, we find new seeded samples from the testing data to pair with samples in an invalid pair.

\end{itemize}

As such, \approach~serves as a generic, generator-agnostic framework that can be seamlessly paired with different perturbation-based generators for fairness testing without requiring substantial modification. In what follows, we will articulate the key designs behind \approach~in detail.




\input{Table/code}
\subsection{Feature Causality Analysis}
\label{sec:fca}

\subsubsection{Causal Graph Extraction}
\label{sec:cg-model}

Using $k$\% of the training data (default to $k=100$), a major step in \approach~is to build a causal model therein for fairness testing, which can be represented as a directed acyclic graph (DAG) where each node can be a feature ($f_i$) or label ($y$) while a path $f_i \rightarrow f_j$ is the causal implication from $f_i$ to $f_j$. Since the causal relationship is unidirectional, we fix a concerned sensitive feature ($f_s$) as the starting node and the label as the ending node, but we permit several possible intermediate features within their causal path, e.g., $f_s \rightarrow f_i \rightarrow y$. All those paths will form a DAG, and our goal is to identify which are the non-sensitive features that are causally affected by changing the value of a sensitive feature, which would then influence the label in the dataset. It is worth noting that, for a dataset, we observe minimal change in the obtained DAG even when using different sensitive features as the starting node. Indeed, there are several methods/models that can be used for the above purpose. In \approach, we leverage LiNGAM~\cite{DBLP:journals/jmlr/ShimizuISHKWHB11} to build the causal graph\footnote{We used LiNGAM under label encoding of our discrete data, for which it has been proved to preserve the same assumption~\cite{DBLP:journals/jmlr/ShimizuISHKWHB11}.}, because:

\begin{itemize}
    \item It has proven effectiveness for a wide range of tasks~\cite{DBLP:journals/jmlr/ShimizuHHK06}.
    \item It is computationally efficient due to its linear and non-Gaussian model. Indeed, this might not generalize to all cases, but we found that it fits the datasets studied well.
    \item It also contains an effective pruning mechanism that reduces the complexity of the causal graph produced.
\end{itemize}

In a nutshell, LiNGAM only needs to take the dataset, starting node, and ending node as inputs, and it then computes the independent components involved by using an ICA model~\cite{DBLP:journals/spl/ErikssonK04}. The output is a linear, non-Gaussian, acyclic model that contains the causal path from the starting node to the ending one, including any intermediate nodes.

\begin{figure}
    \centering
    \includegraphics[width=0.7\columnwidth]{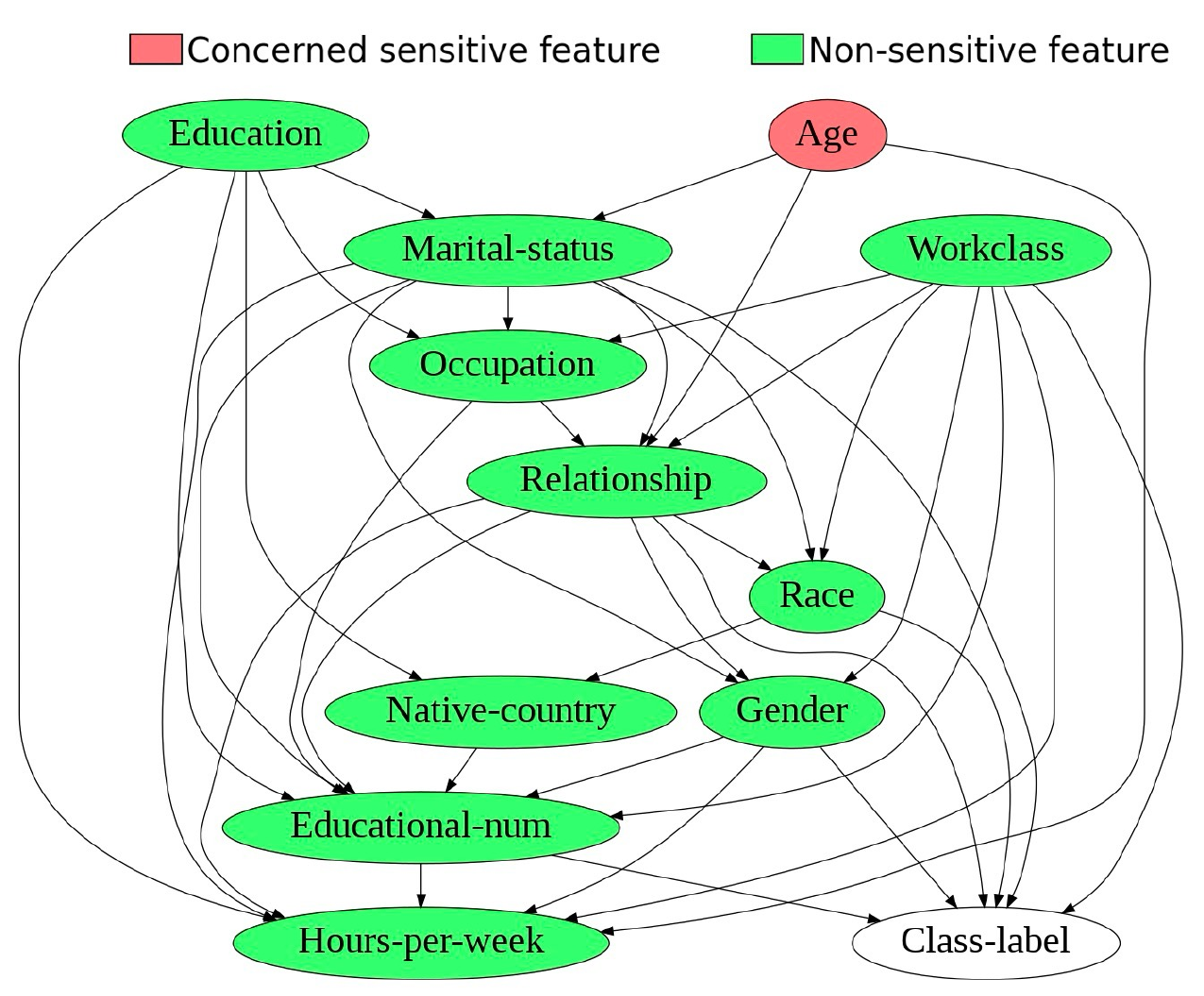}
    \caption{The causal graph built for the \textsc{Adult} dataset (\textit{Age} being the only concerned sensitive feature).}
    \label{fig:causal graph}
\end{figure}

Drawing on the produced/verified causal graph from LiNGAM, \approach~takes two more steps:

\begin{enumerate}
    \item Extract all causal paths starting from the concerned sensitive feature and ending at the label, e.g., $f_s \rightarrow f_i \rightarrow y$.
    \item Within those causal paths, identify all (unique) non-sensitive features that have a \textbf{direct} causal relationship to the sensitive one. For example, if we have a causal path $f_s \rightarrow f_i \rightarrow f_j \rightarrow y$, then only $f_i$ is of interest as $f_j$ exhibits an indirect relationship to $f_s$.
\end{enumerate}


We employ LiNGAM to construct a directed acyclic causal graph (DAG) and identify direct edges from the sensitive attribute to non-sensitive features. Only features with direct causal connections are selected for perturbation, ensuring causal validity. As shown in Figure 3, several causal paths can be extracted, such as \textit{Age} $\rightarrow$ \textit{Marital-status} $\rightarrow$ \textit{Gender} $\rightarrow$ \textit{Class-label} and \textit{Age} $\rightarrow$ \textit{Marital-status} $\rightarrow$ \textit{Class-label}. Thus, \textit{Age} has a direct causal edge to \textit{Marital-status}, while its influence on \textit{Gender} is indirect via \textit{Marital-status}. It is worth noting that the method that builds the causal graph in \approach~can be any off-the-shelf algorithm, hence LiNGAM can be easily replaced with an alternative solution when needed.

To validate the robustness of the causal graphs generated by LiNGAM, we compare it against other causal discovery algorithms, PC~\cite{DBLP:conf/nips/Wu0WT19} and GES~\cite{DBLP:journals/jmlr/Chickering02a}, on a bootstrap stability analysis using the \texttt{Tetrad} package with 1000 bootstraps. We compute the hamming distance on the edge's adjacency matrix of every pair of the graph generated and report the average distance as a measure of stability. We found that:
\begin{itemize}
    \item LiNGAM has 1.6 edge differences.
    \item PC has 2.1 edge differences.
    \item GES has 2.6 edge differences.
\end{itemize}
These results confirm that LiNGAM generates robust and reliable causal structures for \approach.

\subsubsection{Causal Effects Computation}

To compute the causal effects between the concerned sensitive feature $f_s$ and each of its direct causally relevant non-sensitive features $f_n$ as extracted from the causal graph, \approach~follows the steps below:

\begin{enumerate}
    \item Collect $m$ random samples from the $k$\% training data ($m=100$ in this work\footnote{
    Note that $m$ differs from the $k$ used to train/build the causal graph, but for computing the causal effects based on a given graph. Increasing or setting $m$ as $k\%$ of the training data would lead to little gains due to the bootstrapping, while the computational overhead increases exceptionally.}).
    \item Feed the $m$ samples as inputs into the causal graph built to estimate the probabilities of $f_s$ and $f_n$ on affecting the label $y$ under their possible values sampled.
    \item Calculate the causal effect from $f_s$ to $f_n$ on the class label using the \textit{do-calculus} of counterfactual cases via:
    \begin{equation}
    c(f_s,f_n) = 
    {{\theta} \over {n^2}} \sum_{\alpha \in V_{\alpha}} \sum_{\beta \in V_{\beta}} |p(y|do(f_s=\alpha)) - p(y|do(f_n=\beta))| 
    \end{equation}
    where $\alpha$ and $\beta$ are two values for $f_s$ and $f_n$ respectively, each chosen from their permitted set $V_{\alpha}$ and $V_{\beta}$; $\theta$ is the fitted coefficient between $f_s$ and $f_n$ from the causal graph, which represents the strength of the causal relationship between them.
    \item Repeat from 1) for 20 runs following a bootstrapping with replacement.

    \item Compute the median of the causal effect $c(f_s,f_n)$ over the 20 repeats, denoted as $c_n$.
\end{enumerate}


From the above, we can then rank every non-sensitive feature that is directly and causally relevant to the sensitive counterpart using $c_n$; the feature $f_c$ with the highest $c_n$ is what we are seeking. Indeed, it is possible to consider more than one non-sensitive feature. However, we use the most causally relevant non-sensitive feature from those directly related ones in \approach~because:
\begin{itemize}
    \item We discovered that there is always one non-sensitive feature with a significantly higher $c_n$ than the others.
    \item Including more non-sensitive features would have the risk of not being able to explore the search space sufficiently while producing too many invalid samples, as we will discuss in Section~\ref{sec:rq3}.
    \item If a non-sensitive feature is not causally relevant to the sensitive counterpart, then it makes less sense to correlate them, which might lead to misleading perturbation.
\end{itemize}

As a result, we consider only the most causally relevant non-sensitive feature to be a known sensitive one. For example, under the \textsc{Law School} dataset, when \textit{Race} is the concerned sensitive feature, among all its directly and causally relevant non-sensitive features, \textit{Lsat} has the highest causal effect of 4.75, which is significantly higher than the 2.29 of the second highest feature \textit{Decile3} following by the 2.11 of \textit{Decile1b}.

\subsection{Causal Perturbation}

\subsubsection{Common Parts of Perturbation}

Since current test sample generators essentially resemble a search process for finding individual discriminatory instances via perturbing the samples in the testing dataset, there are two important designs therein:

\begin{enumerate}
    \item \textbf{How to define individual discriminatory instances in the perturbation?} This directly influences how a pair can be formed and the direction of the perturbation. 
    
    
    \item \textbf{What features to perturb?} For each pair of samples, this determines the features to be changed during perturbation\footnote{Note that perturbation changes one feature each time and often it changes the same feature of both samples in the pair to a different value.}, which underpins the search space and hence has a major impact on the success of the perturbation.

\end{enumerate}



While distinct generators can be influenced by the above designs to different extents, they generally do not consider the usefulness of the non-sensitive feature and its causal relationship to the concerned sensitive counterpart during perturbation at all. By using the most causally relevant non-sensitive feature identified from the causal analysis in Section~\ref{sec:fca}, we seek to transform the perturbation in a given generator to make it causality aware---a key contribution of this work.

\subsubsection{Seamless Causality Injection}


In \approach, we inject causality into the above two design aspects to transform the perturbation. Specifically, existing generators use the \textbf{true definition} of individual discriminatory instance to influence the perturbation, which only considers the sensitive feature(s) when determining \textbf{how to define individual discriminatory instances in the perturbation}, i.e., $\mathbf{x}_a$ is an individual discriminatory instance in a pair if there is another sample $\mathbf{x}_b$ such that they differ on the concerned sensitive feature $f_s$ but all other feature values are the same while their predictions by the tested AI model ($y_a$ and $y_b$) are different. That is: 
\begin{equation}
    f_{s,a} \neq f_{s,b} \text{ and } \forall f_{i,a} = f_{i,b}: f_{i,a},f_{i,b} \in \mathcal{F'} \text{ while } y_a \neq y_b
\end{equation}
whereby $f_{s,a}$ and $f_{s,b}$ are the values of the concerned sensitive feature for the two samples, respectively; similarly, $f_{i,a}$ and $f_{i,a}$ are the values of any remaining non-sensitive features in those samples from the set $\mathcal{F'}$. As such, for each sample $\mathbf{x}_a$ that is not yet an individual discriminatory instance from the testing data, according to the true definition of individual discriminatory instance, those generators initially build another sample $\mathbf{x}_b$ to form a pair for the perturbation. 

To inject causality into the definition of individual discriminatory instance that influences the perturbation, instead, we temporarily append the direct and most causally relevant non-sensitive feature $f_c$ along with the concerned sensitive one $f_s$ as a new form of ``synthetic sensitive set'', creating a \textbf{relaxed definition} of the individual discriminatory instance. That is, not one but a pair of features would be considered in the sensitive part, which then creates a more diverse sample $\mathbf{x}_b$ that might be different from $\mathbf{x}_a$ on both $f_c$ and $f_s$ for the perturbation and determining when the perturbation of a pair should terminate. Formally, this means that:
   \begin{equation}
     f_{s,a} \neq f_{s,b} \text{ or } f_{c,a} \neq f_{c,b} \text{ and } \forall f_{i,a} = f_{i,b}: f_{i,a},f_{i,b} \in \mathcal{F''} 
   \text{ while } y_a \neq y_b
    \end{equation}
where $\mathcal{F''}$ is the feature set after ruling out $f_s$ and $f_c$; $f_{c,a}$ and $f_{c,b}$ are the values of the most causally relevant non-sensitive feature for the sensitive counterpart under the two samples, respectively. Clearly, a pair that meets the true definition of individual discriminatory instance would certainly meet the relaxed definition, but the reverse might not be true. Thus, pairs that are invalid under the true definition of individual discriminatory instance could still meet the relaxed definition. However, our goal is merely to diversify the perturbation of the most causally relevant sensitive and non-sensitive features, and it is easy to repair this in the end as we will show.


\textbf{\textit{Example:}} Consider a case that \textit{Age} is the concerned sensitive feature at the index of 5, then for a sample $\mathbf{x}_a$, existing generators could randomly generate $\mathbf{x}_b$ to form a pair used for perturbation, as shown below:
\begin{equation}
\label{eq:exist}
   \begin{split}
       &\mathbf{x}_a = \{0,3,4,6,3,\textcolor{blue}{25},8,7,1,1,2\}\\
       &\mathbf{x}_b = \{0,3,4,6,3,\textcolor{blue}{11},8,7,1,1,2\}
   \end{split}  
\end{equation}
Clearly, they only differ in their sensitive features \textit{Age}. Suppose that \approach~ has identified \textit{Hours-per-week} (the index of 7) as the directly and causally relevant non-sensitive feature with the highest causal effects, then, in contrast, \approach~ might generate a pair as below:
\begin{equation}
\label{eq:new}
   \begin{split}
       &\mathbf{x}_a = \{0,3,4,6,3,\textcolor{blue}{25},8,\textcolor{blue}{7},1,1,2\}\\
       &\mathbf{x}_b = \{0,3,4,6,3,\textcolor{blue}{11},8,\textcolor{blue}{1},1,1,2\}
   \end{split}  
\end{equation}
Here, both \textit{Age} and \textit{Hours-per-week} could be different between $\mathbf{x}_a$ and $\mathbf{x}_b$ when the perturbation starts.

For determining \textbf{what features to perturb}, current generators mostly rule out the sensitive one and randomly perturb all the remaining features (e.g.,Zhang et al.~\cite{DBLP:conf/icse/ZhangW0D0WDD20}), or they simply perturb all features in a random manner (e.g.,Aggarwal et al.~\cite{DBLP:conf/sigsoft/AggarwalLNDS19}). In contrast, according to the above ``synthetic sensitive set'', we further reduce the size of the perturbation by fixing both the concerned sensitive feature and its most causally relevant non-sensitive one during perturbation, hence making it less randomized. This would also help to maintain the diversity with respect to those two features across the search space. Naturally, the impact of perturbation might differ based on the generator, but the basic idea remains the same. For instance, for \textsc{ADF}~\cite{DBLP:conf/icse/ZhangW0D0WDD20}, causality would be injected into both the global and local perturbation while for single perturbation generators like \textsc{SG}~\cite{DBLP:conf/sigsoft/AggarwalLNDS19}, this only affects the perturbation once.



\textbf{\textit{Example:}} In the existing generators, a perturbed pair $\mathbf{x'}_a$ and $\mathbf{x'}_b$ from the original $\mathbf{x}_a$ and $\mathbf{x}_b$ at Equation~(\ref{eq:exist}) could be:
\begin{equation}
   \begin{split}
       &\mathbf{x'}_a = \{\textcolor{blue}{1},\textcolor{blue}{2},\textcolor{blue}{3},\textcolor{blue}{7},\textcolor{blue}{5},25,\textcolor{blue}{9},\textcolor{blue}{9},\textcolor{blue}{2},\textcolor{blue}{3},\textcolor{blue}{4}\}\\
       &\mathbf{x'}_b = \{\textcolor{blue}{1},\textcolor{blue}{2},\textcolor{blue}{3},\textcolor{blue}{7},\textcolor{blue}{5},11,\textcolor{blue}{9},\textcolor{blue}{9},\textcolor{blue}{2},\textcolor{blue}{3},\textcolor{blue}{4}\}
   \end{split}  
\end{equation}
where all features except the sensitive feature \textit{Age} might have been changed. In \approach, instead, the perturbed pair from Equation~(\ref{eq:new}) might become:
\begin{equation}
\label{eq:new1}
   \begin{split}
       &\mathbf{x'}_a = \{\textcolor{blue}{1},\textcolor{blue}{2},\textcolor{blue}{3},\textcolor{blue}{7},\textcolor{blue}{5},25,\textcolor{blue}{9},7,\textcolor{blue}{2},\textcolor{blue}{3},\textcolor{blue}{4}\}\\
       &\mathbf{x'}_b = \{\textcolor{blue}{1},\textcolor{blue}{2},\textcolor{blue}{3},\textcolor{blue}{7},\textcolor{blue}{5},11,\textcolor{blue}{9},1,\textcolor{blue}{2},\textcolor{blue}{3},\textcolor{blue}{4}\}
   \end{split}  
\end{equation}
and it is clear that all features could have been changed except \textit{Age} and its directly and most causally relevant non-sensitive feature with the highest causal effects \textit{Hours-per-week}.

For each selected non-sensitive feature, perturbation is performed by randomly sampling a new value within its observed range or valid categories. No fixed increase or decrease direction is enforced, allowing diverse yet causally informed perturbations. We can do so because of the additional mechanism in \approach~ to ensure constraint compliance: all perturbations respect atomic constraints by restricting values to each feature’s observed range or valid categories in the training dataset (e.g., \textit{Age} can only be chosen from $17$ to $90$~\cite{adultdataset2017}), ensuring realistic and valid perturbed samples. As for violating constraints among features that form a realistic sample, \approach~naturally mitigates this by pairing each perturbed sample with another sample from the original testing data (if neither of the samples in the pair is from the testing data), which ensures compliance with the cross-feature constraints\footnote{Note that we do not assume any constraint with respect to the sensitive feature as those constraints themselves, even if exist, can impose unfair discrimination, e.g., in some countries, very talented students can have already graduated from the university at a very young age.}. 


\subsection{Invalidity Repair}


Finally, for pairs that are valid under the relaxed definition but not the true definition, \approach~ performs the following:


\begin{enumerate}
    \item It randomly finds, from other pairs of the original seeded samples of the testing data, the sample that differs from either sample in the invalid pair by only the concerned sensitive feature. \approach~ additionally pairs such a sample with the corresponding sample from the invalid pair to form a new pair.

    \item If a sample in the invalid pair cannot be repaired, we simply do nothing and consider it as a failed sample\footnote{We found that this is unlikely given the large number of samples generated.}.

\end{enumerate}
In this way, we can benefit from the causally perturbed fairness testing while still ensuring that the true definition of individual discriminatory instances can be eventually met.

\textbf{\textit{Example:}} Suppose that in the end, we have the pair $\mathbf{x'}_a$ and $\mathbf{x'}_b$ from Equation~(\ref{eq:new1}) which is clearly invalid under the true definition of individual discriminatory instance. In \approach, we then repair those samples by finding the samples from other pairs of the original seeded samples that differ from $\mathbf{x'}_a$ or $\mathbf{x'}_b$ only on the concerned sensitive feature \textit{Age} at the index of 5 with different predictions by the tested AI model ($y'_a$ and $y''_a$; $y'_b$ and $y''_b$), forming new pairs $\mathbf{x'}_a$ and $\mathbf{x''}_a$ together with $\mathbf{x'}_b$ and $\mathbf{x''}_b$:
\begin{equation}
   \begin{split}
       &\mathbf{x'}_a = \{1,2,3,7,5,\textcolor{blue}{25},9,7,2,3,4\}\text{; } y'_a=\alpha\\
       &\mathbf{x''}_a = \{1,2,3,7,5,\textcolor{blue}{17},9,7,2,3,4\}\text{; } y''_a=\beta
   \end{split}  
\end{equation}
\begin{equation}
   \begin{split}
       &\mathbf{x'}_b = \{1,2,3,7,5,\textcolor{blue}{11},9,1,2,3,4\}\text{; } y'_b=\alpha\\
       &\mathbf{x''}_b = \{1,2,3,7,5,\textcolor{blue}{13},9,1,2,3,4\}\text{; } y''_b=\beta
   \end{split}  
\end{equation}


Through the above, \approach~seamlessly transforms the perturbation with causality knowledge in a base generator.

Note that even if $\mathbf{x'}_a$ and $\mathbf{x'}_b$ form a valid pair initially, \approach~ still needs to verify if either $\mathbf{x'}_a$ or $\mathbf{x'}_b$ exist in the original testing data; if not, we then also try to pair each of them with one from the testing data following the true definition of individual discriminatory instance, which is similar to the case of invalidity repair. This is important to ensure the practicality of the test cases.

It is worth noting that since \approach~ always pairs one of the invalid samples from a perturbed pair with the other original seeded sample of the testing data, rather than the other perturbed pairs, the realism of the results can be improved. An invalid sample would be discarded if it cannot be paired with any original testing data samples.

%% file: Table/code.tex
\SetKwInput{KwDeclare}{Declare}
\definecolor{note_color}{HTML}{FF4F78}

\begin{algorithm}[t]
	\caption{\approach~framework}
	\label{alg:general_framework}
	\SetAlgoLined
	\footnotesize
        \KwIn{Training/testing data $\bm{\mathcal{D}}_{train}$/$\bm{\mathcal{D}}_{test}$; concerned sensitive feature $f_s$; generator \bm{$\mathcal{G}$}; generated size $s$; $k\%$ of $\bm{\mathcal{D}}_{train}$; the AI model under test $\bm{\mathcal{T}}$.}
        \KwDeclare{Causal graph \bm{$\mathcal{M}$}; selected non-sensitive feature $f_c$.}
        \KwOut{A set of generated individual discriminatory instances $\mathcal{S}$.}

        \tcc{\textcolor{blue}{\texttt{Feature causality analysis.}}}
        \bm{$\mathcal{M}$}, $\{f_1,...,f_l\}$ $\leftarrow$ \textsc{doCausalityAnalysis($k\% \times \bm{\mathcal{D}}_{train}$,$f_s$)} \\
        $\{\{f_1,c_1\},...,\{f_1,,c_l\}\}$ $\leftarrow$ \textsc{getCausalEffect($\{f_1,...,f_l\}$,\bm{$\mathcal{M}$})}

        $f_c$ = the direct non-sensitive feature with the largest causal effect\\

          \tcc{\textcolor{blue}{\texttt{Causal perturbation; this might slightly differ according to the base generator.}}}
	\While{$|\mathcal{S}| < s$}{

            $\{\mathbf{x}_a$,$\mathbf{x}_b\}$ $\leftarrow$ randomly get a sample pair from $\bm{\mathcal{D}}_{test}$\\

            \eIf{$\mathbf{x}_a$ and $\mathbf{x}_b$ only differ on $\{f_s,f_c\}$ while their predictions under $\bm{\mathcal{T}}$ are different}{
              $\mathcal{S} = \mathcal{S} \cup$ the new unique sample(s) from $\{\mathbf{x}_a$,$\mathbf{x}_b\}$\\
            } {
               $\mathbf{x}_b \leftarrow$ a random sample differed from $\mathbf{x}_a$ only on $\{f_s,f_c\}$ \\

               $\{\mathbf{x'}_a$,$\mathbf{x'}_b\} \leftarrow$ \textsc{perturbWithDNN($\bm{\mathcal{T}}$,$\bm{\mathcal{G}}$,$f_s$,$f_c$,$\mathbf{x}_a$,$\mathbf{x}_b$)}\\

               
 
               $\mathcal{S} = \mathcal{S} \cup$ the new unique sample(s) from $\{\mathbf{x'}_a$,$\mathbf{x'}_b\}$\\
               \tcc{\textcolor{blue}{\texttt{Note that only a perturbed sample and a sample from $\bm{\mathcal{D}}_{test}$ can be paired.}}}
               \If{neither $\mathbf{x'}_a$ nor $\mathbf{x'}_b$ belongs to $\bm{\mathcal{D}}_{test}$}{
               $\{\mathbf{x'}_a$,$\mathbf{x''}_a\}$, $\{\mathbf{x'}_b$,$\mathbf{x''}_b\} \leftarrow$ find other sample(s) in $\bm{\mathcal{D}}_{test}$, if any, that can be paired with $\mathbf{x'}_a$ and/or $\mathbf{x'}_b$ according to the true IDI definition\\
               }
            }

	}

  \tcc{\textcolor{blue}{\texttt{Invalidity repair.}}}
       \For{$\forall \{\mathbf{x'}_a$,$\mathbf{x'}_b\} \in \mathcal{S}$ that is invalid under the true definition}{
          $\{\mathbf{x'}_a$,$\mathbf{x''}_a\}$, $\{\mathbf{x'}_b$,$\mathbf{x''}_b\} \leftarrow$ find other sample(s) in $\bm{\mathcal{D}}_{test}$, if any, that make $\mathbf{x'}_a$ and/or $\mathbf{x'}_b$ valid under the true definition\\
       }
       \Return $\mathcal{S}$
\end{algorithm}

%% file: study.tex
\section{Experiments}
\label{sec:exp}


We use several research questions (RQs) to evaluate \approach: 

\begin{itemize}
    \item {\textbf{RQ$_1$:} To what extent can \approach~improve state-of-the-art fairness testing generators?}
    \item {\textbf{RQ$_2$:} How does \approach~perform compared to the existing method that ranks the relevance of non-sensitive features via correlation analysis?}
    \item {\textbf{RQ$_3$:} What is the impact of individual discriminatory instance definition relaxation and invalidity repair?} 

    \item {\textbf{RQ$_4$:} What is the fairness improvement by retraining AI model using the test samples generated by \approach?}
    \item {\textbf{RQ$_5$:} How efficient is \approach~in generating the required number of unique samples?}

\end{itemize}

Since \approach~serves as a general framework, in \textbf{RQ$_1$}, we evaluate to what extent it can be beneficial to improve existing generators for fairness testing across three different AI models. \textbf{RQ$_2$} further verifies that the particular causal inference used in \approach~can be important for any testing improvement against existing methods that capture correlation between the features. \textbf{RQ$_3$} evaluates the impact of relaxing the individual discriminatory instance (IDI) definition and performing invalidity repair on the effectiveness and validity of the generated test cases.  \textbf{RQ$_4$} naturally asks how the fairness of AI system/model can be improved if it is re-trained by using the samples generated by \approach. Finally, \textbf{RQ$_5$} examines the efficiency of \approach, as it would be less sensible if any improvements require an extensive amount of overhead to achieve.

\subsection{Datasets}

In this work, we consider eight datasets, containing commonly used binary classification problems for fairness testing, as shown in Table~\ref{tb:dataset}. These datasets are chosen because:

\begin{itemize}
   \item They come from different domains and with a diverse number of samples and search space for perturbation.
\item They contain rich characteristics and diverse demographic groups (e.g., \textit{Age} and \textit{Race}). 


\item They are real-world datasets, which strengthens the practicality of the evaluation.

\item They are publicly available and widely used in prior fairness testing studies~\cite{DBLP:journals/widm/QuyRIZN22,DBLP:journals/tosem/ChenZHHS24}.
\end{itemize}

All datasets come with pre-defined sensitive features. For example, the \textsc{Adult} dataset has \textit{Age}, \textit{Race}, and \textit{Gender}. 



\input{Table/Dataset}

\subsection{Models}

We evaluate \approach~ on four widely used models:
\begin{itemize}
    \item \textbf{Logistic Regression (LR):}
    As a linear and interpretable baseline, LR serves to evaluate \approach's performance on a simple, convex model commonly used in fairness studies. We use the implementation from \texttt{scikit-learn} with L2 regularization ($C=1.0$).

   \item  \textbf{Random Forest (RF):}
     RF provides a non-linear, ensemble-based classical model, useful for evaluating \approach's compatibility with non-differentiable architectures. We also use the \texttt{scikit-learn}'s implementation with 100 estimators, no maximum depth restriction, and default settings otherwise.

    \item \textbf{Six-Layer Fully Connected Deep Neural Network (DNN\textsubscript{6}):}
    A six-layer fully connected neural network (a standard setting \cite{DBLP:conf/issta/ZhangZZ21}) consisting of:
    \begin{itemize}
        \item \textbf{Input layer}: Number of neurons equal to input feature dimensions.
        \item \textbf{Hidden layers}: We follow five hidden layers, each with ReLU activation functions. The number of neurons per hidden layer is empirically tuned per dataset (ranging from 64 to 256) to optimize classification performance while preventing overfitting.
        \item \textbf{Output layer}: A single sigmoid-activated neuron for binary classification tasks.
    \end{itemize}
    We perform optimization using the \texttt{Adam} optimizer with a learning rate of $0.001$ and a batch size of $128$. Early stopping is employed based on validation loss with a patience of $10$ epochs to avoid overfitting. We use the implementation in \texttt{TensorFlow}.
    
    \item \textbf{Five-Layer Fully Connected Deep Neural Network (DNN\textsubscript{5}):} To examine a diverse set of model capacities and investigate robustness under different complexities, we also evaluate a 5-layer fully connected network with the following configuration\footnote{A five-layered DNN is the smallest network used in prior work~\cite{DBLP:conf/issta/ZhangZZ21} for fairness testing.}:
    \begin{itemize}
        \item \textbf{Input layer}: Number of neurons equal to input feature dimensions.
        \item \textbf{Hidden layers}: Five fully connected layers with sizes [256, 256, 128, 64, 32]. Each layer uses ReLU activation. In addition to the six-layer DNN, dropout layers are inserted after the first three hidden layers (with probabilities of 0.3, 0.3, and 0.2, respectively) to reduce overfitting.
        \item \textbf{Output layer}: A single sigmoid-activated neuron for binary classification tasks.
    \end{itemize}
    We perform optimization using the \texttt{Adam} optimizer with a learning rate of $0.001$ and a batch size of $128$. This model is trained for $100$ epochs with $L_2$ regularization, using Binary Cross-Entropy Loss as the objective function. All weights are initialized using Xavier initialization to improve training stability. As before, we use the implementation in \texttt{TensorFlow}.
\end{itemize}

\subsection{Base Generators}

To verify the robustness and comparability of \approach, we consider a wide range of base generators, including both the white-box and black-box generators. In essence, white-box generators exploit and extract properties from the internal working mechanisms of AI model, such as the gradient and activation of the neuron, to guide the perturbation process in test sample generation. Most of those generators differ in terms of the properties they leverage and how the corresponding information is extracted. The black-box generators, in contrast, do not rely on information from the internal structure of AI model but are purely based on sophisticated perturbation designs. In this work, we consider three white-box generators (\texttt{ADF}~\cite{DBLP:conf/icse/ZhangW0D0WDD20}, \texttt{EIDIG}~\cite{DBLP:conf/issta/ZhangZZ21}, and \texttt{NeuronFair}\cite{DBLP:conf/icse/ZhengCD0CJW0C22}) and other three black-box ones, i.e., \texttt{SG}~\cite{DBLP:conf/sigsoft/AggarwalLNDS19}, \texttt{ExpGA}~\cite{DBLP:conf/icse/0002WJY022}, and \texttt{BREAM}~\cite{DBLP:conf/icics/JiangSLWSG23}---all are state-of-the-art approaches with diverse characteristics. To adopt the white-box generators designed for DNN on other AI models, we slightly modify them, e.g., instead of focusing on the neuron activation, they can focus on the coefficient of terms for LR and the Gini importance on RF. Notably, we ruled out certain generators \cite{DBLP:conf/issta/XiaoLL023} because their code is not executable, as such our evaluation seeks to cover representative white-/black-box base generators. We do not claim those to be exhaustive.

It is worth noting that the above generators, either white-box or black-box, leverage the true definition of the individual discriminatory instance and perform the perturbation without the sensitive features (or simply all features). For all cases, we fix a budget of $10,000$ (unique) samples to be generated, which is a standard setting in prior work~\cite{DBLP:conf/issta/ZhangZZ21,DBLP:conf/icse/ZhangW0D0WDD20,DBLP:conf/icse/ZhengCD0CJW0C22}.

\subsection{Metrics}

We evaluate \approach~on both individual and group fairness. For individual fairness, we calculate the ratio between the number of unique individual discriminatory instances under the true definition ($I$) and the size of all of the generated unique samples ($S$), namely IDI ratio ($I \over S$) \cite{DBLP:conf/kbse/UdeshiAC18,DBLP:conf/icse/ZhangW0D0WDD20}. In fairness testing, a higher IDI ratio means more fairness bugs are found. Note that in \approach, only a perturbed sample and a sample from $\bm{\mathcal{D}}_{test}$ can be paired.






For group fairness, we divide all generated test samples into different groups according to the sensitive features. For example, in the \textsc{Adult} dataset, when \textit{Gender} is the concerned sensitive feature, we divide the test samples depending on whether they have 0 (\textit{Male}) or 1 (\textit{Female}) on \textit{Gender}\footnote{For non-binary features such as \textit{Age}, we follow the common way to discretize the values into groups, e.g., splitting age as $[25,60]$ and others.}. Drawing on this, we use two common metrics in the evaluation: 

\begin{itemize}
    \item \textbf{Equal Opportunity Difference (EOD) \cite{DBLP:conf/nips/HardtPNS16}} measures the extent to which the same proportion of each group divided from the sensitive feature receives a favorable outcome. Formally, it is computed as:
    \begin{equation}
    EOD = \left| \mathbb{E}(\hat{y} \mid f_s = \alpha, y = 1) - \mathbb{E}(\hat{y} \mid f_s = \beta, y = 1) \right|
    \end{equation}
    whereby $\hat{y}$ and $y$ are the model-predicted and actual label ($y=1$ means positive label), respectively. $\alpha$ and $\beta$ are values that decide the group for the sensitive feature $f_s$. A higher EOD means finding more fairness bugs.

    \item \textbf{Statistical Parity Difference (SPD) \cite{DBLP:conf/innovations/DworkHPRZ12}} measures the difference that the expected prediction from an AI model made with respect to those groups that differ on the sensitive feature. Formally, it is expressed as:
    \begin{equation}
        SPD = \left| \mathbb{E}(\hat{y} \mid f_s= \alpha) - \mathbb{E}(\hat{y} \mid f_s = \beta) \right|
    \end{equation}
    Again, a higher SPD means a test generator is capable of revealing more fairness bugs.
\end{itemize}







\subsection{Testing Procedure in Experiments}

The testing procedure follows standard practice for fairness testing ~\cite{DBLP:journals/tosem/ChenZHHS24,DBLP:conf/icse/ZhangW0D0WDD20,DBLP:conf/icse/ZhengCD0CJW0C22}, including the following steps:

\begin{enumerate}
    \item Pick a dataset and perform preprocessing. 
    \item Define a concerned sensitive feature and mark all others non-sensitive.
    \item Randomly make 70\%/30\% training and testing data split and use the training data to train three different AI models as used in existing work ~\cite{DBLP:conf/sigsoft/BiswasR20,DBLP:conf/icse/ZhangW0D0WDD20}.
    \item For an approach that can rank non-sensitive features, such as \approach, perform analysis in the training data and select a representative non-sensitive feature.
    \item Pick a test generator and run it based on the 30\% testing data to generate $10,000$ unique test samples, with and without handling the relationship between the selected non-sensitive feature and the sensitive counterpart.  
    \item Evaluate the results using IDI ratio, EOD, and SPD.
    \item Repeat from 5) until all generators have been examined.
    \item Repeat from 4) until \approach~and other state-of-the-art approaches have been evaluated.
    \item Repeat from 3) via bootstrapping (with replacement) for 10 runs.
    \item Repeat from 2) until all possible sensitive features in the dataset have served as the concerned feature once.
    \item Repeat from 1) until all datasets have been used.
\end{enumerate}

\subsection{Statistical Validation}

We use the recommended non-parametric {U-Test}~\cite{wilcoxon1992individual} with $a=0.05$ to verify the significance of pairwise comparisons over 10 runs on each metric~\cite{DBLP:conf/issta/ArcuriB11}. We additionally use $\mathbf{\hat{A}_{12}}$~\cite{Vargha2000ACA} to examine the effect size. According to Vargha and Delaney~\cite{Vargha2000ACA}, $\mathbf{\hat{A}_{12}}\geq 0.56$ (or $\mathbf{\hat{A}_{12}} \leq 0.44$) indicates a non-trivial effect size. In this work, we say the difference is statistically significant only when $\mathbf{\hat{A}_{12}} \geq 0.56$ (or $\mathbf{\hat{A}_{12}} \leq 0.44$) and $p<0.05$; otherwise the deviation is trivial.


%% file: Table/Dataset.tex
\begin{table}
\caption{The real-world datatset used in fairness testing. $|f_s|$ and $|f|$ denote the possible number of sensitive features and the number of all features, respectively.}
\label{tb:dataset}
\begin{adjustbox}{width=0.7\linewidth,center}
\begin{tabular}{l|l|l|l|l|l}
\toprule
    \textbf{Dataset} & \textbf{Domain} & $|f_s|$ & $|f|$ & \textbf{Available Sample Size} & \textbf{Full Sample Size}\\
    \midrule

 \textsc{Adult~\cite{adultdataset2017}} & Finance&3&11&$45,222$&$4.81 \times 10^9$\\ 

 \textsc{Compas~\cite{compasdataset2016}}& Criminology&2&13&$6,172$&$1.45 \times 10^8$\\ 
     
 \textsc{Law School~\cite{lawschooldataset1998}} & Education&2&12&$20,708$&$9.20 \times 10^6$\\ 

  \textsc{Kdd~\cite{misc_census-income_(kdd)_117}} & Criminology&2&19&$284,556$&$4.13 \times 10^{15}$\\ 

 \textsc{Dutch~\cite{dutchcensus2014}}&Finance&2&12&$60,420$&$3.58 \times 10^7$\\ 

 \textsc{Credit~\cite{defaultcreditdataset2016}} & Finance&3&24&$30,000$&$2.01 \times 10^{12}$\\ 
     
 \textsc{Crime~\cite{CommunitiesCrime2011}} & Criminology&2&22&$2,215$&$4.19 \times 10^8$\\ 
 
  \textsc{German~\cite{germancreditdataset1994}} & Finance&2&20&$1,000$&$8.85 \times 10^9$\\

    \bottomrule
\end{tabular}
\end{adjustbox}
\vspace{-0.4cm}
\end{table}

%% file: results.tex
\section{Results}
\label{sec:results}
\input{Table/rq1-4}

\input{Table/rq1-new}

\input{Table/rq1-2}

\input{Table/rq1-3}

\subsection{Improvement over State-of-the-art Generators}

\subsubsection{Method}

In \textbf{RQ$_1$}, to evaluate the generator-agnostic nature of \approach~and its benefits to the state-of-the-art generators for fairness testing, we pair \approach~with all six generators studied and compare it against the vanilla generators without \approach. This evaluation is based on four different AI models, eight datasets, $2$-$3$ possible sensitive features each, and three metrics, with a total of $324$ comparison cases per model. For each comparison, we apply the statistical test and effect size mentioned in Section~\ref{sec:exp}.

\subsubsection{Results}

From Tables~\ref{tb:rq1-dnn5}, ~\ref{tb:rq1}, ~\ref{tb:rq1-1}, and ~\ref{tb:rq1-2}, clearly, we see that \approach~ achieves remarkable improvement over the state-of-the-art generators on all models, finding considerably more fairness bugs in up to $93\%$ of the cases ($1209$ out of $1296$) with statistical significance across all tested models. The above results are consistent regardless of the generator, datasets, and individual/group fairness metrics. In particular, while the figures shown from the metrics might seem small, considering the number of unique test samples generated in our experiments (i.e., $10,000$), the practical difference is large. For example, when \textit{Gender} is the concerned sensitive feature under the \textsc{Dutch} dataset, using \texttt{NeuronFair} with \approach~leads to a mean $0.344$ IDI ratio against the mean of $0.302$ without \approach~ on DNN\textsubscript{6}. If we consider the $10,000$ unique test samples generated, this means that the former actually finds $3440$ individual discriminatory instances while the latter only finds $3020$ ones---pairing with \approach~can, on average, reveal $420$ more unique fairness bugs which is a practically significant improvement. Surprisingly, not only the individual fairness but also the group fairness can be considerably improved by \approach, which further suggests the non-trivial correlation between individual and group fairness metrics. All the above demonstrate the benefit of extracting the most causally relevant non-sensitive features to the concerned sensitive feature and injecting such a relationship to guide the perturbation during fairness testing.

\begin{figure}[t!]
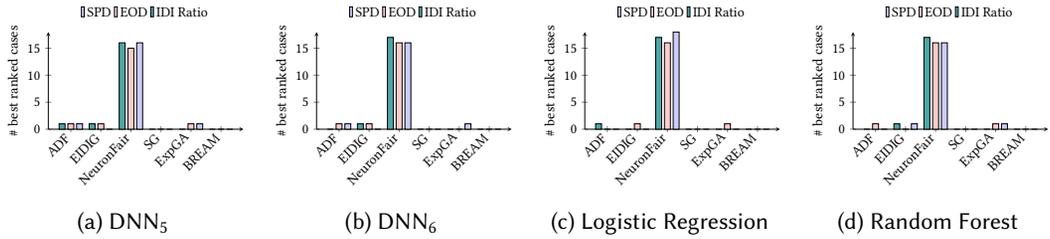

    \centering
     \begin{subfigure}[t!]{0.23\columnwidth}
        \centering        
        \includestandalone[width=\columnwidth]{Figure/dnn_5}
        \caption{DNN\textsubscript{5}} 
    \end{subfigure}
    \hfill
    \begin{subfigure}[t!]{0.23\columnwidth}
        \centering
        \includestandalone[width=\columnwidth]{Figure/dnn}
        \caption{DNN\textsubscript{6}}
    \end{subfigure}
      \hfill
    \begin{subfigure}[t!]{0.23\columnwidth}
        \centering        
        \includestandalone[width=\columnwidth]{Figure/lr}
        \caption{Logistic Regression}      
    \end{subfigure}
    \hfill
     \begin{subfigure}[t!]{0.23\columnwidth}
        \centering        
        \includestandalone[width=\columnwidth]{Figure/rf}
        \caption{Random Forest} 
    \end{subfigure}
     \caption{Comparison of the Scott-Knott test ranks when pairing \approach~with different generators on all models and cases.}
    \label{fig:skrank}
\end{figure}

In addition, as shown in Figure~\ref{fig:skrank}, we see that among the six evaluated base generators (\texttt{ADF}, \texttt{EIDIG}, \texttt{NeuronFair}, \texttt{SG}, \texttt{ExpGA}, and \texttt{Bream}), integrating \approach~ with \texttt{NeuronFair} generally achieves the best performance in revealing fairness bugs across the cases. This is because \texttt{NeuronFair} uses neuron-level adversarial perturbation strategies that directly manipulate hidden model activations to generate fairness-critical inputs. When combined with \approach’s causality-guided test cases generation, the ability of \texttt{NeuronFair} to explore hidden activations can be amplified. This suggests that developers aiming for the best fairness bug discovery ability should prioritize integrating \approach~ with \texttt{NeuronFair}. Overall, we say:
\keystate{
\textit{\textbf{RQ$_1$:} \approach~can considerably improve state-of-the-art generators in finding fairness bugs on approximately $93\%$ of the cases. Working with \texttt{NeuronFair} have enabled \approach~to achieve the generally best results }
}

\subsection{Benefits of Causality Analysis in \approach}

\subsubsection{Method}


Since to the best of our knowledge, we cannot find other work that also considers non-sensitive features to manipulate the perturbation for fairness testing, in \textbf{RQ$_2$}, we compare \approach~with \texttt{FairRF}~\cite{DBLP:conf/wsdm/ZhaoDSW22}, which is a recent work that involves certain steps of similar goal as \approach. Briefly, \texttt{FairRF} aims to find some most correlated non-sensitive features to the concerned sensitive one for training the DNN in order to mitigate fairness bugs. Although it targets fairness mitigation rather than fairness testing, \texttt{FairRF} is similar to \approach~in the sense that it can also rank the non-sensitive feature with respect to the concerned sensitive feature. However, it performs such ranking via correlation analysis rather than understanding the causality as \approach~does. As such, to compare in our fairness testing context, we only leverage the way that \texttt{FairRF} ranks the non-sensitive features and use the most relevant one to be paired together with the concerned sensitive feature. The method that injects such information into the perturbation is the same as the \approach. The same settings for \textbf{RQ$_1$} are used, leading to $324$ cases per model.

\input{Table/rq2-4}

\input{Table/rq2-new}

\input{Table/rq2-2}

\input{Table/rq2-3}

\subsubsection{Results}

The results are shown in Tables~\ref{tb:rq2-dnn5}, ~\ref{tb:rq2}, ~\ref{tb:rq2-1} and ~\ref{tb:rq2-2}. In all models, we can see that under six concerned sensitive features, \approach~and \texttt{FairRF} select the same non-sensitive feature as the most relevant one to its sensitive counterpart, hence they produce the same testing results. This, together with the statistically insignificant comparisons, reveals that both perform similarly in approximately 34\% cases (432 out of 1296) across all tested models. However, on most of the other datasets/concerned sensitive features, they make different choices. In general, \approach~again achieves significantly better outcomes by revealing statistically more fairness bugs over 64\% cases (831 out of 1296) across all tested models. This serves as evidence that, when analyzing the relationships between the concerned sensitive feature and the other non-sensitive ones for guiding the perturbation in fairness testing, it is important to consider causality beyond simple correlation. 
These results indicate that causally guided perturbation not only benefits complex neural networks but also provides advantages for classical models such as LR and RF.
Thus, we conclude that:
\keystate{
\textit{\textbf{RQ$_2$:} The causality analysis in \approach~helps revealing more fairness bugs than current correlation analysis on $64\%$ of the cases.}
}

\subsection{Effect of Relaxation and Invalidity Repair}
\label{sec:rq3}

\subsubsection{Method}
For \textbf{RQ$_3$}, we count the number of pairs generated with the following phases/variants of \approach~ for generating $10,000$ unique test samples\footnote{Note that we only consider the pairs where at least one of the samples matches with the original testing dataset.}:

\begin{itemize}
    \item \textbf{w/o R:} The number of unique pairs generated for the case without relaxation, i.e., no \approach~ but directly using the base generator.
    \item \textbf{w/ R:} The number of unique pairs generated with \approach~ before invalidity repair.
    \item \textbf{I:} The number of unique invalid pairs within those generated by \approach~ before invalidity repair.
    \item \textbf{IR:} The number of unique pairs that have been successfully repaired from those invalid ones generated by \approach~ before invalidity repair.
\end{itemize}

\input{Table/rq5-new}
\subsubsection{Results}
As shown in Table~\ref{tb:rq5-full}, for all cases, the number of pairs generated with \approach~ is much higher than the case when directly using the base generator. This makes sense, since the perturbation in \approach~ with relaxed individual discriminatory instance definition helps to increase the chance of successfully pairing with the original samples in the testing data (even under the true individual discriminatory instance definition), together with the possibility of creating invalid samples. Within those generated pairs, \approach~ produces around $\approx24\%$ invalid ones, which can be successfully repaired by around $\approx81\%$ on average for all cases. This demonstrates the effectiveness of the invalidity repairs that can significantly improve the ability to find fairness bugs. As a result, we conclude that:

\keystate{
\textit{\textbf{RQ$_3$:} The relaxed IDI definition in \approach~can naturally produce more pairs of test samples, within which the invalid ones can be effectively repaired to boost its ability in revealing fairness issues.}
}

\subsection{Fairness Improvement}
\subsubsection{Method}

Although this work focuses on testing, in \textbf{RQ$_4$}, we verify how the samples generated by \approach, after correcting the pair to produce the same outcome, can improve the AI models in better-mitigating fairness bugs, as in prior work~\cite{DBLP:conf/icse/ZhangW0D0WDD20}. 

To that end, for each dataset, we compare \approach~ with a base generator without \approach~ and \texttt{FairRF}, all of which are under their corresponding best generators, in the steps below (for all cases):
\begin{enumerate}
    \item Run a testing approach on the AI models to generate test cases (pairs) that reveal the fairness bugs.
    \item Repair the fairness bugs found, e.g., if $\mathbf{x'}_a$ and $\mathbf{x'}_b$ lead to different labels in the prediction and $\mathbf{x'}_a$ is from the original testing data, we then correct them by changing $\mathbf{x'}_b$ to have the same label as $\mathbf{x'}_a$.
    \item Feed the corrected pairs of samples to retrain the AI model.
    \item The improved model is then tested again by using \approach~ under its best generator again for $10$ runs.
\end{enumerate}
We also compare with the testing results for the AI model before retraining, denoted as \texttt{Before}. If the samples generated by \approach~ can better improve the AI model in handling fairness, the improved DNN should be harder to test, leading to a lower IDI ratio, EOD, and SPD value. In addition, we have also reported on the changes in the model quality via accuracy, F1-score, and AUC, since the fairness can be conflicting with those metrics; though it is worth noting that the model quality might not imply practical usefulness~\cite{chen2025accuracy}.




\input{Table/rq4-4}
\input{Table/rq4-1}
\input{Table/rq4-3}
\input{Table/rq4-2}

\subsubsection{Result}

For the impact on model quality, as can be seen from Tables~\ref{tb:rq4-dnn5}, \ref{tb:rq4-dnn}, \ref{tb:rq4-lr}, and \ref{tb:rq4-rf}, we see that there is indeed a general degradation in the model quality in terms of accuracy and AUC. This makes sense, as it is known that the fairness could conflict with the model quality \cite{DBLP:conf/aistats/ZafarVGG17,DBLP:journals/csur/MehrabiMSLG21,DBLP:conf/fat/Binns20}. However, in general, we see that the degradation is relatively smaller compared with the improvement observed for fairness preservation, e.g., when compared with \texttt{Before} on DNN\textsubscript{6}, the biggest degradations on accuracy and AUC are $0.058$ and $0.052$, respectively, but in return, there is up to $0.290$ improvement on IDI ratio. Interestingly, we also observe an improvement on the F1-score (and sometimes the AUC too). This is because the retraining with fairness-repaired samples, especially those whose labels are corrected to maximize predicted positives, introduces more recall-enhancing positive samples, which helps balance precision and recall, ultimately improving the overall quality.
Thus, we conclude:


\keystate{
\textit{\textbf{RQ$_4$:} \approach~generates test samples that make the model considerably more robust to fairness bugs compared with before (up to $100\%$ cases) and the others (up to $99\%$ cases). At the same time, the degradation on model quality is relatively small.}
}

\subsection{Runtime Efficiency}

\subsubsection{Method}


In \textbf{RQ$_5$}, we report the runtime for pre-testing analysis (e.g., causality analysis) and test sample generation, as the latter mainly depends on the base generator. We showcase the runtime required by the fastest and slowest generators for reaching $10,000$ unique test samples. All statistics cover the results over all sensitive features and runs.


\begin{figure}[t!]
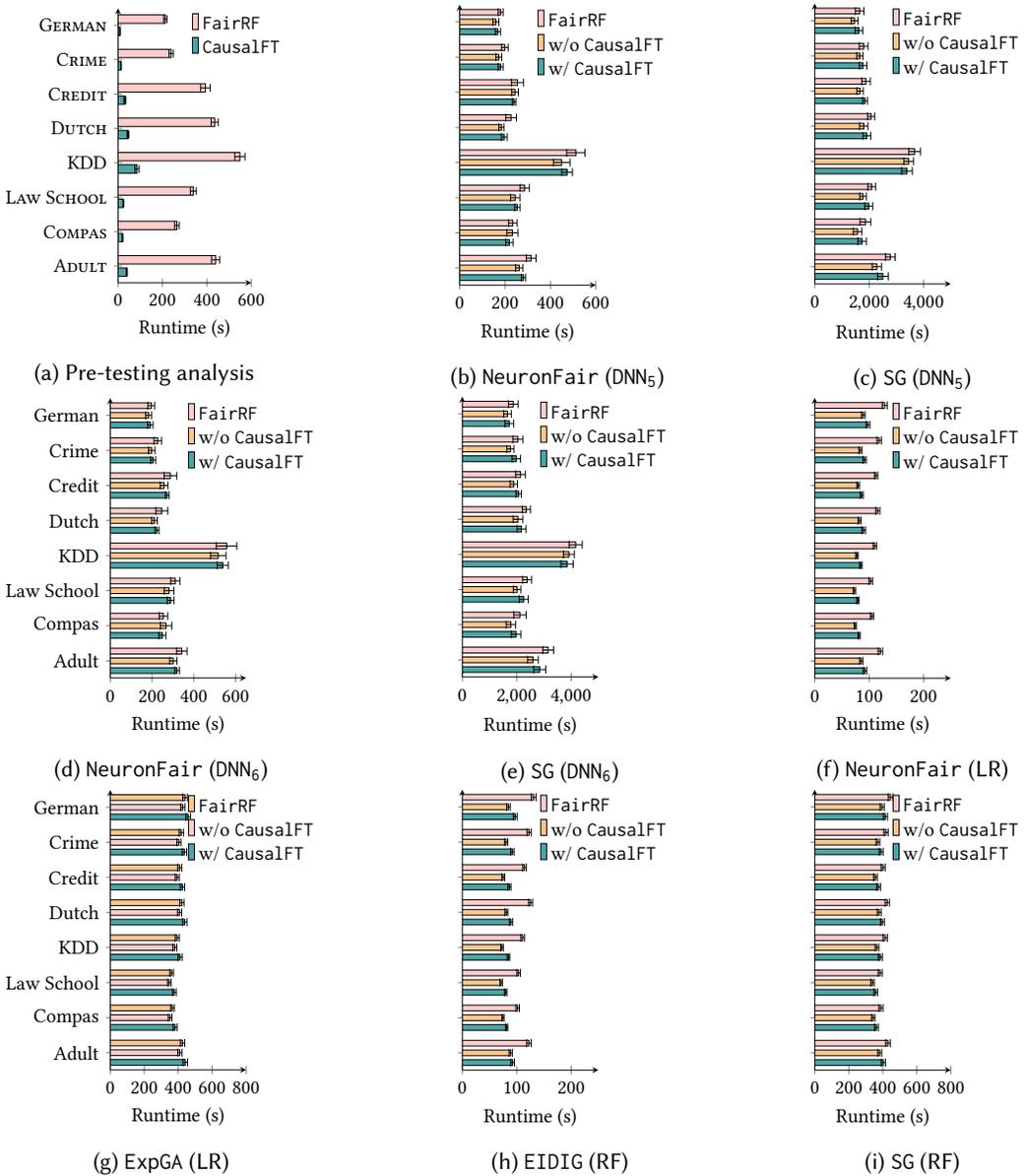

    \centering
    \begin{subfigure}[t!]{0.282\columnwidth}
        \centering
        \includestandalone[width=\columnwidth]{Figure/eff1}
        \subcaption{Pre-testing analysis}
    \end{subfigure}
      \hspace{0.3cm}
      \hfill
    \begin{subfigure}[t!]{0.225\columnwidth}
        \centering        
        \includestandalone[width=\columnwidth]{Figure/eff8}
        \subcaption{\texttt{NeuronFair} (\texttt{DNN\textsubscript{5}})}      
    \end{subfigure}
    \hfill
     \begin{subfigure}[t!]{0.225\columnwidth}
        \centering        
        \includestandalone[width=\columnwidth]{Figure/eff9}
        \subcaption{\texttt{SG} (\texttt{DNN\textsubscript{5}})}      
    \end{subfigure}
    
    \begin{subfigure}[t!]{0.315\columnwidth}
        \centering
        \includestandalone[width=\columnwidth]{Figure/eff2}
        \subcaption{\texttt{NeuronFair} (\texttt{DNN\textsubscript{6}})}
    \end{subfigure}
      \hfill
    \begin{subfigure}[t!]{0.225\columnwidth}
        \centering        
        \includestandalone[width=\columnwidth]{Figure/eff3}
        \subcaption{\texttt{SG} (\texttt{DNN\textsubscript{6}})}      
    \end{subfigure} 
    \hfill
     \begin{subfigure}[t!]{0.225\columnwidth}
        \centering
        \includestandalone[width=\columnwidth]{Figure/eff4}
        \subcaption{\texttt{NeuronFair} (LR)}
    \end{subfigure}
      
    \begin{subfigure}[t!]{0.315\columnwidth}
        \centering        
        \includestandalone[width=\columnwidth]{Figure/eff5}
        \subcaption{\texttt{ExpGA} (LR)}      
    \end{subfigure} 
          \hfill
    \begin{subfigure}[t!]{0.225\columnwidth}
        \centering        
        \includestandalone[width=\columnwidth]{Figure/eff6}
        \subcaption{\texttt{EIDIG} (RF)}      
    \end{subfigure}
    \hfill
     \begin{subfigure}[t!]{0.225\columnwidth}
        \centering        
        \includestandalone[width=\columnwidth]{Figure/eff7}
        \subcaption{\texttt{SG} (RF)}      
    \end{subfigure}
    \caption{Comparing runtime efficiency.\texttt{NeuronFair} (b) and \texttt{SG} (c) are the fastest and slowest generators for \texttt{DNN\textsubscript{5}}; \texttt{NeuronFair} (d) and \texttt{SG} (e) are the fastest and slowest generators for \texttt{DNN\textsubscript{6}}; \texttt{NeuronFair} (f) and \texttt{ExpGA} (g) are the fastest and slowest generators for LR;  \texttt{EIDIG} (h) and \texttt{SG} (i) are the fastest and slowest generators for RF, respectively (mean/deviation over all possible sensitive features/runs).}
    \label{fig:overhead}
\end{figure}

\subsubsection{Result}

From Figure~\ref{fig:overhead}a, we see that \approach~is significantly more efficient than \texttt{FairRF} across all four models (DNN\textsubscript{5}, DNN\textsubscript{6}, LR, RF) in terms of pre-testing analysis, ranging from $8$ to $78$ seconds---up to $400+$ seconds faster than \texttt{FairRF}. This is because when ranking the non-sensitive features, \texttt{FairRF} weights the Pearson correlations and these weights are, by default, learned by an SVM according to the learning rate. Training such an SVM is much more expensive than learning the causal graph in \approach.

For test sample generation (Figures~\ref{fig:overhead}b-\ref{fig:overhead}i), again, we note that \approach~generally requires less time to generate $10,000$ unique test samples than \texttt{FairRF}. This is because \texttt{FairRF} often selects the non-sensitive feature with many possible values, leading to a smaller space of perturbation compared with that of \approach. Thus, \texttt{FairRF} is more likely to generate redundant samples which cause longer runs. Yet, as we have shown, such a further reduced search space does not help \texttt{FairRF} to find more fairness bugs. The results also reveal that, compared with testing without \approach, \approach~needs a slightly longer runtime to reach $10,000$ unique test samples. This makes sense since \approach~further reduces the search space in perturbation from that of testing without \approach, it is more likely to generate redundant samples, especially for datasets with a smaller search space. However, the extra runtime (including causal analysis) will vary depending on the type of model. It merely ranges from a few tens of seconds(under \texttt{NeuronFair} in LR) to a few hundred seconds(under \texttt{SG} in DNN\textsubscript{6}). Overall, the worst case is $\approx 270$ seconds longer under \textsc{Law School} with \texttt{SG} for DNN\textsubscript{6} model. Importantly, as we have shown, \approach~finds, e.g., hundreds more fairness bugs. Thus, we say:

\keystate{
\textit{\textbf{RQ$_5$:} \approach~is much more efficient than \texttt{FairRF} but can cause slightly higher runtime overhead than testing without it. Yet, such extra overhead is acceptable for offline testing, especially considering the benefits it brings.}
}

%% file: Table/rq1-4.tex
\begin{table*}[t!]
\caption{
Comparing different generators with and without \approach~over all cases using the mean (standard deviation) of individual/group fairness metrics under 10 runs. $f_s$ denotes the concerned sensitive feature when testing DNN\textsubscript{5}. \setlength{\fboxsep}{1.5pt}\colorbox{teal!40}{green cells} mean a generator with \approach~leads to higher IDI ratio/EOD/SPD (hence better at finding fairness bugs) with $p<0.05$ and non-trivial $\mathbf{\hat{A}_{12}}$; \setlength{\fboxsep}{1.5pt}\colorbox{red!20}{red cells} denote a generator without \approach~can better reveal fairness bugs with $p<0.05$ and non-trivial $\mathbf{\hat{A}_{12}}$. Statistically insignificant comparisons (i.e., $p \geq 0.05$ or trivial $\mathbf{\hat{A}_{12}}$) have no rendered color regardless of the results.
}

\label{tb:rq1-dnn5}
\setlength{\tabcolsep}{.7mm}
\centering
\begin{adjustbox}{width=\linewidth,center}
\begin{tabular}{lll|rrr|rrr|rrr|rrr|rrr|rrr} 
\toprule
\multirow{2}{*}{\textbf{Dataset}}   & \multirow{2}{*}{\textbf{$f_s$}} & \multicolumn{1}{c|}{\multirow{2}{*}{\textbf{Approach}}} & \multicolumn{3}{c|}{\texttt{ADF}}                                                       & \multicolumn{3}{c|}{\texttt{EIDIG}}                                                     & \multicolumn{3}{c|}{\texttt{NeuronFair}}                                                & \multicolumn{3}{c|}{\texttt{SG}}                                                        & \multicolumn{3}{c|}{\texttt{ExpGA}}                                                     & \multicolumn{3}{c}{\texttt{BREAM}}                                                      \\ 
\cline{4-21}
                           &                                 & \multicolumn{1}{c|}{}                        & \multicolumn{1}{l}{IDI ratio} & \multicolumn{1}{l}{EOD} & \multicolumn{1}{l|}{SPD} & \multicolumn{1}{l}{IDI ratio} & \multicolumn{1}{l}{EOD} & \multicolumn{1}{l|}{SPD} & \multicolumn{1}{l}{IDI ratio} & \multicolumn{1}{l}{EOD} & \multicolumn{1}{l|}{SPD} & \multicolumn{1}{l}{IDI ratio} & \multicolumn{1}{l}{EOD} & \multicolumn{1}{l|}{SPD} & \multicolumn{1}{l}{IDI ratio} & \multicolumn{1}{l}{EOD} & \multicolumn{1}{l|}{SPD} & \multicolumn{1}{l}{IDI ratio} & \multicolumn{1}{l}{EOD} & \multicolumn{1}{l}{SPD}                                            \\
\bottomrule

\end{tabular}
\end{adjustbox}
\arrayrulecolor{black}
\end{table*}

%% file: Table/rq1-new.tex
\begin{table*}[t!]
\caption{Comparing different generators with and without \approach~over all cases using the mean (standard deviation) of individual/group fairness metrics under 10 runs when testing DNN\textsubscript{6}. All formate are the same as Table~\ref{tb:rq1-dnn5}.}

\label{tb:rq1}
\setlength{\tabcolsep}{.7mm}
\centering
\begin{adjustbox}{width=\linewidth,center}
\begin{tabular}{lll|rrr|rrr|rrr|rrr|rrr|rrr} 
\toprule
\multirow{2}{*}{\textbf{Dataset}}   & \multirow{2}{*}{\textbf{$f_s$}} & \multicolumn{1}{c|}{\multirow{2}{*}{\textbf{Approach}}} & \multicolumn{3}{c|}{\texttt{ADF}}                                                       & \multicolumn{3}{c|}{\texttt{EIDIG}}                                                     & \multicolumn{3}{c|}{\texttt{NeuronFair}}                                                & \multicolumn{3}{c|}{\texttt{SG}}                                                        & \multicolumn{3}{c|}{\texttt{ExpGA}}                                                     & \multicolumn{3}{c}{\texttt{BREAM}}                                                      \\ 
\cline{4-21}
                           &                                 & \multicolumn{1}{c|}{}                        & \multicolumn{1}{l}{IDI ratio} & \multicolumn{1}{l}{EOD} & \multicolumn{1}{l|}{SPD} & \multicolumn{1}{l}{IDI ratio} & \multicolumn{1}{l}{EOD} & \multicolumn{1}{l|}{SPD} & \multicolumn{1}{l}{IDI ratio} & \multicolumn{1}{l}{EOD} & \multicolumn{1}{l|}{SPD} & \multicolumn{1}{l}{IDI ratio} & \multicolumn{1}{l}{EOD} & \multicolumn{1}{l|}{SPD} & \multicolumn{1}{l}{IDI ratio} & \multicolumn{1}{l}{EOD} & \multicolumn{1}{l|}{SPD} & \multicolumn{1}{l}{IDI ratio} & \multicolumn{1}{l}{EOD} & \multicolumn{1}{l}{SPD}                                            \\
\bottomrule

\end{tabular}
\end{adjustbox}
\arrayrulecolor{black}
\end{table*}

%% file: Table/rq1-2.tex
\begin{table*}[t!]
\caption{Comparing different generators with and without \approach~over all cases using the mean(standard deviation) of individual/group fairness metrics under 10 runs when testing LR. All formate are the same as Table~\ref{tb:rq1-dnn5}.}


\label{tb:rq1-1}
\setlength{\tabcolsep}{.7mm}
\centering
\begin{adjustbox}{width=\linewidth,center}
\begin{tabular}{lll|rrr|rrr|rrr|rrr|rrr|rrr} 
\toprule
\multirow{2}{*}{\textbf{Dataset}}   & \multirow{2}{*}{\textbf{$f_s$}} & \multicolumn{1}{c|}{\multirow{2}{*}{\textbf{Approach}}} & \multicolumn{3}{c|}{\texttt{ADF}}                                                       & \multicolumn{3}{c|}{\texttt{EIDIG}}                                                     & \multicolumn{3}{c|}{\texttt{NeuronFair}}                                                & \multicolumn{3}{c|}{\texttt{SG}}                                                        & \multicolumn{3}{c|}{\texttt{ExpGA}}                                                     & \multicolumn{3}{c}{\texttt{BREAM}}                                                      \\ 
\cline{4-21}
                           &                                 & \multicolumn{1}{c|}{}                        & \multicolumn{1}{l}{IDI ratio} & \multicolumn{1}{l}{EOD} & \multicolumn{1}{l|}{SPD} & \multicolumn{1}{l}{IDI ratio} & \multicolumn{1}{l}{EOD} & \multicolumn{1}{l|}{SPD} & \multicolumn{1}{l}{IDI ratio} & \multicolumn{1}{l}{EOD} & \multicolumn{1}{l|}{SPD} & \multicolumn{1}{l}{IDI ratio} & \multicolumn{1}{l}{EOD} & \multicolumn{1}{l|}{SPD} & \multicolumn{1}{l}{IDI ratio} & \multicolumn{1}{l}{EOD} & \multicolumn{1}{l|}{SPD} & \multicolumn{1}{l}{IDI ratio} & \multicolumn{1}{l}{EOD} & \multicolumn{1}{l}{SPD}                                            \\
\hline

\end{tabular}
\end{adjustbox}
\arrayrulecolor{black}
\end{table*}

%% file: Table/rq1-3.tex
\begin{table*}[t!]
\caption{Comparing different generators with and without \approach~over all cases using the mean (standard deviation) of individual/group fairness metrics under 10 runs when testing RF. All formate are the same as Table~\ref{tb:rq1-dnn5}.}

\label{tb:rq1-2}
\setlength{\tabcolsep}{.7mm}
\centering
\begin{adjustbox}{width=\linewidth,center}
\begin{tabular}{lll|rrr|rrr|rrr|rrr|rrr|rrr}
\toprule
\multirow{2}{*}{\textbf{Dataset}}   & \multirow{2}{*}{\textbf{$f_s$}} & \multicolumn{1}{c|}{\multirow{2}{*}{\textbf{Approach}}} & \multicolumn{3}{c|}{\texttt{ADF}}                                                       & \multicolumn{3}{c|}{\texttt{EIDIG}}                                                     & \multicolumn{3}{c|}{\texttt{NeuronFair}}                                                & \multicolumn{3}{c|}{\texttt{SG}}                                                        & \multicolumn{3}{c|}{\texttt{ExpGA}}                                                     & \multicolumn{3}{c}{\texttt{BREAM}}                                                      \\
\cline{4-21}
&                                 & \multicolumn{1}{c|}{}                        & \multicolumn{1}{l}{IDI ratio} & \multicolumn{1}{l}{EOD} & \multicolumn{1}{l|}{SPD} & \multicolumn{1}{l}{IDI ratio} & \multicolumn{1}{l}{EOD} & \multicolumn{1}{l|}{SPD} & \multicolumn{1}{l}{IDI ratio} & \multicolumn{1}{l}{EOD} & \multicolumn{1}{l|}{SPD} & \multicolumn{1}{l}{IDI ratio} & \multicolumn{1}{l}{EOD} & \multicolumn{1}{l|}{SPD} & \multicolumn{1}{l}{IDI ratio} & \multicolumn{1}{l}{EOD} & \multicolumn{1}{l|}{SPD} & \multicolumn{1}{l}{IDI ratio} & \multicolumn{1}{l}{EOD} & \multicolumn{1}{l}{SPD}                                            \\
\bottomrule

\end{tabular}
\end{adjustbox}
\arrayrulecolor{black}
\end{table*}

%% file: Figure/dnn_5.tex
\begin{tikzpicture}
\begin{axis}[
    ybar,
    axis x line  = bottom,
    axis y line  = left  ,
    reverse legend,
    bar width=.15cm,
    width=6.5cm,
    height=4cm,
    ymax=18,
    ylabel={\# best ranked cases},
    symbolic x coords={ADF, EIDIG, NeuronFair, SG, ExpGA, BREAM},
    xtick=data,
    xticklabel style={rotate=45, anchor=east},
    enlarge x limits=0.15,
    legend style={at={(0.5,1.2)},draw=none,fill=none,anchor=center,legend columns=3},
    legend cell align={left},
]
\addplot+[fill=teal!70,draw=black] coordinates {(ADF,1) (EIDIG,1) (NeuronFair,16) (SG,0) (ExpGA,0) (BREAM,0)};
\addplot+[fill=red!20,draw=black] coordinates {(ADF,1) (EIDIG,1) (NeuronFair,15) (SG,0) (ExpGA,1) (BREAM,0)};
\addplot+[fill=blue!20,draw=black] coordinates {(ADF,1) (EIDIG,0) (NeuronFair,16) (SG,0) (ExpGA,1) (BREAM,0)};
\legend{IDI Ratio, EOD, SPD}
\end{axis}
\end{tikzpicture}

%% file: Figure/dnn.tex
\begin{tikzpicture}
\begin{axis}[
    ybar,
    axis x line  = bottom,
    axis y line  = left  ,
    reverse legend,
    bar width=.15cm,
    width=6.5cm,
    height=4cm,
    ymax=18,
    ylabel={\# best ranked cases},
    symbolic x coords={ADF, EIDIG, NeuronFair, SG, ExpGA, BREAM},
    xtick=data,
    xticklabel style={rotate=45, anchor=east},
    enlarge x limits=0.15,
    legend style={at={(0.5,1.2)},draw=none,fill=none,anchor=center,legend columns=3},
    legend cell align={left},
]
\addplot+[fill=teal!70,draw=black] coordinates {(ADF,0) (EIDIG,1) (NeuronFair,17) (SG,0) (ExpGA,0) (BREAM,0)};
\addplot+[fill=red!20,draw=black] coordinates {(ADF,1) (EIDIG,1) (NeuronFair,16) (SG,0) (ExpGA,0) (BREAM,0)};
\addplot+[fill=blue!20,draw=black] coordinates {(ADF,1) (EIDIG,0) (NeuronFair,16) (SG,0) (ExpGA,1) (BREAM,0)};
\legend{IDI Ratio, EOD, SPD}
\end{axis}
\end{tikzpicture}

%% file: Figure/lr.tex
\begin{tikzpicture}
\begin{axis}[
    ybar,
    reverse legend,
    axis x line  = bottom,
    axis y line  = left  ,
    bar width=.15cm,
    width=6.5cm,
    height=4cm,
    ymax=18,
    ylabel={\# best ranked cases},
    symbolic x coords={ADF, EIDIG, NeuronFair, SG, ExpGA, BREAM},
    xtick=data,
    xticklabel style={rotate=45, anchor=east},
    enlarge x limits=0.15,
    legend style={at={(0.5,1.2)},draw=none,fill=none,anchor=center,legend columns=3}
]
\addplot+[fill=teal!70,draw=black] coordinates {(ADF,1) (EIDIG,0) (NeuronFair,17) (SG,0) (ExpGA,0) (BREAM,0)};
\addplot+[fill=red!20,draw=black] coordinates {(ADF,0) (EIDIG,1) (NeuronFair,16) (SG,0) (ExpGA,1) (BREAM,0)};
\addplot+[fill=blue!20,draw=black] coordinates {(ADF,0) (EIDIG,0) (NeuronFair,18) (SG,0) (ExpGA,0) (BREAM,0)};

\legend{IDI Ratio, EOD, SPD}
\end{axis}
\end{tikzpicture}

%% file: Figure/rf.tex
\begin{tikzpicture}
\begin{axis}[
    ybar,
    axis x line  = bottom,
    axis y line  = left  ,
    reverse legend,
    bar width=.15cm,
    width=6.5cm,
    height=4cm,
    ymax=18,
    ylabel={\# best ranked cases},
    symbolic x coords={ADF, EIDIG, NeuronFair, SG, ExpGA, BREAM},
    xtick=data,
    xticklabel style={rotate=45, anchor=east},
    enlarge x limits=0.15,
    legend style={at={(0.5,1.2)},draw=none,fill=none,anchor=center,legend columns=3},
    legend cell align={left},
]
\addplot+[fill=teal!70,draw=black] coordinates {(ADF,0) (EIDIG,1) (NeuronFair,17) (SG,0) (ExpGA,0) (BREAM,0)};
\addplot+[fill=red!20,draw=black] coordinates {(ADF,1) (EIDIG,0) (NeuronFair,16) (SG,0) (ExpGA,1) (BREAM,0)};
\addplot+[fill=blue!20,draw=black] coordinates {(ADF,0) (EIDIG,1) (NeuronFair,16) (SG,0) (ExpGA,1) (BREAM,0)};
\legend{IDI Ratio, EOD, SPD}
\end{axis}
\end{tikzpicture}

%% file: Table/rq2-4.tex
\begin{table*}[t!]
\caption{For testing DNN\textsubscript{5}, comparing \approach~with \texttt{FairRF}, which ranks non-sensitive attributes with respect to the sensitive one based on correlation analysis, over all cases using the mean (standard deviation) of all fairness metrics under 10 runs. $f_c$ denotes the direct and most causally relevant non-sensitive feature to the concerned sensitive one. The other format is the same as Table~\ref{tb:rq1-dnn5}.}
\label{tb:rq2-dnn5}
\setlength{\tabcolsep}{.7mm}
\centering
\begin{adjustbox}{width=\linewidth,center}
\begin{tabular}{llll|rrr|rrr|rrr|rrr|rrr|rrr}
\toprule
\multirow{2}{*}{\textbf{Dataset}} & \multirow{2}{*}{\textbf{$f_s$}} & \multicolumn{1}{c}{\multirow{2}{*}{\textbf{Approach}}} & \multirow{2}{*}{\textbf{$f_c$}} &
\multicolumn{3}{c|}{\texttt{ADF}} & \multicolumn{3}{c|}{\texttt{EIDIG}} & \multicolumn{3}{c|}{\texttt{NeuronFair}} &
\multicolumn{3}{c|}{\texttt{SG}} & \multicolumn{3}{c|}{\texttt{ExpGA}} & \multicolumn{3}{c}{\texttt{BREAM}} \\
\cline{5-22}
& & & & \multicolumn{1}{l}{IDI ratio} & \multicolumn{1}{l}{EOD} & \multicolumn{1}{l|}{SPD}
& \multicolumn{1}{l}{IDI ratio} & \multicolumn{1}{l}{EOD} & \multicolumn{1}{l|}{SPD}
& \multicolumn{1}{l}{IDI ratio} & \multicolumn{1}{l}{EOD} & \multicolumn{1}{l|}{SPD}
& \multicolumn{1}{l}{IDI ratio} & \multicolumn{1}{l}{EOD} & \multicolumn{1}{l|}{SPD}
& \multicolumn{1}{l}{IDI ratio} & \multicolumn{1}{l}{EOD} & \multicolumn{1}{l|}{SPD}
& \multicolumn{1}{l}{IDI ratio} & \multicolumn{1}{l}{EOD} & \multicolumn{1}{l}{SPD} \\
\midrule
\multirow{6}{*}{\textsc{Adult}}
& \multirow{2}{*}{\textit{Gender}} & \approach & \textit{Relationship}
& \cellcolor{teal!40}{.284(.003)} & \cellcolor{teal!40}{.137(.003)} & \cellcolor{teal!40}{.099(.003)}
& \cellcolor{teal!40}{.301(.003)} & \cellcolor{teal!40}{.153(.003)} & \cellcolor{teal!40}{.112(.003)}
& \cellcolor{teal!40}{.361(.004)} & \cellcolor{teal!40}{.145(.002)} & \cellcolor{teal!40}{.130(.003)}
& \cellcolor{teal!40}{.248(.003)} & \cellcolor{teal!40}{.120(.003)} & \cellcolor{teal!40}{.123(.003)}
& \cellcolor{teal!40}{.249(.003)} & \cellcolor{teal!40}{.091(.003)} & \cellcolor{teal!40}{.099(.002)}
& \cellcolor{teal!40}{.266(.003)} & \cellcolor{teal!40}{.102(.003)} & \cellcolor{teal!40}{.109(.003)} \\
\hline
\end{tabular}
\end{adjustbox}
\end{table*}

%% file: Table/rq2-new.tex
\begin{table*}[t!]
\caption{When testing DNN\textsubscript{6}, comparing \approach~with \texttt{FairRF}. The other format is the same as Table~\ref{tb:rq2-dnn5}.}

\label{tb:rq2}
\setlength{\tabcolsep}{.7mm}
\centering
\begin{adjustbox}{width=\linewidth,center}
\begin{tabular}{llll|rrr|rrr|rrr|rrr|rrr|rrr} 
\toprule
\multirow{2}{*}{\textbf{Dataset}}   & \multirow{2}{*}{\textbf{$f_s$}} & \multicolumn{1}{c}{\multirow{2}{*}{\textbf{Approach}}}   & \multirow{2}{*}{\textbf{$f_c$}} & \multicolumn{3}{c|}{\texttt{ADF}}                                                       & \multicolumn{3}{c|}{\texttt{EIDIG}}                                                     & \multicolumn{3}{c|}{\texttt{NeuronFair}}                                                & \multicolumn{3}{c|}{\texttt{SG}}                                                        & \multicolumn{3}{c|}{\texttt{ExpGA}}                                                     & \multicolumn{3}{c}{\texttt{BREAM}}                                                      \\ 
\cline{5-22}
                           &           &                      & \multicolumn{1}{c|}{}                        & \multicolumn{1}{l}{IDI ratio} & \multicolumn{1}{l}{EOD} & \multicolumn{1}{l|}{SPD} & \multicolumn{1}{l}{IDI ratio} & \multicolumn{1}{l}{EOD} & \multicolumn{1}{l|}{SPD} & \multicolumn{1}{l}{IDI ratio} & \multicolumn{1}{l}{EOD} & \multicolumn{1}{l|}{SPD} & \multicolumn{1}{l}{IDI ratio} & \multicolumn{1}{l}{EOD} & \multicolumn{1}{l|}{SPD} & \multicolumn{1}{l}{IDI ratio} & \multicolumn{1}{l}{EOD} & \multicolumn{1}{l|}{SPD} & \multicolumn{1}{l}{IDI ratio} & \multicolumn{1}{l}{EOD} & \multicolumn{1}{l}{SPD}                                            \\
\bottomrule

\end{tabular}
\end{adjustbox}
\arrayrulecolor{black}
\end{table*}

%% file: Table/rq2-2.tex
\begin{table*}[t!]
\caption{When testing LR, comparing \approach~with \texttt{FairRF}. The other format is the same as Table~\ref{tb:rq2-dnn5}.}

\label{tb:rq2-1}
\setlength{\tabcolsep}{.7mm}
\centering
\begin{adjustbox}{width=\linewidth,center}
\begin{tabular}{llll|rrr|rrr|rrr|rrr|rrr|rrr} 
\toprule
\multirow{2}{*}{\textbf{Dataset}}   & \multirow{2}{*}{\textbf{$f_s$}} & \multicolumn{1}{c}{\multirow{2}{*}{\textbf{Approach}}}   & \multirow{2}{*}{\textbf{$f_c$}} & \multicolumn{3}{c|}{\texttt{ADF}}                                                       & \multicolumn{3}{c|}{\texttt{EIDIG}}                                                     & \multicolumn{3}{c|}{\texttt{NeuronFair}}                                                & \multicolumn{3}{c|}{\texttt{SG}}                                                        & \multicolumn{3}{c|}{\texttt{ExpGA}}                                                     & \multicolumn{3}{c}{\texttt{BREAM}}                                                      \\ 
\cline{5-22}
                           &           &                      & \multicolumn{1}{c|}{}                        & \multicolumn{1}{l}{IDI ratio} & \multicolumn{1}{l}{EOD} & \multicolumn{1}{l|}{SPD} & \multicolumn{1}{l}{IDI ratio} & \multicolumn{1}{l}{EOD} & \multicolumn{1}{l|}{SPD} & \multicolumn{1}{l}{IDI ratio} & \multicolumn{1}{l}{EOD} & \multicolumn{1}{l|}{SPD} & \multicolumn{1}{l}{IDI ratio} & \multicolumn{1}{l}{EOD} & \multicolumn{1}{l|}{SPD} & \multicolumn{1}{l}{IDI ratio} & \multicolumn{1}{l}{EOD} & \multicolumn{1}{l|}{SPD} & \multicolumn{1}{l}{IDI ratio} & \multicolumn{1}{l}{EOD} & \multicolumn{1}{l}{SPD}                                            \\
\bottomrule

\end{tabular}
\end{adjustbox}
\arrayrulecolor{black}
\vspace{-0.4cm}
\end{table*}

%% file: Table/rq2-3.tex
\begin{table*}[t!]
\caption{When testing RF, comparing \approach~with \texttt{FairRF}. The other format is the same as Table~\ref{tb:rq2-dnn5}.}

\label{tb:rq2-2}
\setlength{\tabcolsep}{.7mm}
\centering
\begin{adjustbox}{width=\linewidth,center}
\begin{tabular}{llll|rrr|rrr|rrr|rrr|rrr|rrr} 
\toprule
\multirow{2}{*}{\textbf{Dataset}}   & \multirow{2}{*}{\textbf{$f_s$}} & \multicolumn{1}{c}{\multirow{2}{*}{\textbf{Approach}}}   & \multirow{2}{*}{\textbf{$f_c$}} & \multicolumn{3}{c|}{\texttt{ADF}}                                                       & \multicolumn{3}{c|}{\texttt{EIDIG}}                                                     & \multicolumn{3}{c|}{\texttt{NeuronFair}}                                                & \multicolumn{3}{c|}{\texttt{SG}}                                                        & \multicolumn{3}{c|}{\texttt{ExpGA}}                                                     & \multicolumn{3}{c}{\texttt{BREAM}}                                                      \\ 
\cline{5-22}
                           &           &                      & \multicolumn{1}{c|}{}                        & \multicolumn{1}{l}{IDI ratio} & \multicolumn{1}{l}{EOD} & \multicolumn{1}{l|}{SPD} & \multicolumn{1}{l}{IDI ratio} & \multicolumn{1}{l}{EOD} & \multicolumn{1}{l|}{SPD} & \multicolumn{1}{l}{IDI ratio} & \multicolumn{1}{l}{EOD} & \multicolumn{1}{l|}{SPD} & \multicolumn{1}{l}{IDI ratio} & \multicolumn{1}{l}{EOD} & \multicolumn{1}{l|}{SPD} & \multicolumn{1}{l}{IDI ratio} & \multicolumn{1}{l}{EOD} & \multicolumn{1}{l|}{SPD} & \multicolumn{1}{l}{IDI ratio} & \multicolumn{1}{l}{EOD} & \multicolumn{1}{l}{SPD}                                            \\
\bottomrule

\end{tabular}
\end{adjustbox}
\arrayrulecolor{black}
\vspace{-0.4cm}
\end{table*}

%% file: Table/rq5-new.tex
\begin{table}[t!]
\caption{Counting the average number of pairs (test cases) generated with the relaxed individual discriminatory instance definition and invalidity repair across all base generators and runs.}
\label{tb:rq5-full}
\centering
\begin{adjustbox}{width=\linewidth,center}
\begin{tabular}{ll|llll|llll|llll|llll}
\toprule

\multirow{2}{*}{\textbf{Dataset}}&\multirow{2}{*}{$f_s$}&\multicolumn{4}{c|}{\textbf{\texttt{DNN\textsubscript{5}}}}&\multicolumn{4}{c|}{\textbf{\texttt{DNN\textsubscript{6}}}} &\multicolumn{4}{c|}{\textbf{\texttt{LR}}}&\multicolumn{4}{c}{\textbf{\texttt{RF}}}\\

\cline{3-18}
& & \multicolumn{1}{l}{\textbf{$\#$w/o R}} & \multicolumn{1}{l}{\textbf{$\#$w/ R}} 
& \multicolumn{1}{l}{\textbf{$\#$I}} & \multicolumn{1}{l|}{\textbf{$\#$IR}}
 & \multicolumn{1}{l}{\textbf{$\#$w/o R}} & \multicolumn{1}{l}{\textbf{$\#$w/ R}} 
& \multicolumn{1}{l}{\textbf{$\#$I}} & \multicolumn{1}{l|}{\textbf{$\#$IR}}
 & \multicolumn{1}{l}{\textbf{$\#$w/o R}} & \multicolumn{1}{l}{\textbf{$\#$w/ R}} 
& \multicolumn{1}{l}{\textbf{$\#$I}} & \multicolumn{1}{l|}{\textbf{$\#$IR}}
& \multicolumn{1}{l}{\textbf{$\#$w/o R}} & \multicolumn{1}{l}{\textbf{$\#$w/ R}} 
& \multicolumn{1}{l}{\textbf{$\#$I}} & \multicolumn{1}{l}{\textbf{$\#$IR}}
 \\

\midrule
\multirow{3}{*}{\textsc{Adult}} & \textit{Gender} 
& 2614.9 & 4314.3 & 649.0 & 475.9
& 2811.7 & 4589.7 & 683.2 & 511.7 
& 1738.3 & 2898.3 & 554.7 & 449.3 
& 1609.3 & 2795.5 & 645.1 & 522.4 
\\
& \textit{Race} 
& 2991.5 & 5301.3 & 893.0 & 721.5
& 3216.7 & 5639.7 & 940.0 & 775.8 
& 1680.0 & 2622.3 & 400.0 & 336.0 
& 1635.0 & 2946.8 & 679.3 & 509.5 
\\
& \textit{Age} 
& 2816.3 & 4877.0 & 940.2 & 713.3
& 3028.3 & 5188.3 & 989.7 & 767.0
& 1566.7 & 2776.5 & 573.0 & 446.9 
& 1531.3 & 2677.0 & 572.6 & 458.1 
\\
\hline

\multirow{2}{*}{\textsc{Compas}} & \textit{Gender} 
& 3072.9 & 5099.2 & 775.9 & 569.6
& 3304.2 & 5424.7 & 816.7 & 612.5 
& 1618.3 & 2662.0 & 462.0 & 378.8 
& 1521.0 & 2526.8 & 541.6 & 442.6 
\\
& \textit{Race}
& 3377.5 & 5483.4 & 1052.7 & 814.1
& 3631.7 & 5833.4 & 1108.1 & 875.4 
& 1578.3 & 2918.8 & 583.8 & 472.9 
& 1464.0 & 2416.2 & 484.0 & 401.7 

\\
\hline

\multirow{2}{*}{\textsc{Law School}} & \textit{Gender} 
& 2823.3 & 4674.6 & 728.5 & 570.5
& 3035.8 & 4973.0 & 766.8 & 613.4 
& 1553.3 & 2790.5 & 627.3 & 483.0 
& 1559.5 & 2467.4 & 502.4 & 414.2 

\\
& \textit{Race} 
& 2814.8 & 5006.7 & 918.7 & 764.5
& 3026.7 & 5326.3 & 967.1 & 822.0 
& 1561.7 & 2703.6 & 431.7 & 345.4 
& 1499.5 & 2397.4 & 483.8 & 398.1 

\\
\hline

\multirow{2}{*}{\textsc{KDD}} & \textit{Gender} 
& 2359.1 & 4191.6 & 610.7 & 495.5
& 2536.7 & 4459.1 & 642.8 & 532.8 
& 1551.7 & 2523.2 & 352.5 & 299.6 
& 1555.0 & 2530.5 & 506.1 & 415.0 
\\
& \textit{Race} 
& 2366.8 & 4283.7 & 721.5 & 561.8
& 2545.0 & 4557.1 & 759.5 & 604.1 
& 1546.7 & 2937.1 & 529.6 & 418.4 
& 1578.2 & 2638.2 & 530.0 & 434.6 
\\
\hline

\multirow{2}{*}{\textsc{Dutch}} & \textit{Gender} 
& 2340.5 & 3716.8 & 485.3 & 380.0
& 2516.7 & 3954.0 & 510.8 & 408.6 
& 1520.0 & 2661.2 & 476.7 & 384.9 
& 1653.3 & 2618.1 & 581.2 & 476.6 

\\
& \textit{Age} 
& 2315.7 & 3666.9 & 546.3 & 438.6
& 2490.0 & 3901.0 & 575.1 & 471.6 
& 1577.0 & 2821.0 & 528.8 & 444.2 
& 1683.8 & 2702.0 & 624.6 & 502.3 
\\
\hline

\multirow{3}{*}{\textsc{Credit}} & \textit{Gender} 
& 2321.9 & 3449.2 & 528.8 & 429.1
& 2496.7 & 3669.4 & 556.6 & 461.4 
& 1577.0 & 2603.8 & 491.4 & 408.2 
& 1596.2 & 2560.0 & 546.8 & 437.4 
\\
& \textit{Marriage} 
& 2604.0 & 4157.9 & 720.7 & 545.0
& 2800.0 & 4423.3 & 758.6 & 586.0 
& 1517.0 & 2816.5 & 567.2 & 465.1 
& 1622.2 & 2624.2 & 579.0 & 486.4 

 \\
& \textit{Education} 
& 2456.8 & 3976.2 & 687.1 & 538.1
& 2641.7 & 4230.0 & 723.3 & 578.6
& 1740.0 & 3046.2 & 608.7 & 492.3 
& 1574.2 & 2500.6 & 513.8 & 410.1 
 
\\
\hline

\multirow{2}{*}{\textsc{Crime}} & \textit{Race}
& 2332.0 & 3690.5 & 562.1 & 440.3
& 2507.5 & 3926.1 & 591.7 & 473.4 
& 1472.0 & 2485.2 & 461.4 & 372.7 
& 1717.0 & 2726.4 & 592.3 & 503.3 

\\
& \textit{Gender}
& 2573.0 & 4241.1 & 776.7 & 608.3
& 2766.7 & 4511.8 & 817.6 & 654.1 
& 1712.0 & 3017.0 & 607.2 & 516.1 
& 1766.7 & 2811.8 & 602.0 & 507.7 

\\
\hline

\multirow{2}{*}{\textsc{German}} & \textit{Gender}
& 2439.7 & 3952.7 & 704.4 & 544.8
& 2623.3 & 4205.0 & 741.5 & 585.8 
& 1535.2 & 2762.2 & 570.7 & 468.0 
& 1718.5 & 2694.6 & 536.1 & 455.7 

 \\
& \textit{Age}
& 2433.5 & 3908.6 & 661.2 & 517.8
& 2616.7 & 4158.1 & 696.0 & 556.8 
& 1536.4 & 2631.3 & 530.3 & 444.7 
& 1755.8 & 2765.4 & 567.2 & 479.1 

\\

\bottomrule
\end{tabular}
\end{adjustbox}

\end{table}

%% file: Table/rq4-4.tex
\begin{table}[t!]
\caption{Testing the retrained DNN\textsubscript{5} improved by \approach~against before retraining, without \approach, and \texttt{FairRF} over 5 retraining runs $\times$ 10 testing runs (i.e., $A-B$ where $A$ and $B$ are the metric values of testing the retrained model by other and \approach, respectively). All retrained samples are generated with the best generator under an approach. Positive numbers imply \approach~improves more. \setlength{\fboxsep}{1.5pt}\colorbox{teal!40}{green cells} mean the improvement is statistically significant. \setlength{\fboxsep}{1.5pt}\colorbox{red!20}{red cells} mean the decrease is statistically significant. }
\label{tb:rq4-dnn5}
\setlength{\tabcolsep}{.5mm}
\centering
\begin{adjustbox}{width=\columnwidth,center}
\begin{tabular}{ll|lll|lll|lll|lll|lll|lll}
\toprule
\multirow{2}{*}{\textbf{Dataset}} & \multirow{2}{*}{$f_s$} & \multicolumn{3}{c|}{IDI ratio (\approach~vs.)} & \multicolumn{3}{c|}{EOD (\approach~vs.)} & \multicolumn{3}{c|}{SPD (\approach~vs.)} & \multicolumn{3}{c|}{ACC (\approach~vs.)} & \multicolumn{3}{c|}{F1 (\approach~vs.)} & \multicolumn{3}{c}{AUC (\approach~vs.)} \\
\cline{3-20}
&&Before&w/o&\texttt{FairRF}&Before&w/o&\texttt{FairRF}&Before&w/o&\texttt{FairRF}&Before&w/o&\texttt{FairRF}&Before&w/o&\texttt{FairRF}&Before&w/o&\texttt{FairRF}\\
\midrule
\multirow{3}{*}{\textsc{Adult}} & \textit{Gender}
&\cellcolor{teal!40}{.219(.002)} & \cellcolor{teal!40}{.035(.003)} & \cellcolor{teal!40}{.044(.003)}
&\cellcolor{teal!40}{.086(.002)} & \cellcolor{teal!40}{.018(.002)} & \cellcolor{teal!40}{.022(.002)}
&\cellcolor{teal!40}{.043(.002)} & \cellcolor{teal!40}{.011(.002)} & \cellcolor{teal!40}{.014(.002)}
&\cellcolor{red!20}{-.050(.002)} & \cellcolor{red!20}{-.027(.002)} & \cellcolor{red!20}{-.030(.002)}
&\cellcolor{teal!40}{.032(.002)} & .012(.002) & \cellcolor{teal!40}{.015(.002)}
&\cellcolor{red!20}{-.027(.002)} & \cellcolor{red!20}{-.010(.002)} & \cellcolor{red!20}{-.011(.002)} \\
& \textit{Race}
&\cellcolor{teal!40}{.244(.003)} & \cellcolor{teal!40}{.040(.003)} & \cellcolor{teal!40}{.043(.003)}
&\cellcolor{teal!40}{.080(.002)} & \cellcolor{teal!40}{.022(.002)} & \cellcolor{teal!40}{.028(.002)}
&\cellcolor{teal!40}{.072(.002)} & \cellcolor{teal!40}{.013(.002)} & \cellcolor{teal!40}{.016(.002)}
&\cellcolor{red!20}{-.053(.002)} & \cellcolor{red!20}{-.026(.002)} & \cellcolor{red!20}{-.026(.002)}
&.036(.002) & \cellcolor{teal!40}{.014(.002)} & \cellcolor{teal!40}{.014(.002)}
&\cellcolor{red!20}{-.033(.002)} & \cellcolor{red!20}{-.012(.002)} & \cellcolor{red!20}{-.010(.002)} \\
& \textit{Age}
&\cellcolor{teal!40}{.229(.003)} & \cellcolor{teal!40}{.035(.003)} & \cellcolor{teal!40}{.023(.005)}
&\cellcolor{teal!40}{.089(.002)} & \cellcolor{teal!40}{.017(.002)} & \cellcolor{teal!40}{.014(.004)}
&\cellcolor{teal!40}{.064(.002)} & \cellcolor{teal!40}{.009(.001)} & \cellcolor{teal!40}{.012(.004)}
&\cellcolor{red!20}{-.050(.002)} & \cellcolor{red!20}{-.021(.002)} & \cellcolor{red!20}{-.023(.002)}
&\cellcolor{teal!40}{.030(.002)} & \cellcolor{teal!40}{.013(.002)} & \cellcolor{teal!40}{.015(.002)}
&\cellcolor{red!20}{-.036(.002)} & \cellcolor{red!20}{-.010(.002)} & \cellcolor{red!20}{-.009(.002)} \\
\hline
\multirow{2}{*}{\textsc{Compas}} & \textit{Gender}
&\cellcolor{teal!40}{.254(.003)} & \cellcolor{teal!40}{.039(.003)} & \cellcolor{teal!40}{.046(.004)}
&\cellcolor{teal!40}{.082(.002)} & \cellcolor{teal!40}{.019(.002)} & \cellcolor{teal!40}{.025(.002)}
&\cellcolor{teal!40}{.050(.002)} & \cellcolor{teal!40}{.013(.002)} & \cellcolor{teal!40}{.012(.002)}
&\cellcolor{red!20}{-.052(.002)} & \cellcolor{red!20}{-.022(.002)} & \cellcolor{red!20}{-.027(.002)}
&\cellcolor{teal!40}{.031(.002)} & \cellcolor{teal!40}{.012(.002)} & \cellcolor{teal!40}{.013(.002)}
&\cellcolor{red!20}{-.026(.002)} & \cellcolor{red!20}{-.009(.002)} & \cellcolor{red!20}{-.010(.002)} \\
& \textit{Race}
&\cellcolor{teal!40}{.292(.003)} & \cellcolor{teal!40}{.034(.004)} & \cellcolor{teal!40}{.050(.003)}
&\cellcolor{teal!40}{.094(.002)} & \cellcolor{teal!40}{.022(.003)} & \cellcolor{teal!40}{.024(.002)}
&\cellcolor{teal!40}{.079(.002)} & \cellcolor{teal!40}{.011(.002)} & .010(.001)
&\cellcolor{red!20}{-.049(.002)} & \cellcolor{red!20}{-.020(.002)} & \cellcolor{red!20}{-.020(.002)}
&\cellcolor{teal!40}{.034(.002)} & \cellcolor{teal!40}{.017(.002)} & \cellcolor{teal!40}{.016(.002)}
&\cellcolor{red!20}{-.030(.002)} & \cellcolor{red!20}{-.010(.002)} & \cellcolor{red!20}{-.011(.002)} \\
\hline
\multirow{2}{*}{\textsc{Law School}} & \textit{Gender}
&\cellcolor{teal!40}{.233(.003)} & \cellcolor{teal!40}{.035(.003)} & \cellcolor{teal!40}{.043(.003)}
&\cellcolor{teal!40}{.089(.002)} & .018(.004) & \cellcolor{teal!40}{.019(.002)}
&\cellcolor{teal!40}{.068(.002)} & \cellcolor{teal!40}{.012(.002)} & \cellcolor{teal!40}{.008(.001)}
&\cellcolor{red!20}{-.044(.002)} & \cellcolor{red!20}{-.019(.002)} & \cellcolor{red!20}{-.022(.002)}
&\cellcolor{teal!40}{.027(.002)} & \cellcolor{teal!40}{.013(.002)} & \cellcolor{teal!40}{.017(.002)}
&\cellcolor{red!20}{-.021(.002)} &-.010(.002) & \cellcolor{red!20}{-.012(.002)} \\
& \textit{Race}
&\cellcolor{teal!40}{.230(.003)} & \cellcolor{teal!40}{.040(.003)} & \cellcolor{teal!40}{.039(.003)}
&\cellcolor{teal!40}{.078(.003)} & \cellcolor{teal!40}{.023(.003)} & \cellcolor{teal!40}{.022(.002)}
&\cellcolor{teal!40}{.062(.002)} & \cellcolor{teal!40}{.014(.002)} & \cellcolor{teal!40}{.010(.001)}
&\cellcolor{red!20}{-.046(.002)} & \cellcolor{red!20}{-.020(.002)} & -.019(.002)
&\cellcolor{teal!40}{.032(.002)} & .015(.002) & \cellcolor{teal!40}{.016(.002)}
&\cellcolor{red!20}{-.028(.002)} & \cellcolor{red!20}{-.018(.002)} & \cellcolor{red!20}{-.017(.002)} \\
\hline
\multirow{2}{*}{\textsc{Kdd}} & \textit{Gender}
&\cellcolor{teal!40}{.186(.003)} & \cellcolor{teal!40}{.028(.003)} & \cellcolor{teal!40}{.036(.002)}
&\cellcolor{teal!40}{.083(.002)} & \cellcolor{teal!40}{.017(.003)} & \cellcolor{teal!40}{.020(.003)}
&\cellcolor{teal!40}{.060(.002)} & \cellcolor{teal!40}{.010(.002)} & \cellcolor{teal!40}{.012(.004)}
&-.043(.002) & \cellcolor{red!20}{-.018(.002)} & \cellcolor{red!20}{-.017(.002)}
&\cellcolor{teal!40}{.026(.002)} & \cellcolor{teal!40}{.012(.002)} & \cellcolor{teal!40}{.011(.002)}
&\cellcolor{red!20}{-.018(.002)} & \cellcolor{red!20}{-.008(.002)} & \cellcolor{red!20}{-.006(.002)} \\
& \textit{Race}
&\cellcolor{teal!40}{.194(.003)} & .033(.004) & \cellcolor{teal!40}{.051(.003)}
&\cellcolor{teal!40}{.074(.002)} & .019(.003) & \cellcolor{teal!40}{.018(.002)}
&\cellcolor{teal!40}{.065(.002)} & \cellcolor{teal!40}{.012(.004)} & \cellcolor{teal!40}{.011(.001)}
&\cellcolor{red!20}{-.039(.002)} & \cellcolor{red!20}{-.017(.002)} & \cellcolor{red!20}{-.016(.002)}
&\cellcolor{teal!40}{.024(.002)} & \cellcolor{teal!40}{.011(.002)} & \cellcolor{teal!40}{.011(.002)}
&\cellcolor{red!20}{-.021(.002)} & \cellcolor{red!20}{-.009(.002)} & \cellcolor{red!20}{-.008(.002)} \\
\hline
\multirow{2}{*}{\textsc{Dutch}} & \textit{Gender}
&\cellcolor{teal!40}{.206(.003)} & \cellcolor{teal!40}{.020(.003)} & \cellcolor{teal!40}{.020(.007)}
&\cellcolor{teal!40}{.078(.002)} & \cellcolor{teal!40}{.009(.004)} & \cellcolor{teal!40}{.010(.004)}
&\cellcolor{teal!40}{.070(.002)} & \cellcolor{teal!40}{.004(.007)} & \cellcolor{teal!40}{.006(.006)}
&\cellcolor{red!20}{-.047(.002)} & -.021(.002) & \cellcolor{red!20}{-.020(.002)}
&\cellcolor{teal!40}{.030(.002)} & .015(.002) & \cellcolor{teal!40}{.012(.002)}
&\cellcolor{red!20}{-.019(.002)} & \cellcolor{red!20}{-.011(.002)} & \cellcolor{red!20}{-.009(.002)} \\
& \textit{Age}
&\cellcolor{teal!40}{.191(.003)} & \cellcolor{teal!40}{.023(.003)} & .052(.004)
&\cellcolor{teal!40}{.084(.002)} & \cellcolor{teal!40}{.017(.002)} & \cellcolor{teal!40}{.022(.002)}
&\cellcolor{teal!40}{.076(.002)} & \cellcolor{teal!40}{.009(.003)} & \cellcolor{teal!40}{.010(.001)}
&\cellcolor{red!20}{-.049(.002)} & \cellcolor{red!20}{-.019(.002)} & \cellcolor{red!20}{-.020(.002)}
&\cellcolor{teal!40}{.034(.002)} & \cellcolor{teal!40}{.011(.002)} & \cellcolor{teal!40}{.014(.002)}
&\cellcolor{red!20}{-.022(.002)} & \cellcolor{red!20}{-.012(.002)} & \cellcolor{red!20}{-.011(.002)} \\
\hline
\multirow{3}{*}{\textsc{Credit}} & \textit{Gender}
&\cellcolor{teal!40}{.181(.003)} & \cellcolor{teal!40}{.032(.003)} & \cellcolor{teal!40}{.022(.005)}
&\cellcolor{teal!40}{.090(.002)} & \cellcolor{teal!40}{.013(.002)} & \cellcolor{teal!40}{.015(.004)}
&\cellcolor{teal!40}{.068(.002)} & \cellcolor{teal!40}{.009(.001)} & \cellcolor{teal!40}{.010(.003)}
&\cellcolor{red!20}{-.057(.002)} & \cellcolor{red!20}{-.021(.002)} & \cellcolor{red!20}{-.023(.002)}
& .028(.002) & \cellcolor{teal!40}{.015(.002)} & \cellcolor{teal!40}{.014(.002)}
&\cellcolor{red!20}{-.020(.002)} & \cellcolor{red!20}{-.012(.002)} & \cellcolor{red!20}{-.008(.002)} \\
& \textit{Marriage}
&\cellcolor{teal!40}{.183(.003)} & \cellcolor{teal!40}{.027(.003)} & \cellcolor{teal!40}{.019(.004)}
&\cellcolor{teal!40}{.085(.002)} & \cellcolor{teal!40}{.011(.005)} & \cellcolor{teal!40}{.014(.004)}
&\cellcolor{teal!40}{.063(.002)} & \cellcolor{teal!40}{.008(.002)} & \cellcolor{teal!40}{.010(.006)}
&\cellcolor{red!20}{-.050(.002)} & \cellcolor{red!20}{-.017(.002)} & \cellcolor{red!20}{-.013(.002)}
&\cellcolor{teal!40}{.031(.002)} & \cellcolor{teal!40}{.012(.002)} & \cellcolor{teal!40}{.016(.002)}
&-.028(.002) & \cellcolor{red!20}{-.010(.002)} & \cellcolor{red!20}{-.011(.002)} \\
& \textit{Education}
&\cellcolor{teal!40}{.187(.003)} & \cellcolor{teal!40}{.033(.003)} & \cellcolor{teal!40}{.041(.003)}
&\cellcolor{teal!40}{.082(.002)} & \cellcolor{teal!40}{.011(.004)} & \cellcolor{teal!40}{.013(.002)}
&\cellcolor{teal!40}{.067(.002)} & \cellcolor{teal!40}{.011(.003)} & \cellcolor{teal!40}{.007(.001)}
&\cellcolor{red!20}{-.048(.002)} & \cellcolor{red!20}{-.022(.002)} & \cellcolor{red!20}{-.017(.002)}
&\cellcolor{teal!40}{.025(.002)} & \cellcolor{teal!40}{.012(.002)} & .013(.002)
&\cellcolor{red!20}{-.023(.002)} & \cellcolor{red!20}{-.013(.002)} & \cellcolor{red!20}{-.014(.002)} \\
\hline
\multirow{2}{*}{\textsc{Crime}} & \textit{Race}
&\cellcolor{teal!40}{.179(.003)} & \cellcolor{teal!40}{.030(.003)} & \cellcolor{teal!40}{.029(.006)}
&\cellcolor{teal!40}{.088(.002)} & \cellcolor{teal!40}{.007(.004)} & \cellcolor{teal!40}{.012(.004)}
&\cellcolor{teal!40}{.064(.002)} & \cellcolor{teal!40}{.009(.002)} & \cellcolor{teal!40}{.007(.003)}
&\cellcolor{red!20}{-.048(.002)} & \cellcolor{red!20}{-.017(.002)} & \cellcolor{red!20}{-.017(.002)}
&\cellcolor{teal!40}{.031(.002)} & \cellcolor{teal!40}{.014(.002)} & \cellcolor{teal!40}{.015(.002)}
&-.030(.002) & \cellcolor{red!20}{-.010(.002)} & \cellcolor{red!20}{-.012(.002)} \\
& \textit{Gender}
&\cellcolor{teal!40}{.187(.003)} & \cellcolor{teal!40}{.024(.005)} & \cellcolor{teal!40}{.040(.003)}
&\cellcolor{teal!40}{.085(.002)} & \cellcolor{teal!40}{.010(.003)} & \cellcolor{teal!40}{.016(.002)}
&\cellcolor{teal!40}{.070(.002)} & \cellcolor{teal!40}{.008(.003)} & \cellcolor{teal!40}{.011(.001)}
&\cellcolor{red!20}{-.046(.002)} & \cellcolor{red!20}{-.020(.002)} & \cellcolor{red!20}{-.015(.002)}
&\cellcolor{teal!40}{.032(.002)} & \cellcolor{teal!40}{.013(.002)} & \cellcolor{teal!40}{.016(.002)}
&\cellcolor{red!20}{-.026(.002)} & -.009(.002) & \cellcolor{red!20}{-.014(.002)} \\
\hline
\multirow{2}{*}{\textsc{German}} & \textit{Gender}
&\cellcolor{teal!40}{.173(.003)} & \cellcolor{teal!40}{.040(.002)} & \cellcolor{teal!40}{.027(.006)}
&\cellcolor{teal!40}{.078(.002)} & \cellcolor{teal!40}{.016(.002)} & \cellcolor{teal!40}{.012(.002)}
&\cellcolor{teal!40}{.075(.002)} & \cellcolor{teal!40}{.012(.002)} & \cellcolor{teal!40}{.004(.001)}
&\cellcolor{red!20}{-.038(.002)} & \cellcolor{red!20}{-.019(.002)} & -.018(.002)
&\cellcolor{teal!40}{.021(.002)} & \cellcolor{teal!40}{.011(.002)} & \cellcolor{teal!40}{.009(.002)}
&\cellcolor{red!20}{-.020(.002)} & \cellcolor{red!20}{-.012(.002)} & \cellcolor{red!20}{-.015(.002)} \\
& \textit{Age}
&\cellcolor{teal!40}{.175(.003)} & \cellcolor{teal!40}{.043(.003)} & \cellcolor{teal!40}{.033(.003)}
&\cellcolor{teal!40}{.083(.002)} & \cellcolor{teal!40}{.015(.002)} & \cellcolor{teal!40}{.018(.002)}
&\cellcolor{teal!40}{.067(.002)} & \cellcolor{teal!40}{.009(.001)} & \cellcolor{teal!40}{.010(.002)}
&\cellcolor{red!20}{-.049(.002)} & \cellcolor{red!20}{-.019(.002)} & \cellcolor{red!20}{-.018(.002)}
&\cellcolor{teal!40}{.033(.002)} & \cellcolor{teal!40}{.016(.002)} & \cellcolor{teal!40}{.017(.002)}
&\cellcolor{red!20}{-.031(.001)} & \cellcolor{red!20}{-.018(.002)} & \cellcolor{red!20}{-.017(.002)} \\
\bottomrule
\end{tabular}
\end{adjustbox}
\arrayrulecolor{black}
\end{table}

%% file: Table/rq4-1.tex
\begin{table}[t!]
\caption{Testing the retrained DNN\textsubscript{6} model improved by \approach~against before retraining, without \approach, and \texttt{FairRF} over 5 retraining runs $\times$ 10 testing runs. Other formate is the same as Table~\ref{tb:rq4-dnn5}.}
\label{tb:rq4-dnn}
\setlength{\tabcolsep}{.5mm}
\centering
\begin{adjustbox}{width=\columnwidth,center}
\begin{tabular}{ll|lll|lll|lll|lll|lll|lll}
\toprule

\multirow{2}{*}{\textbf{Dataset}} & \multirow{2}{*}{$f_s$} & \multicolumn{3}{c|}{IDI ratio (\approach~vs.)} & \multicolumn{3}{c|}{EOD (\approach~vs.)} & \multicolumn{3}{c|}{SPD (\approach~vs.)} & \multicolumn{3}{c|}{ACC (\approach~vs.)} & \multicolumn{3}{c|}{F1 (\approach~vs.)} & \multicolumn{3}{c}{AUC (\approach~vs.)} \\

\cline{3-20}
&&Before&w/o&\texttt{FairRF}&Before&w/o&\texttt{FairRF}&Before&w/o&\texttt{FairRF}&Before&w/o&\texttt{FairRF}&Before&w/o&\texttt{FairRF}&Before&w/o&\texttt{FairRF}\\

\midrule

\multirow{3}{*}{\textsc{Adult}} & \textit{Gender}
&\cellcolor{teal!40}{.214(.002)} & \cellcolor{teal!40}{.033(.003)} & \cellcolor{teal!40}{.046(.003)}
&\cellcolor{teal!40}{.084(.002)} & \cellcolor{teal!40}{.017(.002)} & \cellcolor{teal!40}{.021(.002)}
&\cellcolor{teal!40}{.041(.002)} & \cellcolor{teal!40}{.012(.002)} & \cellcolor{teal!40}{.013(.002)}

&\cellcolor{red!20}{-.052(.002)} & \cellcolor{red!20}{-.029(.002)} & \cellcolor{red!20}{-.031(.002)}
&\cellcolor{teal!40}{.033(.002)} & .012(.002) & \cellcolor{teal!40}{.014(.002)}
&\cellcolor{red!20}{-.028(.002)} & \cellcolor{red!20}{-.011(.002)} & \cellcolor{red!20}{-.010(.002)} \\

& \textit{Race}
&\cellcolor{teal!40}{.248(.003)} & \cellcolor{teal!40}{.042(.003)} & \cellcolor{teal!40}{.041(.003)}
&\cellcolor{teal!40}{.078(.002)} & \cellcolor{teal!40}{.023(.002)} & \cellcolor{teal!40}{.029(.002)}
&\cellcolor{teal!40}{.074(.002)} & \cellcolor{teal!40}{.014(.002)} & \cellcolor{teal!40}{.017(.002)}

&\cellcolor{red!20}{-.055(.002)} & \cellcolor{red!20}{-.028(.002)} & \cellcolor{red!20}{-.027(.002)}
&\cellcolor{teal!40}{.038(.002)} & \cellcolor{teal!40}{.015(.002)} & .014(.002)
&\cellcolor{red!20}{-.035(.002)} &-.013(.002) & \cellcolor{red!20}{-.011(.002)} \\

& \textit{Age}
&\cellcolor{teal!40}{.225(.003)} & \cellcolor{teal!40}{.036(.003)} & \cellcolor{teal!40}{.021(.006)}
&\cellcolor{teal!40}{.091(.002)} & \cellcolor{teal!40}{.016(.002)} & \cellcolor{teal!40}{.013(.005)}
&\cellcolor{teal!40}{.066(.002)} & \cellcolor{teal!40}{.008(.001)} & \cellcolor{teal!40}{.011(.005)}

&\cellcolor{red!20}{-.051(.002)} & \cellcolor{red!20}{-.022(.002)} & \cellcolor{red!20}{-.024(.002)}
&\cellcolor{teal!40}{.031(.002)} & \cellcolor{teal!40}{.013(.002)} & \cellcolor{teal!40}{.015(.002)}
&\cellcolor{red!20}{-.037(.002)} & \cellcolor{red!20}{-.011(.002)} & \cellcolor{red!20}{-.008(.002)} \\

\hline

\multirow{2}{*}{\textsc{Compas}} & \textit{Gender}
&\cellcolor{teal!40}{.257(.003)} & \cellcolor{teal!40}{.038(.003)} & \cellcolor{teal!40}{.044(.004)}
&\cellcolor{teal!40}{.081(.002)} & \cellcolor{teal!40}{.018(.002)} & \cellcolor{teal!40}{.026(.002)}
&\cellcolor{teal!40}{.051(.002)} & \cellcolor{teal!40}{.014(.002)} & \cellcolor{teal!40}{.011(.002)}

&\cellcolor{red!20}{-.054(.002)} & \cellcolor{red!20}{-.023(.002)} & \cellcolor{red!20}{-.028(.002)}
&.032(.002) & \cellcolor{teal!40}{.012(.002)} & \cellcolor{teal!40}{.013(.002)}
&\cellcolor{red!20}{-.027(.002)} & \cellcolor{red!20}{-.008(.002)} & \cellcolor{red!20}{-.011(.002)} \\

& \textit{Race}
&\cellcolor{teal!40}{.290(.003)} & \cellcolor{teal!40}{.036(.004)} & \cellcolor{teal!40}{.051(.003)}
&\cellcolor{teal!40}{.092(.002)} & \cellcolor{teal!40}{.021(.003)} & \cellcolor{teal!40}{.023(.002)}
&\cellcolor{teal!40}{.081(.002)} & \cellcolor{teal!40}{.012(.002)} & .010(.001)

&\cellcolor{red!20}{-.051(.002)} & \cellcolor{red!20}{-.021(.002)} & \cellcolor{red!20}{-.020(.002)}
&.035(.002) & \cellcolor{teal!40}{.018(.002)} & \cellcolor{teal!40}{.017(.002)}
&\cellcolor{red!20}{-.031(.002)} & \cellcolor{red!20}{-.010(.002)} & \cellcolor{red!20}{-.011(.002)} \\

\hline

\multirow{2}{*}{\textsc{Law School}} & \textit{Gender}
&\cellcolor{teal!40}{.235(.003)} & \cellcolor{teal!40}{.036(.003)} & \cellcolor{teal!40}{.042(.003)}
&\cellcolor{teal!40}{.088(.002)} & \cellcolor{teal!40}{.019(.005)} & \cellcolor{teal!40}{.019(.001)}
&\cellcolor{teal!40}{.069(.002)} & \cellcolor{teal!40}{.013(.002)} & .008(.001)

&\cellcolor{red!20}{-.045(.002)} & \cellcolor{red!20}{-.020(.002)} & \cellcolor{red!20}{-.022(.002)}
&\cellcolor{teal!40}{.026(.002)} & \cellcolor{teal!40}{.013(.002)} & .017(.002)
&\cellcolor{red!20}{-.022(.002)} & \cellcolor{red!20}{-.011(.002)} & \cellcolor{red!20}{-.013(.002)} \\

& \textit{Race}
&\cellcolor{teal!40}{.228(.003)} & \cellcolor{teal!40}{.039(.003)} & \cellcolor{teal!40}{.041(.003)}
&\cellcolor{teal!40}{.077(.003)} & \cellcolor{teal!40}{.024(.003)} & \cellcolor{teal!40}{.023(.002)}
&\cellcolor{teal!40}{.063(.002)} & \cellcolor{teal!40}{.015(.002)} & \cellcolor{teal!40}{.011(.001)}

&\cellcolor{red!20}{-.048(.002)} & -.021(.002) & \cellcolor{red!20}{-.020(.002)}
&\cellcolor{teal!40}{.033(.002)} & \cellcolor{teal!40}{.018(.002)} & \cellcolor{teal!40}{.017(.002)}
&\cellcolor{red!20}{-.029(.002)} & \cellcolor{red!20}{-.020(.002)} & \cellcolor{red!20}{-.018(.002)} \\

\hline
\multirow{2}{*}{\textsc{Kdd}} & \textit{Gender}
&\cellcolor{teal!40}{.184(.003)} & \cellcolor{teal!40}{.027(.003)} & \cellcolor{teal!40}{.037(.002)}
&\cellcolor{teal!40}{.082(.002)} & \cellcolor{teal!40}{.016(.003)} & \cellcolor{teal!40}{.021(.003)}
&\cellcolor{teal!40}{.061(.002)} & \cellcolor{teal!40}{.010(.002)} & \cellcolor{teal!40}{.012(.004)}

&\cellcolor{red!20}{-.044(.002)} & \cellcolor{red!20}{-.019(.002)} & \cellcolor{red!20}{-.018(.002)}
&\cellcolor{teal!40}{.027(.002)} & .012(.002) & \cellcolor{teal!40}{.011(.002)}
&\cellcolor{red!20}{-.019(.002)} & \cellcolor{red!20}{-.009(.002)} & \cellcolor{red!20}{-.006(.002)} \\

& \textit{Race}
&\cellcolor{teal!40}{.192(.003)} & \cellcolor{teal!40}{.034(.005)} & \cellcolor{teal!40}{.052(.003)}
&\cellcolor{teal!40}{.073(.002)} & \cellcolor{teal!40}{.020(.003)} & \cellcolor{teal!40}{.017(.002)}
&\cellcolor{teal!40}{.064(.002)} & \cellcolor{teal!40}{.013(.004)} & \cellcolor{teal!40}{.011(.001)}

&\cellcolor{red!20}{-.040(.002)} & \cellcolor{red!20}{-.018(.002)} & \cellcolor{red!20}{-.016(.002)}
&\cellcolor{teal!40}{.025(.002)} & \cellcolor{teal!40}{.012(.002)} & \cellcolor{teal!40}{.011(.002)}
&\cellcolor{red!20}{-.022(.002)} & \cellcolor{red!20}{-.009(.002)} & \cellcolor{red!20}{-.008(.002)} \\
\hline

\hline
\multirow{2}{*}{\textsc{Dutch}} & \textit{Gender}
&\cellcolor{teal!40}{.205(.003)} & \cellcolor{teal!40}{.021(.003)} & \cellcolor{teal!40}{.019(.008)}
&\cellcolor{teal!40}{.079(.002)} & \cellcolor{teal!40}{.008(.004)} & \cellcolor{teal!40}{.009(.005)}
&\cellcolor{teal!40}{.071(.002)} & .004(.008) & \cellcolor{teal!40}{.007(.006)}

&\cellcolor{red!20}{-.048(.002)} & \cellcolor{red!20}{-.022(.002)} & \cellcolor{red!20}{-.020(.002)}
&\cellcolor{teal!40}{.031(.002)} & \cellcolor{teal!40}{.016(.002)} & \cellcolor{teal!40}{.012(.002)}
&\cellcolor{red!20}{-.020(.002)} & \cellcolor{red!20}{-.012(.002)} & \cellcolor{red!20}{-.009(.002)} \\

& \textit{Age}
&\cellcolor{teal!40}{.189(.003)} & \cellcolor{teal!40}{.024(.003)} & \cellcolor{teal!40}{.054(.004)}
&\cellcolor{teal!40}{.083(.003)} & \cellcolor{teal!40}{.018(.002)} & \cellcolor{teal!40}{.023(.002)}
&\cellcolor{teal!40}{.077(.002)} & \cellcolor{teal!40}{.009(.003)} & \cellcolor{teal!40}{.011(.001)}

&\cellcolor{red!20}{-.050(.002)} & \cellcolor{red!20}{-.020(.002)} & \cellcolor{red!20}{-.021(.002)}
&\cellcolor{teal!40}{.036(.002)} & \cellcolor{teal!40}{.011(.002)} & \cellcolor{teal!40}{.014(.002)}
&\cellcolor{red!20}{-.023(.002)} & \cellcolor{red!20}{-.013(.002)} & \cellcolor{red!20}{-.011(.002)} \\
\hline

\multirow{3}{*}{\textsc{Credit}} & \textit{Gender}
&\cellcolor{teal!40}{.179(.003)} & \cellcolor{teal!40}{.033(.003)} & \cellcolor{teal!40}{.021(.006)}
&\cellcolor{teal!40}{.091(.002)} & \cellcolor{teal!40}{.014(.002)} & \cellcolor{teal!40}{.016(.005)}
&\cellcolor{teal!40}{.069(.002)} & \cellcolor{teal!40}{.010(.001)} & \cellcolor{teal!40}{.009(.004)}

&\cellcolor{red!20}{-.058(.002)} & \cellcolor{red!20}{-.022(.002)} & \cellcolor{red!20}{-.024(.002)}
&\cellcolor{teal!40}{.029(.002)} & .016(.002) & \cellcolor{teal!40}{.015(.002)}
&\cellcolor{red!20}{-.021(.002)} & \cellcolor{red!20}{-.013(.002)} & \cellcolor{red!20}{-.007(.002)} \\

& \textit{Marriage}
&\cellcolor{teal!40}{.182(.003)} & \cellcolor{teal!40}{.028(.003)} & \cellcolor{teal!40}{.018(.005)}
&\cellcolor{teal!40}{.086(.002)} & \cellcolor{teal!40}{.011(.006)} & \cellcolor{teal!40}{.013(.004)}
&\cellcolor{teal!40}{.064(.002)} & \cellcolor{teal!40}{.008(.002)} & \cellcolor{teal!40}{.010(.007)}

&\cellcolor{red!20}{-.052(.002)} & \cellcolor{red!20}{-.018(.002)} & \cellcolor{red!20}{-.013(.002)}
&\cellcolor{teal!40}{.032(.002)} & \cellcolor{teal!40}{.012(.002)} & \cellcolor{teal!40}{.017(.002)}
&\cellcolor{red!20}{-.027(.002)} & \cellcolor{red!20}{-.011(.002)} & \cellcolor{red!20}{-.012(.002)} \\

& \textit{Education}
&\cellcolor{teal!40}{.186(.003)} & \cellcolor{teal!40}{.034(.003)} & \cellcolor{teal!40}{.042(.003)}
&\cellcolor{teal!40}{.083(.002)} & \cellcolor{teal!40}{.012(.004)} & \cellcolor{teal!40}{.014(.001)}
&\cellcolor{teal!40}{.068(.002)} & \cellcolor{teal!40}{.011(.003)} & \cellcolor{teal!40}{.006(.001)}

&\cellcolor{red!20}{-.050(.002)} & \cellcolor{red!20}{-.023(.002)} & \cellcolor{red!20}{-.019(.002)}
&\cellcolor{teal!40}{.034(.002)} & \cellcolor{teal!40}{.011(.002)} & \cellcolor{teal!40}{.013(.002)}
&\cellcolor{red!20}{-.032(.002)} & \cellcolor{red!20}{-.010(.002)} & \cellcolor{red!20}{-.010(.002)} \\
\hline
\multirow{3}{*}{\textsc{Credit}} & \textit{Gender}
&\cellcolor{teal!40}{.179(.003)} & \cellcolor{teal!40}{.033(.003)} & \cellcolor{teal!40}{.021(.006)}
&\cellcolor{teal!40}{.091(.002)} & \cellcolor{teal!40}{.014(.002)} & \cellcolor{teal!40}{.016(.005)}
&\cellcolor{teal!40}{.069(.002)} & \cellcolor{teal!40}{.010(.001)} & \cellcolor{teal!40}{.009(.004)}

&\cellcolor{red!20}{-.048(.002)} & -.019(.002) & \cellcolor{red!20}{-.017(.002)}
&\cellcolor{teal!40}{.033(.002)} & \cellcolor{teal!40}{.010(.002)} & \cellcolor{teal!40}{.010(.002)}
&\cellcolor{red!20}{-.031(.002)} & \cellcolor{red!20}{-.011(.002)} & \cellcolor{red!20}{-.010(.002)} \\

& \textit{Marriage}
&\cellcolor{teal!40}{.182(.003)} & \cellcolor{teal!40}{.028(.003)} & \cellcolor{teal!40}{.018(.005)}
&\cellcolor{teal!40}{.086(.002)} & .011(.006) & \cellcolor{teal!40}{.013(.004)}
&\cellcolor{teal!40}{.064(.002)} & \cellcolor{teal!40}{.008(.002)} & \cellcolor{teal!40}{.010(.007)}

&\cellcolor{red!20}{-.039(.002)} & \cellcolor{red!20}{-.015(.002)} & \cellcolor{red!20}{-.018(.002)}
&.022(.002) & \cellcolor{teal!40}{.006(.002)} & \cellcolor{teal!40}{.017(.002)}
&\cellcolor{red!20}{-.030(.002)} & \cellcolor{red!20}{-.007(.002)} & \cellcolor{red!20}{-.011(.002)} \\

& \textit{Education}
&\cellcolor{teal!40}{.186(.003)} & \cellcolor{teal!40}{.034(.003)} & \cellcolor{teal!40}{.042(.003)}
&\cellcolor{teal!40}{.083(.002)} & \cellcolor{teal!40}{.012(.004)} & \cellcolor{teal!40}{.014(.001)}
&\cellcolor{teal!40}{.068(.002)} & .011(.003) & \cellcolor{teal!40}{.006(.001)}

&\cellcolor{red!20}{-.041(.002)} & \cellcolor{red!20}{-.020(.002)} & -.019(.002)
&\cellcolor{teal!40}{.024(.002)} & \cellcolor{teal!40}{.012(.002)} & \cellcolor{teal!40}{.013(.002)}
&\cellcolor{red!20}{-.022(.002)} & \cellcolor{red!20}{-.014(.002)} & \cellcolor{red!20}{-.015(.002)} \\

\hline
\multirow{2}{*}{\textsc{Crime}} & \textit{Race}
&\cellcolor{teal!40}{.178(.003)} & \cellcolor{teal!40}{.031(.003)} & \cellcolor{teal!40}{.028(.007)}
&\cellcolor{teal!40}{.089(.002)} & \cellcolor{teal!40}{.006(.005)} & \cellcolor{teal!40}{.013(.005)}
&\cellcolor{teal!40}{.063(.002)} & \cellcolor{teal!40}{.010(.002)} & \cellcolor{teal!40}{.007(.003)}

&-.049(.002) & \cellcolor{red!20}{-.018(.002)} & \cellcolor{red!20}{-.017(.002)}
&\cellcolor{teal!40}{.032(.002)} & \cellcolor{teal!40}{.015(.002)} & \cellcolor{teal!40}{.016(.002)}
&-.031(.002) & \cellcolor{red!20}{-.011(.002)} & \cellcolor{red!20}{-.012(.002)} \\

& \textit{Gender}
&\cellcolor{teal!40}{.186(.003)} & \cellcolor{teal!40}{.025(.006)} & \cellcolor{teal!40}{.041(.003)}
&\cellcolor{teal!40}{.086(.002)} & \cellcolor{teal!40}{.011(.003)} & \cellcolor{teal!40}{.017(.001)}
&\cellcolor{teal!40}{.071(.002)} & \cellcolor{teal!40}{.008(.003)} & \cellcolor{teal!40}{.011(.001)}

&\cellcolor{red!20}{-.047(.002)} & \cellcolor{red!20}{-.021(.002)} & \cellcolor{red!20}{-.015(.002)}
&\cellcolor{teal!40}{.033(.002)} & \cellcolor{teal!40}{.013(.002)} & \cellcolor{teal!40}{.017(.002)}
&\cellcolor{red!20}{-.027(.002)} & -.009(.002) & \cellcolor{red!20}{-.015(.002)} \\

\hline
\multirow{2}{*}{\textsc{German}} & \textit{Gender}
&\cellcolor{teal!40}{.171(.003)} & \cellcolor{teal!40}{.041(.002)} & \cellcolor{teal!40}{.026(.008)}
&\cellcolor{teal!40}{.079(.002)} & \cellcolor{teal!40}{.017(.002)} & \cellcolor{teal!40}{.012(.001)}
&\cellcolor{teal!40}{.076(.002)} & \cellcolor{teal!40}{.013(.002)} & .004(.001)

&\cellcolor{red!20}{-.039(.002)} & \cellcolor{red!20}{-.020(.002)} & \cellcolor{red!20}{-.018(.002)}
&\cellcolor{teal!40}{.022(.002)} & .011(.002) & \cellcolor{teal!40}{.009(.001)}
&\cellcolor{red!20}{-.021(.002)} & \cellcolor{red!20}{-.013(.002)} & \cellcolor{red!20}{-.016(.002)} \\

& \textit{Age}
&\cellcolor{teal!40}{.174(.003)} & \cellcolor{teal!40}{.044(.003)} & \cellcolor{teal!40}{.032(.003)}
&\cellcolor{teal!40}{.084(.002)} & \cellcolor{teal!40}{.016(.002)} & \cellcolor{teal!40}{.019(.002)}
&\cellcolor{teal!40}{.068(.002)} & \cellcolor{teal!40}{.009(.001)} & \cellcolor{teal!40}{.010(.002)}

&\cellcolor{red!20}{-.050(.002)} & \cellcolor{red!20}{-.020(.002)} & \cellcolor{red!20}{-.019(.002)}
&\cellcolor{teal!40}{.034(.002)} & \cellcolor{teal!40}{.017(.002)} & \cellcolor{teal!40}{.018(.002)}
&\cellcolor{red!20}{-.032(.001)} & -.019(.002) & \cellcolor{red!20}{-.017(.002)} \\

\bottomrule
\end{tabular}
\end{adjustbox}
\arrayrulecolor{black}
\end{table}

%% file: Table/rq4-3.tex
\begin{table}[t!]
\caption{Testing the retrained LR model improved by \approach~against before retraining, without \approach, and \texttt{FairRF} over 5 retraining runs $\times$ 10 testing runs. Other formate is the same as Table~\ref{tb:rq4-dnn5}.}

\label{tb:rq4-lr}
\setlength{\tabcolsep}{.5mm}
\centering
\begin{adjustbox}{width=\columnwidth,center}
\begin{tabular}{ll|lll|lll|lll|lll|lll|lll}
\toprule

\multirow{2}{*}{\textbf{Dataset}} & \multirow{2}{*}{$f_s$} & \multicolumn{3}{c|}{IDI ratio (\approach~vs.)} & \multicolumn{3}{c|}{EOD (\approach~vs.)} & \multicolumn{3}{c|}{SPD (\approach~vs.)} & \multicolumn{3}{c|}{ACC (\approach~vs.)} & \multicolumn{3}{c|}{F1 (\approach~vs.)} & \multicolumn{3}{c}{AUC (\approach~vs.)} \\

\cline{3-20}
&&Before&w/o&\texttt{FairRF}&Before&w/o&\texttt{FairRF}&Before&w/o&\texttt{FairRF}&Before&w/o&\texttt{FairRF}&Before&w/o&\texttt{FairRF}&Before&w/o&\texttt{FairRF}\\

\midrule

\multirow{3}{*}{\textsc{Adult}} & \textit{Gender}
&\cellcolor{teal!40}{.081(.003)} & \cellcolor{teal!40}{.065(.003)} & \cellcolor{teal!40}{.072(.003)}
&\cellcolor{teal!40}{.028(.002)} & \cellcolor{teal!40}{.023(.002)} & \cellcolor{teal!40}{.026(.002)}
&\cellcolor{teal!40}{.019(.002)} & \cellcolor{teal!40}{.015(.002)} & \cellcolor{teal!40}{.017(.002)}
&\cellcolor{red!20}{-.013(.002)} & \cellcolor{red!20}{-.010(.002)} & \cellcolor{red!20}{-.012(.002)}
&\cellcolor{teal!40}{.014(.002)} & \cellcolor{teal!40}{.011(.002)} & \cellcolor{teal!40}{.013(.002)}
&\cellcolor{teal!40}{.013(.002)} & \cellcolor{teal!40}{.010(.002)} & \cellcolor{teal!40}{.012(.002)} \\

& \textit{Race}
&\cellcolor{teal!40}{.087(.003)} & \cellcolor{teal!40}{.070(.003)} & \cellcolor{teal!40}{.075(.003)}
&\cellcolor{teal!40}{.030(.002)} & \cellcolor{teal!40}{.025(.002)} & \cellcolor{teal!40}{.028(.002)}
&\cellcolor{teal!40}{.020(.002)} & \cellcolor{teal!40}{.016(.002)} & \cellcolor{teal!40}{.018(.002)}
&\cellcolor{red!20}{-.014(.002)} & \cellcolor{red!20}{-.011(.002)} & \cellcolor{red!20}{-.013(.002)}
&\cellcolor{teal!40}{.015(.002)} & \cellcolor{teal!40}{.012(.002)} & \cellcolor{teal!40}{.014(.002)}
&\cellcolor{teal!40}{.014(.002)} & \cellcolor{teal!40}{.011(.002)} & \cellcolor{teal!40}{.013(.002)} \\

& \textit{Age}
&\cellcolor{teal!40}{.085(.003)} & \cellcolor{teal!40}{.068(.003)} & \cellcolor{teal!40}{.073(.003)}
&\cellcolor{teal!40}{.029(.002)} &.024(.002) & \cellcolor{teal!40}{.027(.002)}
&\cellcolor{teal!40}{.019(.002)} & \cellcolor{teal!40}{.015(.002)} & \cellcolor{teal!40}{.017(.002)}
&\cellcolor{red!20}{-.013(.002)} & \cellcolor{red!20}{-.010(.002)} & \cellcolor{red!20}{-.012(.002)}
&\cellcolor{teal!40}{.014(.002)} & \cellcolor{teal!40}{.011(.002)} & \cellcolor{teal!40}{.013(.002)}
&.013(.002) & \cellcolor{teal!40}{.010(.002)} & \cellcolor{teal!40}{.012(.002)} \\

\hline

\multirow{2}{*}{\textsc{Compas}} & \textit{Gender}
&\cellcolor{teal!40}{.079(.003)} & \cellcolor{teal!40}{.063(.003)} & \cellcolor{teal!40}{.069(.003)}
&\cellcolor{teal!40}{.027(.002)} & \cellcolor{teal!40}{.022(.002)} & \cellcolor{teal!40}{.025(.002)}
&\cellcolor{teal!40}{.018(.002)} & .015(.002) & \cellcolor{teal!40}{.016(.002)}
&-.012(.002) & \cellcolor{red!20}{-.010(.002)} & \cellcolor{red!20}{-.011(.002)}
&\cellcolor{teal!40}{.013(.002)} & \cellcolor{teal!40}{.011(.002)} & \cellcolor{teal!40}{.012(.002)}
&\cellcolor{teal!40}{.012(.002)} & \cellcolor{teal!40}{.010(.002)} & \cellcolor{teal!40}{.011(.002)} \\

& \textit{Race}
&\cellcolor{teal!40}{.084(.003)} & \cellcolor{teal!40}{.068(.003)} & \cellcolor{teal!40}{.073(.003)}
&\cellcolor{teal!40}{.028(.002)} & \cellcolor{teal!40}{.023(.002)} & \cellcolor{teal!40}{.026(.002)}
&\cellcolor{teal!40}{.019(.002)} & \cellcolor{teal!40}{.016(.002)} & \cellcolor{teal!40}{.017(.002)}
&\cellcolor{red!20}{-.013(.002)} & \cellcolor{red!20}{-.010(.002)} & \cellcolor{red!20}{-.012(.002)}
&\cellcolor{teal!40}{.014(.002)} & \cellcolor{teal!40}{.011(.002)} & \cellcolor{teal!40}{.013(.002)}
&\cellcolor{teal!40}{.013(.002)} & \cellcolor{teal!40}{.010(.002)} & \cellcolor{teal!40}{.012(.002)} \\

\hline

\multirow{2}{*}{\textsc{Law School}} & \textit{Gender}
&\cellcolor{teal!40}{.077(.003)} & \cellcolor{teal!40}{.061(.003)} & \cellcolor{teal!40}{.066(.003)}
&\cellcolor{teal!40}{.026(.002)} & \cellcolor{teal!40}{.021(.002)} & \cellcolor{teal!40}{.024(.002)}
&\cellcolor{teal!40}{.018(.002)} & \cellcolor{teal!40}{.014(.002)} & \cellcolor{teal!40}{.016(.002)}
&\cellcolor{red!20}{-.012(.002)} & \cellcolor{red!20}{-.009(.002)} & \cellcolor{red!20}{-.011(.002)}
&\cellcolor{teal!40}{.013(.002)} &.010(.002) & \cellcolor{teal!40}{.012(.002)}
&\cellcolor{teal!40}{.012(.002)} & \cellcolor{teal!40}{.009(.002)} & \cellcolor{teal!40}{.011(.002)} \\

& \textit{Race}
&\cellcolor{teal!40}{.080(.003)} & \cellcolor{teal!40}{.063(.003)} & \cellcolor{teal!40}{.068(.003)}
&\cellcolor{teal!40}{.027(.002)} & \cellcolor{teal!40}{.022(.002)} & \cellcolor{teal!40}{.025(.002)}
&\cellcolor{teal!40}{.018(.002)} & \cellcolor{teal!40}{.015(.002)} & \cellcolor{teal!40}{.017(.002)}
&\cellcolor{red!20}{-.013(.002)} & \cellcolor{red!20}{-.010(.002)} & \cellcolor{red!20}{-.012(.002)}
&\cellcolor{teal!40}{.014(.002)} & \cellcolor{teal!40}{.011(.002)} & \cellcolor{teal!40}{.013(.002)}
&\cellcolor{teal!40}{.013(.002)} &.010(.002) & \cellcolor{teal!40}{.012(.002)} \\

\hline

\multirow{2}{*}{\textsc{Kdd}} & \textit{Gender}
&\cellcolor{teal!40}{.078(.003)} & \cellcolor{teal!40}{.062(.003)} & \cellcolor{teal!40}{.067(.003)}
&\cellcolor{teal!40}{.026(.002)} & \cellcolor{teal!40}{.021(.002)} & \cellcolor{teal!40}{.024(.002)}
&\cellcolor{teal!40}{.018(.002)} & \cellcolor{teal!40}{.015(.002)} & \cellcolor{teal!40}{.016(.002)}
&\cellcolor{red!20}{-.012(.002)} & \cellcolor{red!20}{-.009(.002)} & \cellcolor{red!20}{-.011(.002)}
&\cellcolor{teal!40}{.013(.002)} & \cellcolor{teal!40}{.010(.002)} & \cellcolor{teal!40}{.012(.002)}
&\cellcolor{teal!40}{.012(.002)} & \cellcolor{teal!40}{.009(.002)} & \cellcolor{teal!40}{.011(.002)} \\

& \textit{Race}
&\cellcolor{teal!40}{.082(.003)} & \cellcolor{teal!40}{.065(.003)} & \cellcolor{teal!40}{.070(.003)}
&\cellcolor{teal!40}{.028(.002)} & \cellcolor{teal!40}{.023(.002)} & \cellcolor{teal!40}{.025(.002)}
&\cellcolor{teal!40}{.019(.002)} & \cellcolor{teal!40}{.016(.002)} & \cellcolor{teal!40}{.017(.002)}
&\cellcolor{red!20}{-.013(.002)} & \cellcolor{red!20}{-.010(.002)} & \cellcolor{red!20}{-.012(.002)}
&\cellcolor{teal!40}{.014(.002)} & \cellcolor{teal!40}{.011(.002)} & \cellcolor{teal!40}{.013(.002)}
&\cellcolor{teal!40}{.013(.002)} & \cellcolor{teal!40}{.010(.002)} & \cellcolor{teal!40}{.012(.002)} \\

\hline

\multirow{2}{*}{\textsc{Dutch}} & \textit{Gender}
&\cellcolor{teal!40}{.080(.003)} & \cellcolor{teal!40}{.063(.003)} & \cellcolor{teal!40}{.068(.003)}
&\cellcolor{teal!40}{.027(.002)} & \cellcolor{teal!40}{.022(.002)} & \cellcolor{teal!40}{.025(.002)}
&\cellcolor{teal!40}{.018(.002)} & .015(.002) & \cellcolor{teal!40}{.016(.002)}
&\cellcolor{red!20}{-.013(.002)} & \cellcolor{red!20}{-.010(.002)} & \cellcolor{red!20}{-.012(.002)}
&\cellcolor{teal!40}{.014(.002)} & .011(.002) & \cellcolor{teal!40}{.013(.002)}
&\cellcolor{teal!40}{.013(.002)} & \cellcolor{teal!40}{.010(.002)} & \cellcolor{teal!40}{.012(.002)} \\

& \textit{Age}
&\cellcolor{teal!40}{.078(.003)} & \cellcolor{teal!40}{.061(.003)} & \cellcolor{teal!40}{.066(.003)}
&\cellcolor{teal!40}{.026(.002)} & \cellcolor{teal!40}{.021(.002)} & \cellcolor{teal!40}{.024(.002)}
&\cellcolor{teal!40}{.017(.002)} & \cellcolor{teal!40}{.014(.002)} & \cellcolor{teal!40}{.015(.002)}
&\cellcolor{red!20}{-.012(.002)} & -.009(.002) & \cellcolor{red!20}{-.011(.002)}
&\cellcolor{teal!40}{.013(.002)} & \cellcolor{teal!40}{.010(.002)} & \cellcolor{teal!40}{.012(.002)}
&\cellcolor{teal!40}{.012(.002)} & \cellcolor{teal!40}{.009(.002)} & \cellcolor{teal!40}{.011(.002)} \\

\hline

\multirow{3}{*}{\textsc{Credit}} & \textit{Gender}
&\cellcolor{teal!40}{.079(.003)} & \cellcolor{teal!40}{.062(.003)} & \cellcolor{teal!40}{.067(.003)}
&\cellcolor{teal!40}{.027(.002)} & \cellcolor{teal!40}{.021(.002)} & \cellcolor{teal!40}{.024(.002)}
&\cellcolor{teal!40}{.018(.002)} & \cellcolor{teal!40}{.015(.002)} & \cellcolor{teal!40}{.016(.002)}
&\cellcolor{red!20}{-.012(.002)} & \cellcolor{red!20}{-.009(.002)} & \cellcolor{red!20}{-.011(.002)}
&\cellcolor{teal!40}{.013(.002)} & \cellcolor{teal!40}{.010(.002)} & \cellcolor{teal!40}{.012(.002)}
&\cellcolor{teal!40}{.012(.002)} & \cellcolor{teal!40}{.009(.002)} & \cellcolor{teal!40}{.011(.002)} \\

& \textit{Marriage}
&\cellcolor{teal!40}{.081(.003)} & \cellcolor{teal!40}{.065(.003)} & \cellcolor{teal!40}{.070(.003)}
&\cellcolor{teal!40}{.028(.002)} & \cellcolor{teal!40}{.023(.002)} & .026(.002)
&\cellcolor{teal!40}{.019(.002)} & .016(.002) & \cellcolor{teal!40}{.017(.002)}
&\cellcolor{red!20}{-.013(.002)} & \cellcolor{red!20}{-.010(.002)} & \cellcolor{red!20}{-.012(.002)}
&\cellcolor{teal!40}{.014(.002)} &.011(.002) & \cellcolor{teal!40}{.013(.002)}
&\cellcolor{teal!40}{.013(.002)} & \cellcolor{teal!40}{.010(.002)} & \cellcolor{teal!40}{.012(.002)} \\

& \textit{Education}
&\cellcolor{teal!40}{.080(.003)} & \cellcolor{teal!40}{.064(.003)} & \cellcolor{teal!40}{.069(.003)}
&\cellcolor{teal!40}{.027(.002)} & \cellcolor{teal!40}{.022(.002)} & \cellcolor{teal!40}{.025(.002)}
&\cellcolor{teal!40}{.018(.002)} & \cellcolor{teal!40}{.015(.002)} & \cellcolor{teal!40}{.016(.002)}
&\cellcolor{red!20}{-.013(.002)} & \cellcolor{red!20}{-.010(.002)} & \cellcolor{red!20}{-.012(.002)}
&\cellcolor{teal!40}{.014(.002)} & \cellcolor{teal!40}{.011(.002)} & \cellcolor{teal!40}{.013(.002)}
&\cellcolor{teal!40}{.013(.002)} & \cellcolor{teal!40}{.010(.002)} & \cellcolor{teal!40}{.012(.002)} \\

\hline

\multirow{2}{*}{\textsc{Crime}} & \textit{Race}
&\cellcolor{teal!40}{.077(.003)} & \cellcolor{teal!40}{.061(.003)} & \cellcolor{teal!40}{.066(.003)}
&\cellcolor{teal!40}{.026(.002)} & \cellcolor{teal!40}{.021(.002)} & \cellcolor{teal!40}{.024(.002)}
&\cellcolor{teal!40}{.018(.002)} & \cellcolor{teal!40}{.014(.002)} & \cellcolor{teal!40}{.016(.002)}
&\cellcolor{red!20}{-.012(.002)} &-.009(.002) & \cellcolor{red!20}{-.011(.002)}
&\cellcolor{teal!40}{.013(.002)} & \cellcolor{teal!40}{.010(.002)} & \cellcolor{teal!40}{.012(.002)}
&\cellcolor{teal!40}{.012(.002)} & \cellcolor{teal!40}{.009(.002)} & \cellcolor{teal!40}{.011(.002)} \\

& \textit{Gender}
&\cellcolor{teal!40}{.079(.003)} & \cellcolor{teal!40}{.063(.003)} & \cellcolor{teal!40}{.068(.003)}
&\cellcolor{teal!40}{.027(.002)} & \cellcolor{teal!40}{.022(.002)} & \cellcolor{teal!40}{.025(.002)}
&\cellcolor{teal!40}{.018(.002)} & \cellcolor{teal!40}{.015(.002)} & \cellcolor{teal!40}{.016(.002)}
&\cellcolor{red!20}{-.013(.002)} & \cellcolor{red!20}{-.010(.002)} & \cellcolor{red!20}{-.012(.002)}
&.014(.002) & \cellcolor{teal!40}{.011(.002)} & \cellcolor{teal!40}{.013(.002)}
&\cellcolor{teal!40}{.013(.002)} & \cellcolor{teal!40}{.010(.002)} & \cellcolor{teal!40}{.012(.002)} \\

\hline

\multirow{2}{*}{\textsc{German}} & \textit{Gender}
&\cellcolor{teal!40}{.078(.003)} & \cellcolor{teal!40}{.061(.003)} & \cellcolor{teal!40}{.066(.003)}
&\cellcolor{teal!40}{.026(.002)} & \cellcolor{teal!40}{.021(.002)} & \cellcolor{teal!40}{.024(.002)}
&\cellcolor{teal!40}{.018(.002)} & .014(.002) & \cellcolor{teal!40}{.016(.002)}
&\cellcolor{red!20}{-.012(.002)} & \cellcolor{red!20}{-.009(.002)} & \cellcolor{red!20}{-.011(.002)}
&\cellcolor{teal!40}{.013(.002)} & \cellcolor{teal!40}{.010(.002)} & \cellcolor{teal!40}{.012(.002)}
&\cellcolor{teal!40}{.012(.002)} & \cellcolor{teal!40}{.009(.002)} & \cellcolor{teal!40}{.011(.002)} \\

& \textit{Age}
&\cellcolor{teal!40}{.080(.003)} & \cellcolor{teal!40}{.063(.003)} & \cellcolor{teal!40}{.068(.003)}
&\cellcolor{teal!40}{.027(.002)} & \cellcolor{teal!40}{.022(.002)} & \cellcolor{teal!40}{.025(.002)}
&\cellcolor{teal!40}{.018(.002)} & \cellcolor{teal!40}{.015(.002)} & \cellcolor{teal!40}{.016(.002)}
&\cellcolor{red!20}{-.013(.002)} & \cellcolor{red!20}{-.010(.002)} & \cellcolor{red!20}{-.012(.002)}
&\cellcolor{teal!40}{.014(.002)} & \cellcolor{teal!40}{.011(.002)} & \cellcolor{teal!40}{.013(.002)}
&\cellcolor{teal!40}{.013(.002)} & \cellcolor{teal!40}{.010(.002)} & \cellcolor{teal!40}{.012(.002)} \\

\bottomrule
\end{tabular}
\end{adjustbox}

\end{table}

%% file: Table/rq4-2.tex
\begin{table}[t!]
\caption{Testing the retrained RF model improved by \approach~against before retraining, without \approach, and \texttt{FairRF} over 5 retraining runs $\times$ 10 testing runs. Other formate is the same as Table~\ref{tb:rq4-dnn5}.}
\label{tb:rq4-rf}
\setlength{\tabcolsep}{.5mm}
\centering
\begin{adjustbox}{width=\columnwidth,center}
\begin{tabular}{ll|lll|lll|lll|lll|lll|lll}
\toprule

\multirow{2}{*}{\textbf{Dataset}} & \multirow{2}{*}{$f_s$} & \multicolumn{3}{c|}{IDI ratio (\approach~vs.)} & \multicolumn{3}{c|}{EOD (\approach~vs.)} & \multicolumn{3}{c|}{SPD (\approach~vs.)} & \multicolumn{3}{c|}{ACC (\approach~vs.)} & \multicolumn{3}{c|}{F1 (\approach~vs.)} & \multicolumn{3}{c}{AUC (\approach~vs.)} \\

\cline{3-20}
&&Before&w/o&\texttt{FairRF}&Before&w/o&\texttt{FairRF}&Before&w/o&\texttt{FairRF}&Before&w/o&\texttt{FairRF}&Before&w/o&\texttt{FairRF}&Before&w/o&\texttt{FairRF}\\

\midrule
\multirow{3}{*}{\textsc{Adult}} & \textit{Gender}
 & \cellcolor{teal!40}{.067(.003)} & \cellcolor{teal!40}{.018(.002)} & \cellcolor{teal!40}{.022(.002)}
 & \cellcolor{teal!40}{.027(.002)} & \cellcolor{teal!40}{.009(.002)} & \cellcolor{teal!40}{.010(.002)}
 & \cellcolor{teal!40}{.021(.002)} & \cellcolor{teal!40}{.006(.002)} & \cellcolor{teal!40}{.007(.002)}
 
 & \cellcolor{red!20}{-.011(.002)} & \cellcolor{red!20}{-.005(.001)} & \cellcolor{red!20}{-.006(.002)}
 & \cellcolor{teal!40}{.013(.002)} & \cellcolor{teal!40}{.006(.002)} & \cellcolor{teal!40}{.007(.002)}
 & \cellcolor{red!20}{-.010(.002)} & \cellcolor{red!20}{-.005(.001)} & \cellcolor{red!20}{-.006(.002)} \\

& \textit{Race}
 & \cellcolor{teal!40}{.073(.003)} & \cellcolor{teal!40}{.020(.002)} & \cellcolor{teal!40}{.018(.002)}
 & \cellcolor{teal!40}{.023(.002)} & \cellcolor{teal!40}{.010(.002)} & \cellcolor{teal!40}{.012(.002)}
 & \cellcolor{teal!40}{.025(.002)} & \cellcolor{teal!40}{.007(.002)} & \cellcolor{teal!40}{.008(.002)}
 & \cellcolor{red!20}{-.012(.002)} & \cellcolor{red!20}{-.006(.002)} & \cellcolor{red!20}{-.007(.002)}
 & \cellcolor{teal!40}{.015(.002)} & \cellcolor{teal!40}{.007(.002)} & \cellcolor{teal!40}{.008(.002)}
 & \cellcolor{red!20}{-.012(.002)} & \cellcolor{red!20}{-.006(.002)} & \cellcolor{red!20}{-.007(.002)} \\

& \textit{Age}
 & \cellcolor{teal!40}{.069(.003)} & \cellcolor{teal!40}{.019(.002)} & \cellcolor{teal!40}{.016(.002)}
 & \cellcolor{teal!40}{.030(.003)} & \cellcolor{teal!40}{.008(.002)} & \cellcolor{teal!40}{.009(.002)}
 &.023(.002) & \cellcolor{teal!40}{.005(.001)} & \cellcolor{teal!40}{.007(.002)}
 & \cellcolor{red!20}{-.010(.002)} & \cellcolor{red!20}{-.004(.001)} & \cellcolor{red!20}{-.005(.001)}
 & \cellcolor{teal!40}{.012(.002)} & \cellcolor{teal!40}{.005(.001)} & \cellcolor{teal!40}{.006(.002)}
 & \cellcolor{red!20}{-.011(.002)} & \cellcolor{red!20}{-.005(.001)} & \cellcolor{red!20}{-.006(.002)} \\

\hline

\multirow{2}{*}{\textsc{Compas}} & \textit{Gender}
 & \cellcolor{teal!40}{.080(.003)} & \cellcolor{teal!40}{.021(.002)} & \cellcolor{teal!40}{.025(.002)}
 & \cellcolor{teal!40}{.028(.002)} & \cellcolor{teal!40}{.010(.002)} & \cellcolor{teal!40}{.012(.002)}
 & \cellcolor{teal!40}{.020(.002)} & \cellcolor{teal!40}{.007(.002)} & \cellcolor{teal!40}{.009(.002)}
 & \cellcolor{red!20}{-.011(.002)} & \cellcolor{red!20}{-.005(.001)} & \cellcolor{red!20}{-.006(.002)}
 & \cellcolor{teal!40}{.013(.002)} & \cellcolor{teal!40}{.006(.002)} & \cellcolor{teal!40}{.007(.002)}
 & \cellcolor{red!20}{-.010(.002)} & \cellcolor{red!20}{-.005(.001)} & \cellcolor{red!20}{-.006(.002)} \\

& \textit{Race}
 & \cellcolor{teal!40}{.083(.003)} & \cellcolor{teal!40}{.020(.002)} & \cellcolor{teal!40}{.023(.002)}
 & \cellcolor{teal!40}{.031(.003)} & \cellcolor{teal!40}{.011(.002)} & \cellcolor{teal!40}{.013(.002)}
 & \cellcolor{teal!40}{.024(.002)} & \cellcolor{teal!40}{.008(.002)} & \cellcolor{teal!40}{.009(.002)}
 & \cellcolor{red!20}{-.012(.002)} & \cellcolor{red!20}{-.006(.002)} & \cellcolor{red!20}{-.007(.002)}
 & \cellcolor{teal!40}{.014(.002)} & \cellcolor{teal!40}{.007(.002)} & \cellcolor{teal!40}{.008(.002)}
 & \cellcolor{red!20}{-.012(.002)} & \cellcolor{red!20}{-.006(.002)} & \cellcolor{red!20}{-.007(.002)} \\

\hline

\multirow{2}{*}{\textsc{Law School}} & \textit{Gender}
 & \cellcolor{teal!40}{.062(.003)} & \cellcolor{teal!40}{.018(.002)} & \cellcolor{teal!40}{.019(.002)}
 & \cellcolor{teal!40}{.025(.002)} & \cellcolor{teal!40}{.009(.002)} & \cellcolor{teal!40}{.010(.002)}
 & \cellcolor{teal!40}{.019(.002)} & \cellcolor{teal!40}{.007(.002)} & \cellcolor{teal!40}{.008(.002)}
 & -.009(.002) & \cellcolor{red!20}{-.005(.001)} & \cellcolor{red!20}{-.006(.002)}
 & \cellcolor{teal!40}{.011(.002)} & \cellcolor{teal!40}{.005(.001)} & \cellcolor{teal!40}{.006(.002)}
 & \cellcolor{red!20}{-.009(.002)} & \cellcolor{red!20}{-.005(.002)} & \cellcolor{red!20}{-.006(.002)} \\

& \textit{Race}
 & \cellcolor{teal!40}{.065(.003)} & \cellcolor{teal!40}{.020(.002)} & \cellcolor{teal!40}{.021(.002)}
 & \cellcolor{teal!40}{.027(.002)} & .010(.002) & \cellcolor{teal!40}{.011(.002)}
 & \cellcolor{teal!40}{.021(.002)} & \cellcolor{teal!40}{.007(.002)} & \cellcolor{teal!40}{.008(.002)}
 & \cellcolor{red!20}{-.010(.002)} & \cellcolor{red!20}{-.005(.002)} & \cellcolor{red!20}{-.006(.002)}
 & \cellcolor{teal!40}{.012(.002)} & \cellcolor{teal!40}{.006(.002)} & \cellcolor{teal!40}{.007(.002)}
 & \cellcolor{red!20}{-.010(.002)} & \cellcolor{red!20}{-.005(.001)} & \cellcolor{red!20}{-.006(.002)} \\

\hline

\multirow{2}{*}{\textsc{Kdd}} & \textit{Gender}
 & \cellcolor{teal!40}{.071(.003)} & \cellcolor{teal!40}{.019(.002)} & \cellcolor{teal!40}{.023(.002)}
 & \cellcolor{teal!40}{.026(.002)} & \cellcolor{teal!40}{.009(.002)} & \cellcolor{teal!40}{.011(.002)}
 & \cellcolor{teal!40}{.020(.002)} & .009(.002) & \cellcolor{teal!40}{.008(.002)}
 & \cellcolor{red!20}{-.010(.002)} & \cellcolor{red!20}{-.005(.001)} & \cellcolor{red!20}{-.006(.001)}
 & \cellcolor{teal!40}{.012(.002)} & \cellcolor{teal!40}{.006(.002)} & \cellcolor{teal!40}{.007(.002)}
 & \cellcolor{red!20}{-.010(.002)} & \cellcolor{red!20}{-.005(.001)} & \cellcolor{red!20}{-.006(.002)} \\

& \textit{Race}
 & \cellcolor{teal!40}{.075(.003)} & \cellcolor{teal!40}{.020(.002)} & .022(.002)
 & \cellcolor{teal!40}{.029(.002)} & .010(.002) & \cellcolor{teal!40}{.012(.002)}
 & \cellcolor{teal!40}{.022(.002)} & \cellcolor{teal!40}{.007(.002)} & \cellcolor{teal!40}{.008(.002)}
 & \cellcolor{red!20}{-.011(.002)} & \cellcolor{red!20}{-.005(.001)} & \cellcolor{red!20}{-.006(.002)}
 & \cellcolor{teal!40}{.013(.002)} & \cellcolor{teal!40}{.006(.002)} & \cellcolor{teal!40}{.007(.002)}
 & \cellcolor{red!20}{-.011(.002)} & \cellcolor{red!20}{-.005(.001)} & \cellcolor{red!20}{-.006(.002)} \\

\hline

\multirow{2}{*}{\textsc{Dutch}} & \textit{Gender}
 & \cellcolor{teal!40}{.064(.003)} & \cellcolor{teal!40}{.017(.002)} & \cellcolor{teal!40}{.019(.002)}
 & \cellcolor{teal!40}{.026(.002)} & \cellcolor{teal!40}{.008(.002)} & \cellcolor{teal!40}{.009(.002)}
 & \cellcolor{teal!40}{.020(.002)} & \cellcolor{teal!40}{.006(.001)} & \cellcolor{teal!40}{.007(.002)}
 & \cellcolor{red!20}{-.010(.002)} & \cellcolor{red!20}{-.005(.001)} & \cellcolor{red!20}{-.006(.002)}
 & \cellcolor{teal!40}{.012(.002)} & \cellcolor{teal!40}{.006(.002)} & \cellcolor{teal!40}{.007(.002)}
 & -.010(.002) & \cellcolor{red!20}{-.005(.001)} & \cellcolor{red!20}{-.006(.002)} \\

& \textit{Age}
 & \cellcolor{teal!40}{.060(.003)} & \cellcolor{teal!40}{.018(.002)} & \cellcolor{teal!40}{.020(.002)}
 & \cellcolor{teal!40}{.027(.002)} & \cellcolor{teal!40}{.009(.002)} & \cellcolor{teal!40}{.010(.002)}
 & \cellcolor{teal!40}{.021(.002)} & \cellcolor{teal!40}{.007(.002)} & \cellcolor{teal!40}{.008(.002)}
 & \cellcolor{red!20}{-.011(.002)} & \cellcolor{red!20}{-.005(.001)} & \cellcolor{red!20}{-.006(.002)}
 & \cellcolor{teal!40}{.013(.002)} & \cellcolor{teal!40}{.006(.002)} & \cellcolor{teal!40}{.007(.002)}
 & \cellcolor{red!20}{-.011(.002)} & \cellcolor{red!20}{-.005(.001)} & \cellcolor{red!20}{-.006(.002)} \\

\hline

\multirow{3}{*}{\textsc{Credit}} & \textit{Gender}
 & \cellcolor{teal!40}{.058(.003)} & \cellcolor{teal!40}{.015(.002)} & \cellcolor{teal!40}{.017(.002)}
 & \cellcolor{teal!40}{.025(.002)} & \cellcolor{teal!40}{.008(.002)} & \cellcolor{teal!40}{.009(.002)}
 & \cellcolor{teal!40}{.019(.002)} & \cellcolor{teal!40}{.006(.001)} & \cellcolor{teal!40}{.007(.002)}
 & \cellcolor{red!20}{-.009(.002)} & \cellcolor{red!20}{-.004(.001)} & -.005(.001)
 & \cellcolor{teal!40}{.011(.002)} & \cellcolor{teal!40}{.005(.001)} & \cellcolor{teal!40}{.006(.002)}
 & \cellcolor{red!20}{-.009(.002)} & \cellcolor{red!20}{-.005(.001)} & \cellcolor{red!20}{-.006(.002)} \\

& \textit{Marriage}
 & \cellcolor{teal!40}{.061(.003)} & \cellcolor{teal!40}{.016(.002)} & \cellcolor{teal!40}{.018(.002)}
 & \cellcolor{teal!40}{.024(.002)} & \cellcolor{teal!40}{.007(.002)} & \cellcolor{teal!40}{.009(.002)}
 & \cellcolor{teal!40}{.020(.002)} & \cellcolor{teal!40}{.006(.001)} & \cellcolor{teal!40}{.007(.002)}
 & \cellcolor{red!20}{-.010(.002)} & \cellcolor{red!20}{-.005(.001)} & \cellcolor{red!20}{-.006(.002)}
 & \cellcolor{teal!40}{.012(.002)} & .006(.002) & \cellcolor{teal!40}{.007(.002)}
 & \cellcolor{red!20}{-.010(.002)} & \cellcolor{red!20}{-.005(.001)} & \cellcolor{red!20}{-.006(.002)} \\

& \textit{Education}
 & \cellcolor{teal!40}{.065(.003)} & \cellcolor{teal!40}{.017(.002)} & \cellcolor{teal!40}{.019(.002)}
 & \cellcolor{teal!40}{.026(.002)} & \cellcolor{teal!40}{.008(.002)} & \cellcolor{teal!40}{.009(.002)}
 & \cellcolor{teal!40}{.021(.002)} & \cellcolor{teal!40}{.007(.002)} & \cellcolor{teal!40}{.008(.002)}
 & \cellcolor{red!20}{-.011(.002)} & \cellcolor{red!20}{-.005(.001)} & \cellcolor{red!20}{-.006(.002)}
 & \cellcolor{teal!40}{.013(.002)} & .006(.002) & \cellcolor{teal!40}{.007(.002)}
 & \cellcolor{red!20}{-.011(.002)} & -.005(.001) & \cellcolor{red!20}{-.006(.002)} \\

\hline

\multirow{2}{*}{\textsc{Crime}} & \textit{Race}
 & \cellcolor{teal!40}{.056(.003)} & \cellcolor{teal!40}{.015(.002)} & \cellcolor{teal!40}{.017(.002)}
 & \cellcolor{teal!40}{.023(.002)} & \cellcolor{teal!40}{.007(.002)} & \cellcolor{teal!40}{.008(.002)}
 & \cellcolor{teal!40}{.018(.002)} & \cellcolor{teal!40}{.005(.001)} & \cellcolor{teal!40}{.006(.002)}
 & \cellcolor{red!20}{-.009(.002)} & \cellcolor{red!20}{-.004(.001)} & \cellcolor{red!20}{-.005(.001)}
 & \cellcolor{teal!40}{.011(.002)} & \cellcolor{teal!40}{.005(.001)} & \cellcolor{teal!40}{.006(.002)}
 & \cellcolor{red!20}{-.009(.002)} & \cellcolor{red!20}{-.005(.001)} & \cellcolor{red!20}{-.006(.002)} \\

& \textit{Gender}
 & \cellcolor{teal!40}{.059(.003)} & \cellcolor{teal!40}{.016(.002)} & \cellcolor{teal!40}{.018(.002)}
 & \cellcolor{teal!40}{.024(.002)} & \cellcolor{teal!40}{.008(.002)} & \cellcolor{teal!40}{.009(.002)}
 & \cellcolor{teal!40}{.019(.002)} & \cellcolor{teal!40}{.006(.001)} & \cellcolor{teal!40}{.007(.002)}
 & \cellcolor{red!20}{-.010(.002)} & \cellcolor{red!20}{-.005(.001)} & \cellcolor{red!20}{-.006(.002)}
 & \cellcolor{teal!40}{.012(.002)} & \cellcolor{teal!40}{.006(.002)} & \cellcolor{teal!40}{.007(.002)}
 & \cellcolor{red!20}{-.010(.002)} & \cellcolor{red!20}{-.005(.001)} & \cellcolor{red!20}{-.006(.002)} \\

\hline

\multirow{2}{*}{\textsc{German}} & \textit{Gender}
 & \cellcolor{teal!40}{.053(.003)} & \cellcolor{teal!40}{.014(.002)} & \cellcolor{teal!40}{.016(.002)}
 & .022(.002) & \cellcolor{teal!40}{.007(.002)} & \cellcolor{teal!40}{.008(.002)}
 & \cellcolor{teal!40}{.017(.002)} & \cellcolor{teal!40}{.005(.001)} & \cellcolor{teal!40}{.006(.002)}
 & \cellcolor{red!20}{-.009(.002)} & \cellcolor{red!20}{-.004(.001)} & \cellcolor{red!20}{-.005(.001)}
 & \cellcolor{teal!40}{.011(.002)} & \cellcolor{teal!40}{.005(.001)} & \cellcolor{teal!40}{.006(.002)}
 & \cellcolor{red!20}{-.009(.002)} & \cellcolor{red!20}{-.005(.001)} & \cellcolor{red!20}{-.006(.002)} \\

& \textit{Age}
 & \cellcolor{teal!40}{.055(.003)} & \cellcolor{teal!40}{.015(.002)} & \cellcolor{teal!40}{.017(.002)}
 & \cellcolor{teal!40}{.023(.002)} & \cellcolor{teal!40}{.008(.002)} & \cellcolor{teal!40}{.009(.002)}
 & \cellcolor{teal!40}{.018(.002)} & \cellcolor{teal!40}{.006(.001)} & \cellcolor{teal!40}{.007(.002)}
 & \cellcolor{red!20}{-.009(.002)} & \cellcolor{red!20}{-.004(.001)} & \cellcolor{red!20}{-.005(.001)}
 & \cellcolor{teal!40}{.011(.002)} & \cellcolor{teal!40}{.005(.001)} & \cellcolor{teal!40}{.006(.002)}
 & \cellcolor{red!20}{-.009(.002)} & \cellcolor{red!20}{-.005(.001)} & \cellcolor{red!20}{-.006(.002)} \\

\bottomrule
\end{tabular}
\end{adjustbox}

\end{table}

%% file: Figure/eff1.tex
    \begin{tikzpicture}
        \begin{axis}[
            xbar,
            axis x line  = bottom,
            axis y line  = left  ,
             reverse legend,
            bar width=.15cm,
            width=4cm,
            height=6.5cm,
               enlarge y limits=0.07,
            xlabel={Runtime (s)},
            symbolic y coords={\textsc{Adult}, \textsc{Compas}, \textsc{Law School}, \textsc{KDD}, \textsc{Dutch}, \textsc{Credit}, \textsc{Crime}, \textsc{German}},
            ytick=data,
            xmin=0,
            xmax=600,
            legend style={at={(0.9,1)},draw=none,fill=none,anchor=north,legend columns=1},
            legend cell align={left},
        ]
        \addplot+[fill=teal!70,draw=black,error bars/.cd,
                x dir=both,
                x explicit,
                error bar style={black}
            ] plot coordinates {
            (39.37,\textsc{Adult}) +- (2.50,0)
            (20.27,\textsc{Compas}) +- (2.15,0)
            (23.59,\textsc{Law School}) +- (1.65,0)
            (85.29,\textsc{KDD}) +- (9.70,0)
            (44.78,\textsc{Dutch}) +- (3.81,0)
            (30.43,\textsc{Credit}) +- (4.40,0)
            (13.67,\textsc{Crime}) +- (1.29,0)
            (9.17,\textsc{German}) +- (1.46,0) 
        };

   \addplot+[fill=red!20,draw=black,error bars/.cd,
                x dir=both,
                x explicit,
                error bar style={black}
            ] plot coordinates {
            (439.27,\textsc{Adult}) +- (18.18,0)
            (265.19,\textsc{Compas}) +- (10.41,0) 
            (338.85,\textsc{Law School}) +- (13.13,0)
            (548.76,\textsc{KDD}) +- (22.79,0)
            (435.22,\textsc{Dutch}) +- (16.12,0)
            (393.44,\textsc{Credit}) +- (21.57,0)
            (239.48,\textsc{Crime}) +- (9.81,0)
            (212.98,\textsc{German}) +- (7.60,0) 
        };
         \legend{\approach,\texttt{FairRF}}
        \end{axis}

    \end{tikzpicture}

%% file: Figure/eff8.tex
\begin{tikzpicture}
    \begin{axis}[
        xbar,
        axis x line  = bottom,
        axis y line  = left,
        reverse legend,
        bar width=.1cm,
        width=4cm,
        height=6.5cm,
        enlarge y limits=0.07,
        xlabel={Runtime (s)},
        symbolic y coords={Adult, Compas, Law School, KDD, Dutch, Credit, Crime, German},
        yticklabel=\empty,
        ytick=data,
        xmin=0,
        xmax=600,
        legend style={at={(1.05,1)},draw=none,fill=none,anchor=north,legend columns=1},
        legend cell align={left},
    ]
    \addplot+[fill=teal!70,draw=black,error bars/.cd,
            x dir=both,
            x explicit,
            error bar style={black}
        ] plot coordinates {
        (281.37,Adult) +- (9.38,0)
        (219.42,Compas) +- (16.14,0)
        (254.20,Law School) +- (11.76,0)
        (473.70,KDD) +- (23.63,0)
        (196.52,Dutch) +- (12.89,0)
        (240.05,Credit) +- (8.22,0)
        (180.89,Crime) +- (11.09,0)
        (168.56,German) +- (11.52,0)
    };

    \addplot+[fill=orange!45,draw=black,error bars/.cd,
            x dir=both,
            x explicit,
            error bar style={black}
        ] plot coordinates {
        (262.44,Adult) +- (15.83,0)
        (232.43,Compas) +- (24.42,0)
        (244.89,Law School) +- (20.76,0)
        (449.69,KDD) +- (36.44,0)
        (184.31,Dutch) +- (11.21,0)
        (244.06,Credit) +- (14.51,0)
        (172.76,Crime) +- (12.88,0)
        (159.91,German) +- (13.04,0)
    };

    \addplot+[fill=red!20,draw=black,error bars/.cd,
            x dir=both,
            x explicit,
            error bar style={black}
        ] plot coordinates {
        (315.16,Adult) +- (21.29,0)
        (234.16,Compas) +- (18.78,0)
        (286.51,Law School) +- (20.57,0)
        (512.16,KDD) +- (40.34,0)
        (226.80,Dutch) +- (23.33,0)
        (254.72,Credit) +- (26.75,0)
        (199.30,Crime) +- (14.82,0)
        (180.74,German) +- (11.20,0)
    };

    \legend{w/ \approach,w/o \approach,\texttt{FairRF}}
    \end{axis}
\end{tikzpicture}

%% file: Figure/eff9.tex
\begin{tikzpicture}
\begin{axis}[
      xbar,
    axis x line  = bottom,
    axis y line  = left,
    reverse legend,
    bar width=.1cm,
    width=4cm,
    height=6.5cm,
    enlarge y limits=0.07,
    xlabel={Runtime (s)},
    symbolic y coords={Adult, Compas, Law School, KDD, Dutch, Credit, Crime, German},
    yticklabel=\empty,
    ytick=data,
    xmin=0,
    xmax=5000,
    legend style={at={(1.05,1)},draw=none,fill=none,anchor=north,legend columns=1},
    legend cell align={left},
]
\addplot+[fill=teal!70,draw=black,error bars/.cd,
        x dir=both, x explicit, error bar style={black}
    ] plot coordinates {
    (2501.72,Adult)      +- (194.74,0)
    (1738.88,Compas)     +- (155.35,0)
    (1983.82,Law School) +- (148.60,0)
    (3386.31,KDD)        +- (196.89,0)
    (1912.76,Dutch)      +- (144.01,0)
    (1842.62,Credit)     +- (88.69,0)
    (1773.24,Crime)      +- (136.86,0)
    (1619.66,German)     +- (140.43,0)
};

\addplot+[fill=orange!45,draw=black,error bars/.cd,
        x dir=both, x explicit, error bar style={black}
    ] plot coordinates {
    (2282.95,Adult)      +- (173.03,0)
    (1571.26,Compas)     +- (152.86,0)
    (1770.88,Law School) +- (122.25,0)
    (3447.09,KDD)        +- (176.67,0)
    (1798.62,Dutch)      +- (156.23,0)
    (1662.94,Credit)     +- (119.22,0)
    (1654.80,Crime)      +- (114.07,0)
    (1461.69,German)     +- (121.99,0)
};

\addplot+[fill=red!20,draw=black,error bars/.cd,
        x dir=both, x explicit, error bar style={black}
    ] plot coordinates {
    (2774.75,Adult)      +- (175.10,0)
    (1862.27,Compas)     +- (198.24,0)
    (2094.15,Law School) +- (144.88,0)
    (3664.59,KDD)        +- (213.68,0)
    (2070.97,Dutch)      +- (132.77,0)
    (1882.04,Credit)     +- (156.93,0)
    (1796.51,Crime)      +- (160.81,0)
    (1647.60,German)     +- (155.87,0)
};

\legend{w/ \approach,w/o \approach,\texttt{FairRF}}
\end{axis}
\end{tikzpicture}

%% file: Figure/eff2.tex
    \begin{tikzpicture}
        \begin{axis}[
            xbar,
            axis x line  = bottom,
            axis y line  = left  ,
              reverse legend,
            bar width=.1cm,
            width=4cm,
            height=6.5cm,
               enlarge y limits=0.07,
            xlabel={Runtime (s)},
            symbolic y coords={Adult, Compas, Law School, KDD, Dutch, Credit, Crime, German},
            ytick=data,
            xmin=0,
            xmax=650,
            legend style={at={(1.05,1)},draw=none,fill=none,anchor=north,legend columns=1},
            legend cell align={left},
        ]
        \addplot+[fill=teal!70,draw=black,error bars/.cd,
                x dir=both,
                x explicit,
                error bar style={black}
            ] plot coordinates {
            (319.74,Adult) +- (11.53,0)
            (249.34,Compas) +- (16.82,0)
            (288.86,Law School) +- (16.4,0)
            (538.3,KDD) +- (26.26,0)
            (223.32,Dutch) +- (9.88,0)
            (272.78,Credit) +- (9.13,0)
            (205.56,Crime) +- (12.32,0)
            (191.54,German) +- (12.8,0) 
        };

        \addplot+[fill=orange!45,draw=black,error bars/.cd,
                x dir=both,
                x explicit,
                error bar style={black}
            ] plot coordinates {
            (301.65,Adult) +- (17.59,0)
            (267.16,Compas) +- (27.13,0)
            (281.48,Law School) +- (23.07,0)
            (516.89,KDD) +- (37.15,0)
            (211.85,Dutch) +- (13.57,0)
            (257.54,Credit) +- (18.34,0)
            (198.58,Crime) +- (15.42,0)
            (183.81,German) +- (14.49,0) 
        };

  \addplot+[fill=red!20,draw=black,error bars/.cd,
                x dir=both,
                x explicit,
                error bar style={black}
            ] plot coordinates {
            (342.56,Adult) +- (25.32,0)
            (254.52,Compas) +- (20.41,0)
            (311.42,Law School) +- (22.36,0)
            (556.7,KDD) +- (49.28,0)
            (246.52,Dutch) +- (28.62,0)
            (287.74,Credit) +- (30.16,0)
            (227.5,Crime) +- (18.28,0)
            (196.46,German) +- (16.52,0) 
        };
        
         \legend{w/ \approach,w/o \approach,\texttt{FairRF}}
        \end{axis}

    \end{tikzpicture}

%% file: Figure/eff3.tex
    \begin{tikzpicture}
        \begin{axis}[
              xbar,
            axis x line  = bottom,
            axis y line  = left  ,
              reverse legend,
            bar width=.1cm,
            width=4cm,
            height=6.5cm,
               enlarge y limits=0.07,
            xlabel={Runtime (s)},
            symbolic y coords={Adult, Compas, Law School, KDD, Dutch, Credit, Crime, German},
            yticklabel=\empty,
            ytick=data,
            xmin=0,
            xmax=5000,
            legend style={at={(1.05,1)},draw=none,fill=none,anchor=north,legend columns=1},
            legend cell align={left},
        ]
        \addplot+[fill=teal!70,draw=black,error bars/.cd,
                x dir=both,
                x explicit,
                error bar style={black}
            ] plot coordinates {
            (2841.73,Adult) +- (221.29,0)
            (1974.86,Compas) +- (176.53,0)
            (2254.34,Law School) +- (168.6,0)
            (3848.08,KDD) +- (223.74,0)
            (2172.68,Dutch) +- (163.65,0)
            (2071.16,Credit) +- (100.78,0)
            (1981.86,Crime) +- (155.53,0)
            (1726.88,German) +- (159.58,0) 
        };

 \addplot+[fill=orange!45,draw=black,error bars/.cd,
                x dir=both,
                x explicit,
                error bar style={black}
            ] plot coordinates {
            (2594.26,Adult) +- (196.63,0)
            (1785.52,Compas) +- (173.71,0)
            (2012.36,Law School) +- (138.92,0)
            (3917.16,KDD) +- (200.76,0)
            (2043.89,Dutch) +- (177.54,0)
            (1889.70,Credit) +- (135.48,0)
            (1767.27,Crime) +- (129.62,0)
            (1661.47,German) +- (138.63,0) 
        };

   \addplot+[fill=red!20,draw=black,error bars/.cd,
                x dir=both,
                x explicit,
                error bar style={black}
            ] plot coordinates {
            (3153.13,Adult) +- (198.97,0)
            (2116.21,Compas) +- (225.27,0)
            (2379.72,Law School) +- (164.64,0)
            (4164.31,KDD) +- (242.82,0)
            (2353.37,Dutch) +- (150.88,0)
            (2139.82,Credit) +- (178.33,0)
            (2041.49,Crime) +- (182.74,0)
            (1871.13,German) +- (177.13,0) 
        };

         \legend{w/ \approach,w/o \approach,\texttt{FairRF}}
        \end{axis}

    \end{tikzpicture}

%% file: Figure/eff4.tex
\begin{tikzpicture}

\begin{axis}[
    xbar,
    axis x line=bottom,
    axis y line=left,
    reverse legend,
    bar width=.1cm,
    width=4cm,
    height=6.5cm,
    enlarge y limits=0.07,
    xlabel={Runtime (s)},
    symbolic y coords={Adult, Compas, Law School, KDD, Dutch, Credit, Crime, German},
    yticklabel=\empty,
    ytick=data,
    xmin=0,
    xmax=250,
    legend style={at={(1.05,1)},draw=none,fill=none,anchor=north,legend columns=1},
    legend cell align={left},
]

\addplot+[fill=teal!70,draw=black,error bars/.cd,
    x dir=both, x explicit,
    error bar style={black}
] plot coordinates {
    (92.1,Adult) +- (3.2,0)
    (81.4,Compas) +- (2.1,0)
    (79.2,Law School) +- (2.3,0)
    (84.6,KDD) +- (2.5,0)
    (89.7,Dutch) +- (3.4,0)
    (86.3,Credit) +- (2.8,0)
    (91.5,Crime) +- (3.1,0)
    (97.8,German) +- (3.6,0)
};

\addplot+[fill=orange!45,draw=black,error bars/.cd,
    x dir=both, x explicit,
    error bar style={black}
] plot coordinates {
    (85.2,Adult) +- (3.0,0)
    (74.5,Compas) +- (2.0,0)
    (72.8,Law School) +- (2.2,0)
    (77.4,KDD) +- (2.4,0)
    (82.3,Dutch) +- (2.9,0)
    (79.6,Credit) +- (2.5,0)
    (83.9,Crime) +- (3.0,0)
    (89.1,German) +- (3.4,0)
};

\addplot+[fill=red!20,draw=black,error bars/.cd,
    x dir=both, x explicit,
    error bar style={black}
] plot coordinates {
    (120.6,Adult) +- (4.2,0)
    (105.3,Compas) +- (3.1,0)
    (102.8,Law School) +- (3.3,0)
    (110.5,KDD) +- (3.4,0)
    (115.9,Dutch) +- (3.8,0)
    (112.7,Credit) +- (3.2,0)
    (118.2,Crime) +- (4.0,0)
    (128.4,German) +- (4.5,0)
};

 \legend{w/ \approach,w/o \approach,\texttt{FairRF}}
 \end{axis}
 
\end{tikzpicture}

%% file: Figure/eff5.tex
\begin{tikzpicture}
\begin{axis}[
    xbar,
    axis x line=bottom,
    axis y line=left,
    reverse legend,
    bar width=.1cm,
    width=4cm,
    height=6.5cm,
    enlarge y limits=0.07,
    xlabel={Runtime (s)},
    symbolic y coords={Adult, Compas, Law School, KDD, Dutch, Credit, Crime, German},
    ytick=data,
    xmin=0,
    xmax=800,
    legend style={at={(1.05,1)},draw=none,fill=none,anchor=north,legend columns=1},
    legend cell align={left},
]

\addplot+[fill=teal!70,draw=black,error bars/.cd,
    x dir=both, x explicit,
    error bar style={black}
] plot coordinates {
    (442.1,Adult) +- (13.5,0)
    (382.6,Compas) +- (11.7,0)
    (377.8,Law School) +- (11.5,0)
    (411.5,KDD) +- (12.5,0)
    (439.1,Dutch) +- (13.3,0) 
    (425.6,Credit) +- (12.8,0)
    (436.8,Crime) +- (13.2,0)
    (459.2,German) +- (13.9,0)
};

\addplot+[fill=red!20,draw=black,error bars/.cd,
    x dir=both, x explicit,
    error bar style={black}
] plot coordinates {
    (410.2,Adult) +- (12.5,0)
    (352.7,Compas) +- (10.8,0)
    (347.9,Law School) +- (10.6,0)
    (380.5,KDD) +- (11.6,0)
    (408.6,Dutch) +- (12.4,0) 
    (394.8,Credit) +- (11.9,0)
    (405.1,Crime) +- (12.3,0)
    (427.4,German) +- (12.9,0)
};

\addplot+[fill=orange!45,draw=black,error bars/.cd,
    x dir=both, x explicit,
    error bar style={black}
] plot coordinates {
    (425.7,Adult) +- (13.0,0)
    (366.4,Compas) +- (11.2,0)
    (361.4,Law School) +- (11.0,0)
    (394.9,KDD) +- (12.0,0) 
    (422.3,Dutch) +- (12.8,0) 
    (409.2,Credit) +- (12.3,0)
    (418.7,Crime) +- (12.7,0)
    (440.8,German) +- (13.4,0)
};

\legend{w/ \approach,w/o \approach,\texttt{FairRF}}
\end{axis}
\end{tikzpicture}

%% file: Figure/eff6.tex
\begin{tikzpicture}
\begin{axis}[
    xbar,
    axis x line=bottom,
    axis y line=left,
    reverse legend,
    bar width=.1cm,
    width=4cm,
    height=6.5cm,
    enlarge y limits=0.07,
    xlabel={Runtime (s)},
    symbolic y coords={Adult, Compas, Law School, KDD, Dutch, Credit, Crime, German},
    yticklabel=\empty,
    ytick=data,
    xmin=0,
    xmax=250,
    legend style={at={(1.05,1)},draw=none,fill=none,anchor=north,legend columns=1},
    legend cell align={left},
]

\addplot+[fill=teal!70,draw=black,error bars/.cd,
    x dir=both, x explicit,
    error bar style={black}
] plot coordinates {
    (92.4,Adult) +- (3.3,0)
    (81.6,Compas) +- (2.2,0)
    (79.8,Law School) +- (2.4,0)
    (84.9,KDD) +- (2.5,0)
    (89.5,Dutch) +- (3.1,0)
    (86.7,Credit) +- (2.9,0)
    (91.9,Crime) +- (3.2,0)
    (97.2,German) +- (3.5,0)
};

\addplot+[fill=orange!45,draw=black,error bars/.cd,
    x dir=both, x explicit,
    error bar style={black}
] plot coordinates {
    (88.9,Adult) +- (3.0,0)
    (74.8,Compas) +- (2.1,0)
    (71.5,Law School) +- (2.3,0)
    (73.1,KDD) +- (2.4,0)
    (81.1,Dutch) +- (2.8,0)
    (75.3,Credit) +- (2.6,0)
    (80.5,Crime) +- (3.0,0)
    (84.7,German) +- (3.3,0)
};

\addplot+[fill=red!20,draw=black,error bars/.cd,
    x dir=both, x explicit,
    error bar style={black}
] plot coordinates {
    (122.2,Adult) +- (4.1,0)
    (101.6,Compas) +- (3.2,0)
    (103.4,Law School) +- (3.3,0)
    (110.9,KDD) +- (3.5,0)
    (125.4,Dutch) +- (3.7,0)
    (114.3,Credit) +- (3.4,0)
    (122.7,Crime) +- (4.0,0)
    (131.1,German) +- (4.4,0)
};

\legend{w/ \approach,w/o \approach,\texttt{FairRF}}
\end{axis}
\end{tikzpicture}

%% file: Figure/eff7.tex
\begin{tikzpicture}
\begin{axis}[
    xbar,
    axis x line=bottom,
    axis y line=left,
    reverse legend,
    bar width=.1cm,
    width=4cm,
    height=6.5cm,
    enlarge y limits=0.07,
    xlabel={Runtime (s)},
    symbolic y coords={Adult, Compas, Law School, KDD, Dutch, Credit, Crime, German},
    yticklabel=\empty,
    ytick=data,
    xmin=0,
    xmax=800,
    legend style={at={(1.05,1)},draw=none,fill=none,anchor=north,legend columns=1},
    legend cell align={left},
]

\addplot+[fill=teal!70,draw=black,error bars/.cd,
    x dir=both, x explicit,
    error bar style={black}
] plot coordinates {
    (402.7,Adult) +- (12.3,0)
    (362.1,Compas) +- (11.1,0)
    (358.5,Law School) +- (10.9,0)
    (385.4,KDD) +- (11.7,0)
    (398.2,Dutch) +- (12.1,0)
    (374.6,Credit) +- (11.4,0)
    (389.8,Crime) +- (11.9,0)
    (415.3,German) +- (12.8,0)
};

\addplot+[fill=orange!45,draw=black,error bars/.cd,
    x dir=both, x explicit,
    error bar style={black}
] plot coordinates {
    (382.1,Adult) +- (11.7,0)
    (343.5,Compas) +- (10.5,0)
    (339.2,Law School) +- (10.2,0)
    (365.8,KDD) +- (11.2,0)
    (378.6,Dutch) +- (11.6,0)
    (356.4,Credit) +- (10.8,0)
    (371.5,Crime) +- (11.3,0)
    (395.7,German) +- (12.2,0)
};

\addplot+[fill=red!20,draw=black,error bars/.cd,
    x dir=both, x explicit,
    error bar style={black}
] plot coordinates {
    (430.6,Adult) +- (13.5,0)
    (388.9,Compas) +- (12.2,0)
    (384.2,Law School) +- (11.9,0)
    (413.8,KDD) +- (12.7,0)
    (426.4,Dutch) +- (13.1,0)
    (401.7,Credit) +- (12.0,0)
    (418.6,Crime) +- (12.5,0)
    (445.2,German) +- (13.8,0)
};

\legend{w/ \approach,w/o \approach,\texttt{FairRF}}
\end{axis}
\end{tikzpicture}

%% file: discussion.tex
\section{Discussion}
\label{sec:dis}

\subsection{Modifying Base Generator}

A strength of \approach~is that it can be paired with different base generators, benefiting from their diverse perturbation strategies in fairness testing. 

Specifically, \approach~can benefit and integrate with base generators based on their designs:
\begin{itemize}
    \item \textbf{Partial perturbation-based generators:} This includes, e.g., \texttt{ADF} and \texttt{EIDIG}, which naturally perturb all the non-sensitive features. Here, \approach~can simply ``trick'' them to believe that both the concerned sensitive feature and the most causally relevant non-sensitive feature are part of the sensitive proportion, and hence to keep the other procedure unchanged. That is to say, \approach~ only seeds their local instances to form an expanded global candidate pool.
    \item \textbf{Full perturbation-based generators:} This includes, e.g., \textsc{SG}, such that all features are perturbed. In this case, we can simply feed the features other than the concerned sensitive feature and the most causally relevant non-sensitive feature into the generator for perturbation.
\end{itemize}

Integrating \approach~ with an arbitrary base generator requires minimal engineering effort. In our implementation, \approach~operates as a wrapper around the generator’s perturbation function, injecting causally guided perturbation strategies before calling the base generator. The same applies to partial perturbation-based generators such as \texttt{ADF} or full perturbation ones, e.g., \texttt{SG}. This requires approximately 50–100 lines of code changes depending on the generator's API, with integration time typically under 2 hours for a standard Python-based generator, including the time to comprehend its code structure. As such, \approach~ requires an acceptable amount of effort to be integrated into a new base generator.

Indeed, some base generators require training data for their feature distribution learning~\cite{DBLP:conf/sigsoft/AggarwalLNDS19,DBLP:conf/icse/0002WJY022,DBLP:conf/kbse/UdeshiAC18,DBLP:conf/sigsoft/GalhotraBM17}. In those cases, they often leverage the training in their initialization stage while \approach~ benefits them at the perturbation stage. As such, using test data for fairness testing does not negatively affect the performance of these generators.




\subsection{Sensitivity to $k$}

\begin{figure}[t!]
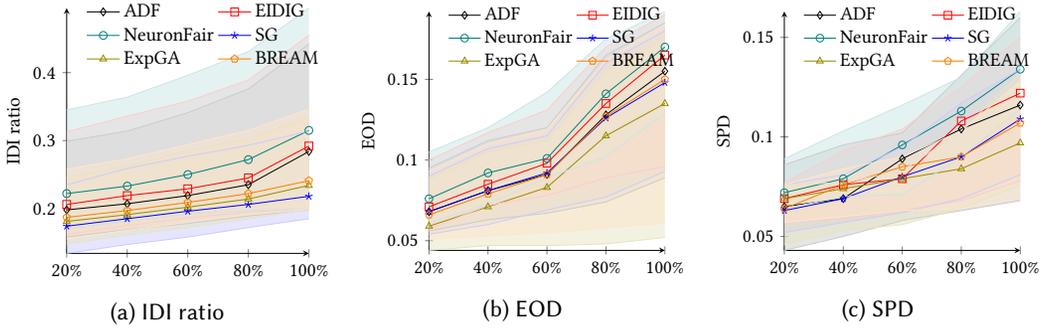

    \centering
    \begin{subfigure}[t!]{0.32\columnwidth}
        \centering
        \includestandalone[width=\columnwidth]{Figure/3-1}
        \caption{IDI ratio}
    \end{subfigure}
     \hfill
    \begin{subfigure}[t!]{0.32\columnwidth}
        \centering        
        \includestandalone[width=\columnwidth]{Figure/3-2}
        \caption{EOD}      
    \end{subfigure}
        \hfill
     \begin{subfigure}[t!]{0.32\columnwidth}
        \centering        
        \includestandalone[width=\columnwidth]{Figure/3-3}
        \caption{SPD}      
    \end{subfigure}
    \caption{Sensitivity of \approach~to $k$ over all models, datasets, possible sensitive features, generators, and runs.}
    \label{fig:sen}
\end{figure}

By default, we set $k=100$, meaning that the causal graph in \approach~is built by using $100\%$ training data. From Figure~\ref{fig:sen}, we see that although a smaller $k$ would certainly lead to better efficiency, increasing $k$ enables \approach~to find more fairness bugs since a more reliable causal graph can be built, leading to more accurately estimated causal relationships/causal effects and hence better fairness testing.

One practical challenge is that some generators, such as ADF and FairRF, require access to the full training dataset to construct their internal models. As our focus is fairness testing on unseen data, we used test data to generate IDIs while ensuring no data leakage into model retraining or causal graph extraction. Empirically, we observed that using test data as the input for seed generation did not negatively affect base generator performance because their perturbation strategies operate independently of model training, and \approach~ focuses on input generation rather than model learning.

Key engineering tricks include aligning feature encoding schemas between \approach~ and base generators, and batching seed inputs to minimize invocation overhead when integrating with generators designed for large-batch sampling.

\subsection{Multiple Sensitive Features}

While considering one concerned sensitive feature at a time is common \cite{DBLP:conf/icse/ZhangW0D0WDD20,DBLP:conf/issta/ZhangZZ21,DBLP:conf/icse/ZhengCD0CJW0C22,DBLP:conf/sigsoft/AggarwalLNDS19,DBLP:conf/icse/0002WJY022,DBLP:conf/icics/JiangSLWSG23,DBLP:conf/issta/XiaoLL023,DBLP:conf/innovations/DworkHPRZ12,DBLP:conf/aies/ZhangLM18}, extending \approach~with multiple sensitive features is also easy. There could be two ways for \approach~to handle such a scenario with multiple sensitive features: the \textbf{direct} and \textbf{indirect} extension.

\subsubsection{Indirect Extension}

Since the presence of multiple sensitive features merely influences how discrimination instances are defined while \approach~operates at the perturbation level of fairness testing, for the \textbf{indirect} way, one can perform the following:

\begin{enumerate}
    \item Pick a sensitive feature and run \approach~as if that feature is the only sensitive one that is of concern.
    \item Repeat from 1) to cover all sensitive features.
    \item Combine all the samples generated and break the pairs; find the discrimination instances that fit the definition of multiple sensitive features.
\end{enumerate}

\subsubsection{Direct Extension}

The \textbf{indirect} extension clearly requires no change on the internal mechanism of \approach, but relies on re-pairing the samples generated. Alternatively, we can also guide the perturbation from the beginning, i.e., the \textbf{direct} extension:

\begin{enumerate}
    \item Adjust the causal graph construction with multiple sensitive features as starting points.
    \item Modify the perturbation rules and the relaxed different fairness definition to accommodate combinations of sensitive features.
\end{enumerate}
The above would enable \approach~to uncover nuanced biases stemming from intersections of sensitive features, thereby providing a more comprehensive fairness testing therein.

It is clear that the key difference between the above two extensions in handling multiple sensitive features differs in terms of whether we conduct multiple runs, each targeting one sensitive feature individually and combine the results; or run once for all multiple sensitive features. Given sufficient budget, both ways should lead to similar outcomes. Yet, when the number of (unique) samples to be generated is small, the \textbf{direct} extension could be better in detecting hidden fairness bugs related to the interaction of all sensitive features.

\subsection{Qualitative Analysis of \approach}

Perturbation with the relaxed definition in \approach~ has enabled it to reveal many fairness bugs that are otherwise difficult to find. For example, when testing DNN under the \textsc{Adult} dataset, \approach~ finds the pair below which has never been identified by any other methods:
\[
\mathbf{x'}_a = \{5,0 \text{ (Male)},10,3 \text{ (Separated)},4,2,1,0,0,45,0,0,0,1\}
\]
\[
\mathbf{x'}_b = \{5,1 \text{ (Female)},10,4 \text{ (Married-civ-spouse)},4,2,1,0,0,45,0,0,0,1\}
\]  
Clearly, it is an invalid pair under the true definition, since both the sensitive feature \texttt{Gender} (first index) and its most relevant non-sensitive feature \texttt{Marital-status} (third index) have different values. However, \approach~ then repairs by successfully finding another sample ($\mathbf{x''}_b$) from the testing dataset as:
\[
\mathbf{x'}_b = \{5,1 \text{ (Female)},10,4,4,2,1,0,0,45,0,0,0,1\}
\]
\[
\mathbf{x''}_b = \{5,0 \text{ (Male)},10,4,4,2,1,0,0,45,0,0,0,1\}
\] 

In this corrected pair, only the \texttt{Gender} attribute differs, while all other features remain identical. Nevertheless, the model produces different predictions, indicating an individual discriminatory instance.

The above would be very difficult for the other approaches where only all the non-sensitive features (or all features) are to be perturbed, since the fairness bug can be more easily revealed if the value combination of both the sensitive feature and its most relevant non-sensitive counterpart is explicitly considered, which subsequently reduces the space of perturbation. Indeed, in practical scenarios, it is not uncommon that the gender of a person might strongly interact with his/her marital status to influence the decision made by an AI system.

\subsection{Evidence-based Guidance for Generator Selection}
Drawing on the findings from the experiments, we observe that although \approach can significantly improve different base generators, the choice of which generator can indeed affect the results. In particular, \texttt{NeuronFair} is generally the best generator across the different models (all cases for DNN$_5$ and DNN$_6$; 16/18 cases for RF; 15/18 cases for LR). This is because \texttt{NeuronFair} leverages internal neuron activations and gradient signals to explore fine-grained decision boundaries and trace latent bias pathways, which are especially prevalent in complex and highly non-linear architectures such as deep neural networks. Even in statistical models such as RF and LR, these gradient-informed perturbations help navigate local decision regions more effectively and expose subtle fairness violations that heuristic search alone might miss.

However, there are some discrepancies between the models: \texttt{NeuronFair} is consistently the best across all datasets for DNN, while this might not always be the case for LR and RF. The key reason is that the generators are inherently influenced by the interplay between model and dataset features. Deep neural networks exhibit highly non-linear and fine-grained decision boundaries, which make them more sensitive to neuron-level perturbations and allow \texttt{NeuronFair} to exploit internal representations effectively. In contrast, models like RF and LR often construct monotonic decision surfaces, especially on datasets with lower feature dimensionality and strong feature correlations (e.g., \textsc{German}, \textsc{Crime}), where decision regions are easier to traverse. In such cases, the advantages of neuron-level guidance diminish, and gradient-agnostic strategies, e.g., \texttt{EIDIG}, may occasionally achieve comparable or even slightly better discriminatory instance discovery.

Drawing on the above evidence and findings, we outline a brief guideline for selecting the base generator while pairing with ~\approach as follows:

\begin{itemize}
    \item In general, using \texttt{NeuronFair} is more reliable for DNN in all cases.
    \item \texttt{NeuronFair} is also generally the most robust choice for statistical machine learning models such as RF and LR when the characteristics of the target dataset are unknown.
    \item When it is possible to analyze the dataset, for models such as RF and LR, it is preferable to use \texttt{EIDIG} when the datasets have low feature dimensionality and strong feature correlations.     
\end{itemize}

%% file: Figure/3-1.tex
   \begin{tikzpicture}
  \begin{axis}[
  width = 6cm,
  height = 6cm,
  xtick={1,2,3,4,5},
  xticklabels={20\%,40\%,60\%,80\%,100\%},
  ylabel={IDI ratio},
  axis x line  = bottom,
  axis y line  = left ,
  legend cell align={left},
  legend columns=2,
  legend style={draw=none,fill=none,at={(1.08,1.05)}},
  ]
\addplot [mark=diamond,black,name path=F] plot coordinates {
(1,0.198)
(2,0.207)
(3,0.219)
(4,0.235)
(5,0.284)
};

\addplot [mark=square,red,name path=I] plot coordinates {
(1,0.206)
(2,0.219)
(3,0.229)
(4,0.245)
(5,0.292)
};

\addplot [mark=o,teal,name path=L] plot coordinates {
(1,0.222)
(2,0.233)
(3,0.250)
(4,0.272)
(5,0.315)
};

\addplot [mark=star,blue,name path=C] plot coordinates {
(1,0.174)
(2,0.185)
(3,0.196)
(4,0.206)
(5,0.218)
};

\addplot [mark=triangle,olive,name path=Z] plot coordinates {
(1,0.181)
(2,0.191)
(3,0.202)
(4,0.214)
(5,0.234)
};

\addplot [mark=pentagon,orange,name path=X] plot coordinates {
(1,0.187)
(2,0.197)
(3,0.209)
(4,0.222)
(5,0.241)
};

\addplot [no marks,black!20,name path=D] plot coordinates {
(1,0.299)
(2,0.314)
(3,0.341)
(4,0.376)
(5,0.442)
};
\addplot [no marks,black!20,name path=E] plot coordinates {
(1,0.158)
(2,0.169)
(3,0.178)
(4,0.189)
(5,0.198)
};
\addplot [no marks,red!20,name path=G] plot coordinates {
(1,0.313)
(2,0.336)
(3,0.358)
(4,0.389)
(5,0.455)
};
\addplot [no marks,red!20,name path=H] plot coordinates {
(1,0.162)
(2,0.175)
(3,0.186)
(4,0.195)
(5,0.199)
};
\addplot [no marks,teal!20,name path=J] plot coordinates {
(1,0.345)
(2,0.364)
(3,0.396)
(4,0.431)
(5,0.495)
};
\addplot [no marks,teal!20,name path=K] plot coordinates {
(1,0.170)
(2,0.181)
(3,0.192)
(4,0.197)
(5,0.199)
};
\addplot [no marks,blue!20,name path=A] plot coordinates {
(1,0.235)
(2,0.259)
(3,0.277)
(4,0.293)
(5,0.314)
};
\addplot [no marks,blue!20,name path=B] plot coordinates {
(1,0.134)
(2,0.147)
(3,0.158)
(4,0.172)
(5,0.185)
};
 \addplot [no marks,olive!20,name path=T] plot coordinates {
(1,0.246)
(2,0.262)
(3,0.285)
(4,0.303)
(5,0.334)
};
\addplot [no marks,olive!20,name path=S] plot coordinates {
(1,0.148)
(2,0.161)
(3,0.172)
(4,0.185)
(5,0.199)
};
 \addplot [no marks,orange!20,name path=Y] plot coordinates {
(1,0.258)
(2,0.273)
(3,0.294)
(4,0.315)
(5,0.346)
};
 \addplot [no marks,orange!20,name path=W] plot coordinates {
(1,0.155)
(2,0.168)
(3,0.180)
(4,0.191)
(5,0.199)
  };
\addplot[blue!20,opacity=0.5] fill between[of=B and C];
\addplot[blue!20,opacity=0.5] fill between[of=A and C];
    
 \addplot[black!20,opacity=0.5] fill between[of=E and F];
 \addplot[black!20,opacity=0.5] fill between[of=D and F];
          
\addplot[red!20,opacity=0.5] fill between[of=H and I];
    \addplot[red!20,opacity=0.5] fill between[of=G and I];
    
         \addplot[teal!20,opacity=0.5] fill between[of=K and L];
    \addplot[teal!20,opacity=0.5] fill between[of=J and L];
    
         \addplot[olive!20,opacity=0.5] fill between[of=S and Z];
    \addplot[olive!20,opacity=0.5] fill between[of=T and Z];
    
          \addplot[orange!20,opacity=0.5] fill between[of=W and X];
    \addplot[orange!20,opacity=0.5] fill between[of=Y and X];

\addlegendentry{ADF}
\addlegendentry{EIDIG}
\addlegendentry{NeuronFair}
\addlegendentry{SG}
\addlegendentry{ExpGA}
\addlegendentry{BREAM}

\end{axis}
\end{tikzpicture}

%% file: Figure/3-2.tex
   \begin{tikzpicture}
  \begin{axis}[
  width = 6cm,
  height = 6cm,
  xtick={1,2,3,4,5},
  xticklabels={20\%,40\%,60\%,80\%,100\%},
  ylabel={EOD},
  axis x line  = bottom,
  axis y line  = left ,
  legend cell align={left},
  tick label style={/pgf/number format/fixed},
    legend columns=2,
  legend style={draw=none,fill=none,at={(1.08,1.05)}},
  ]
   \addplot [mark=diamond,black,name path=F] plot coordinates {
(1,0.068)
(2,0.081)
(3,0.091)
(4,0.128)
(5,0.155)
 };

\addplot [mark=square,red,name path=I] plot coordinates {
(1,0.071)
(2,0.085)
(3,0.098)
(4,0.135)
(5,0.165)
 };

\addplot [mark=o,teal,name path=L] plot coordinates {
(1,0.076)
(2,0.092)
(3,0.101)
(4,0.141)
(5,0.170)
 };

\addplot [mark=star,blue,name path=C] plot coordinates {
(1,0.068)
(2,0.081)
(3,0.092)
(4,0.126)
(5,0.148)
 };

 \addplot [mark=triangle,olive,name path=Z] plot coordinates {
(1,0.059)
(2,0.071)
(3,0.083)
(4,0.115)
(5,0.135)
 };

 \addplot [mark=pentagon,orange,name path=X] plot coordinates {
(1,0.066)
(2,0.079)
(3,0.091)
(4,0.127)
(5,0.150)
 };

\addplot [no marks,black!20,name path=D] plot coordinates {
(1,0.095)
(2,0.112)
(3,0.120)
(4,0.165)
(5,0.185)
 };
\addplot [no marks,black!20,name path=E] plot coordinates {
(1,0.056)
(2,0.063)
(3,0.067)
(4,0.074)
(5,0.089)
 };

\addplot [no marks,red!20,name path=G] plot coordinates {
(1,0.100)
(2,0.117)
(3,0.131)
(4,0.170)
(5,0.190)
 };
 \addplot [no marks,red!20,name path=H] plot coordinates {
(1,0.057)
(2,0.072)
(3,0.078)
(4,0.089)
(5,0.096)
 }; 
 \addplot [no marks,teal!20,name path=J] plot coordinates {
(1,0.105)
(2,0.120)
(3,0.142)
(4,0.175)
(5,0.192)
 };
 \addplot [no marks,teal!20,name path=K] plot coordinates {
(1,0.061)
(2,0.073)
(3,0.089)
(4,0.102)
(5,0.129)
 };
  \addplot [no marks,blue!20,name path=A] plot coordinates {
(1,0.090)
(2,0.107)
(3,0.115)
(4,0.160)
(5,0.180)
 };
 \addplot [no marks,blue!20,name path=B] plot coordinates {
(1,0.054)
(2,0.060)
(3,0.069)
(4,0.077)
(5,0.093)
 };
  \addplot [no marks,olive!20,name path=T] plot coordinates {
 (1,0.088)
(2,0.105)
(3,0.112)
(4,0.155)
(5,0.175)
 };
 \addplot [no marks,olive!20,name path=S] plot coordinates {
 (1,0.044)
(2,0.047)
(3,0.047)
(4,0.048)
(5,0.052)
 };
   \addplot [no marks,orange!20,name path=Y] plot coordinates {
(1,0.093)
(2,0.110)
(3,0.119)
(4,0.162)
(5,0.184)
 }; \addplot [no marks,orange!20,name path=W] plot coordinates {
(1,0.051)
(2,0.053)
(3,0.055)
(4,0.058)
(5,0.061)
 };
\addplot[blue!20,opacity=0.5] fill between[of=B and C];
\addplot[blue!20,opacity=0.5] fill between[of=A and C];
    
 \addplot[black!20,opacity=0.5] fill between[of=E and F];
 \addplot[black!20,opacity=0.5] fill between[of=D and F];
    
           \addplot[red!20,opacity=0.5] fill between[of=H and I];
    \addplot[red!20,opacity=0.5] fill between[of=G and I];
    
         \addplot[teal!20,opacity=0.5] fill between[of=K and L];
    \addplot[teal!20,opacity=0.5] fill between[of=J and L];
    
         \addplot[olive!20,opacity=0.5] fill between[of=S and Z];
    \addplot[olive!20,opacity=0.5] fill between[of=T and Z];
          \addplot[orange!20,opacity=0.5] fill between[of=W and X];
    \addplot[orange!20,opacity=0.5] fill between[of=Y and X];
    
\addlegendentry{ADF}
\addlegendentry{EIDIG}
\addlegendentry{NeuronFair}
\addlegendentry{SG}
\addlegendentry{ExpGA}
\addlegendentry{BREAM}
  \end{axis}
\end{tikzpicture}

%% file: Figure/3-3.tex
   \begin{tikzpicture}
  \begin{axis}[
  width = 6cm,
  height = 6cm,
  xtick={1,2,3,4,5},
   xticklabels={20\%,40\%,60\%,80\%,100\%},
  ylabel={SPD},
  axis x line  = bottom,
  axis y line  = left ,
  legend cell align={left},
  tick label style={/pgf/number format/fixed},
   legend columns=2,
  legend style={draw=none,fill=none,at={(1.08,1.05)}},
  ]
\addplot [mark=diamond,black,name path=F] plot coordinates {
(1,0.065)
(2,0.069)
(3,0.089)
(4,0.104)
(5,0.116)
};

\addplot [mark=square,red,name path=I] plot coordinates {
(1,0.069)
(2,0.076)
(3,0.079)
(4,0.108)
(5,0.122)
};

 \addplot [mark=o,teal,name path=L] plot coordinates {
(1,0.072)
(2,0.079)
(3,0.096)
(4,0.113)
(5,0.134)
};

 \addplot [mark=star,blue,name path=C] plot coordinates {
(1,0.063)
(2,0.069)
(3,0.080)
(4,0.090)
(5,0.109)
};

\addplot [mark=triangle,olive,name path=Z] plot coordinates {
(1,0.069)
(2,0.074)
(3,0.079)
(4,0.084)
(5,0.097)
};

\addplot [mark=pentagon,orange,name path=X] plot coordinates {
(1,0.064)
(2,0.076)
(3,0.085)
(4,0.090)
(5,0.107)
};

\addplot [no marks,black!20,name path=D] plot coordinates {
(1,0.086)
(2,0.096)
(3,0.102)
(4,0.131)
(5,0.160)
};
\addplot [no marks,black!20,name path=E] plot coordinates {
(1,0.043)
(2,0.050)
(3,0.058)
(4,0.063)
(5,0.068)
};
\addplot [no marks,red!20,name path=G] plot coordinates {
(1,0.077)
(2,0.093)
(3,0.104)
(4,0.125)
(5,0.150)
};
\addplot [no marks,red!20,name path=H] plot coordinates {
(1,0.057)
(2,0.059)
(3,0.063)
(4,0.068)
(5,0.078)
};

 \addplot [no marks,blue!20,name path=A] plot coordinates {
(1,0.076)
(2,0.081)
(3,0.095)
(4,0.117)
(5,0.134)
};
\addplot [no marks,blue!20,name path=B] plot coordinates {
(1,0.052)
(2,0.057)
(3,0.062)
(4,0.069)
(5,0.081)
};
 \addplot [no marks,teal!20,name path=J] plot coordinates {
(1,0.089)
(2,0.103)
(3,0.116)
(4,0.130)
(5,0.163)
};
\addplot [no marks,teal!20,name path=K] plot coordinates {
(1,0.064)
(2,0.070)
(3,0.071)
(4,0.090)
(5,0.114)
};
 \addplot [no marks,olive!20,name path=T] plot coordinates {
(1,0.076)
(2,0.077)
(3,0.083)
(4,0.103)
(5,0.132)
};
\addplot [no marks,olive!20,name path=S] plot coordinates {
(1,0.046)
(2,0.053)
(3,0.056)
(4,0.064)
(5,0.076)
};
 \addplot [no marks,orange!20,name path=Y] plot coordinates {
(1,0.076)
(2,0.084)
(3,0.090)
(4,0.103)
(5,0.128)
}; \addplot [no marks,orange!20,name path=W] plot coordinates {
(1,0.057)
(2,0.060)
(3,0.063)
(4,0.064)
(5,0.069)
};
\addplot[blue!20,opacity=0.5] fill between[of=B and C];
\addplot[blue!20,opacity=0.5] fill between[of=A and C];
    
 \addplot[black!20,opacity=0.5] fill between[of=E and F];
 \addplot[black!20,opacity=0.5] fill between[of=D and F];
    
           \addplot[red!20,opacity=0.5] fill between[of=H and I];
    \addplot[red!20,opacity=0.5] fill between[of=G and I];
    
         \addplot[teal!20,opacity=0.5] fill between[of=K and L];
    \addplot[teal!20,opacity=0.5] fill between[of=J and L];
    
         \addplot[olive!20,opacity=0.5] fill between[of=S and Z];
    \addplot[olive!20,opacity=0.5] fill between[of=T and Z];
          \addplot[orange!20,opacity=0.5] fill between[of=W and X];
    \addplot[orange!20,opacity=0.5] fill between[of=Y and X];
\addlegendentry{ADF}
\addlegendentry{EIDIG}
\addlegendentry{NeuronFair}
\addlegendentry{SG}
\addlegendentry{ExpGA}
\addlegendentry{BREAM}
  \end{axis}
\end{tikzpicture}

%% file: threats_to_validity.tex
\section{Threats to validity}
\label{sec:tov}
\textbf{Internal threats} concern parameter settings. In this work, we follow the same settings from prior work~\cite{DBLP:conf/wsdm/ZhaoDSW22,DBLP:conf/icse/ZhangW0D0WDD20,DBLP:conf/issta/ZhangZZ21,DBLP:conf/sigsoft/AggarwalLNDS19}, including those for the generators, \texttt{FairRF}, and the fairness testing procedure. As for \approach, we examine its sensitivity to the most critical parameter $k$. Yet, unintended omission of information or options is always possible. 

While we evaluated \approach~on classical ML models and fully connected neural networks, the omission of other deep architectures (e.g., CNNs, RNNs) is a potential threat to external validity. However, it is worth noting that the modalities of such other DNNs, e.g., CNN and RNN can be highly different, i.e., the definition of image features could be different from tubular data. As such, applying the causality concept to those might require slight amendments to the definition, which might require future investigation. Indeed, fully connected neural networks remain the standard in prior fairness testing work~\cite{DBLP:conf/sigsoft/AggarwalLNDS19,DBLP:conf/icse/0002WJY022,DBLP:conf/icics/JiangSLWSG23,DBLP:conf/icse/MonjeziTTT23,DBLP:conf/issta/ZhangZZ21,DBLP:conf/icse/ZhangW0D0WDD20,DBLP:conf/icse/ZhengCD0CJW0C22}.


\textbf{Construct threats} are related to the choice of metrics. We mitigate this by exploiting three individual/group fairness metrics of different types and covering diverse aspects. To ensure statistical significance, we use U-Test~\cite{wilcoxon1992individual} and $\mathbf{\hat{A}_{12}}$~\cite{Vargha2000ACA} to verify the results. Indeed, a more exhaustive study of a wider range of metrics can be part of our future work.


Finally, \textbf{external threats} to validity can come from the generalizability of the conclusions. To address this, we examine the eight most widely used datasets and six base generators with diverse characteristics. Those, together with the metrics considered, lead to $324$ cases per model. Nevertheless, we agree that examining more diverse scenarios may prove fruitful.



%% file: related_work.tex
\section{Related Work}
\label{sec:related}

\subsection{Causal Inference for Fairness} 
\subsubsection{Causal Inference Approaches for Fairness Testing}
There exists prior work that adopts causal inference fairness testing in AI systems~\cite{DBLP:conf/nips/RussellKLS17,DBLP:conf/sigir/LiCXGZ21,DBLP:conf/bigdataconf/ZhaoBRK23}. Among others, Russell et al.~\cite{DBLP:conf/nips/RussellKLS17} build a causal graph to estimate how sensitive features influence the label outcome. \texttt{FairQuant}\cite{DBLP:journals/corr/abs-2409-03220} is an approach that aims to certify and quantify individual fairness of deep neural networks using symbolic interval analysis, focusing on altering different hyperparameter settings. Similarly, \texttt{Parfait-ML}\cite{DBLP:conf/icse/Tizpaz-NiariKT022} relies on search-based software to explore different hyperparameter configurations in machine learning libraries, identifying settings that improve fairness without significantly compromising accuracy. However, both of them search in the space of hyperparameters while \approach~focuses on perturbation in the space of model input samples as the test cases. 

\texttt{DICE} \cite{DBLP:conf/icse/MonjeziTTT23} is a tool that adopts an information-theoretic framework to quantify fairness defects by measuring the amount of protected information leakage using Shannon entropy and minimum entropy between the layers of a DNN model, which also follows a similar concept of causal inferences. Despite this, we see \texttt{DICE} as complementary to \approach: it can well serve as a base generator, benefiting from the advanced perturbation in \approach~that preserves the interactions between sensitive features and the most causally relevant non-sensitive counterpart, similar to other base generators such as \texttt{ADF} and \texttt{EIDIG} which rely on the identical two-level search as \texttt{DICE}.

Overall, a key concept that distinguishes the above work and \approach~is that they analyze the causal relationship between a (sensitive) feature and the label, or the in-between layers of a DNN model, while \approach~extracts the causal information between sensitive and non-sensitive feature, capturing their interaction in jointly influencing the label, which is then used to guide perturbation. 

\subsubsection{Causal Inference Approaches for Fairness Measurement and Improvement}

Recent studies have integrated causal inference into fairness analysis~\cite{DBLP:journals/corr/abs-2406-03064,DBLP:journals/tkdd/SuYZWWZD24,DBLP:journals/corr/abs-2305-13057}. Among these, Wu et al.~\cite{DBLP:conf/nips/Wu0WT19} propose a unified framework to quantify fairness via path-specific causal effects derived from causal graphs constructed using the PC algorithm. Its primary goal is to measure and explain unfairness by analyzing the causal pathways between sensitive attributes and outcomes, rather than guiding test generation via perturbations.   \texttt{FairCFS}~\cite{DBLP:journals/tkdd/LingXZDYW24} is a causal feature selection method that constructs localized causal graphs to identify Markov blankets of class and sensitive features, thereby selecting features that block sensitive information transmission while preserving predictive performance. Zhang et al.~\cite{DBLP:conf/sigsoft/Zhang022} propose an adaptive fairness improvement framework leveraging causal graphs to inform retraining and data modification strategies for mitigating unfair outcomes. Their work focuses on fairness mitigation after model training though.


However, the above approaches directly focus on fairness measurement, explanation, or mitigation. In contrast, \approach~ is fundamentally different in its goal, where we aim for test case generation that reveals fairness bugs. This is achieved by integrating causal inference to identify non-sensitive features causally related to sensitive ones and systematically incorporating these into perturbation-based fairness testing.

\subsection{Generator for Fairness Testing} 

A vast number of test sample generators have been proposed in recent years for fairness testing, mainly relying on different perturbation mechanisms. White-box generators are popular as they leverage internal information about AI model. \texttt{ADF}~\cite{DBLP:conf/icse/ZhangW0D0WDD20} is a generator that meticulously probes the model's sensitivity to input features with perturbation that leads to a biased outcome. Similarly, \texttt{EIDIG}~\cite{DBLP:conf/issta/ZhangZZ21} improves \texttt{ADF} with gradient information to guide the generation. \texttt{NeuronFair}~\cite{DBLP:conf/icse/ZhengCD0CJW0C22} analyzes the biased neurons, from which it gains insight into the fundamental sources causing model bias. On the other hand, black-box generators can benefit from better exploration and diversity in testing generation. For example, \texttt{SG}~\cite{DBLP:conf/sigsoft/AggarwalLNDS19} generates test samples based on the symmetry principle to guide the perturbation. \texttt{ExpGA}~\cite{DBLP:conf/icse/0002WJY022} combines model interpretation methods and genetic algorithms to search for biased regions. \texttt{BREAM}~\cite{DBLP:conf/icics/JiangSLWSG23} builds a shadow model to simulate the behavior of the target AI model under testing and guides perturbation.

Galhotra et al.~\cite{DBLP:conf/sigsoft/GalhotraBM17} detect individual discrimination by generating test cases differing only on sensitive features. While seminal for fairness testing, their approach does not utilize causal information to guide input generation, treating all non-sensitive features equally.

However, those generators either perturb the non-sensitive features or all features. \approach, in contrast, identifies the direct and most causally relevant non-sensitive feature to join with the concerned sensitive one, which is excluded in the perturbation to ensure their diversity and complex interactions are preserved. In particular, those generators are complementary to \approach~such that they can serve as the underlying base generators.


\subsection{Accounting Non-sensitive Features for Fairness Testing} 

Not much work has considered the usefulness of non-sensitive features in detecting fairness bugs. The most notable ones are \texttt{FairRF}~\cite{DBLP:conf/wsdm/ZhaoDSW22} and \texttt{FairWS}~\cite{DBLP:journals/ijon/ZhuDLW23}, where the non-sensitive features are ranked based on correlation analysis and some fixed weights, which are then used to manipulate their contributions in AI model training to mitigate bias. Yet, they differ on how the ranking is exploited. Yan et al.~\cite{DBLP:conf/cikm/YanKF20} use clustering to divide datasets into groups, from which newly formed non-sensitive features can be used to retrain an AI model for better fairness.



Yet, the above ignores the causality in the datasets, particularly between non-sensitive and sensitive features, which, as we have shown, can be highly beneficial for fairness testing.

%% file: conclusion.tex
\section{Conclusion}
\label{sec:conclusion}

This paper proposes \approach, a generic, higher-level fairness testing framework using causal perturbation. This is achieved by extracting the most directly and causally relevant non-sensitive feature with respect to the concerned sensitive counterpart and injecting such information into the perturbation to guide the test sample generation. \approach~is unique in the sense that it can be seamlessly paired with any perturbation-based generators. Extensive experiments on eight datasets, six generators, and three metrics demonstrate that \approach:

\begin{itemize}
    \item significantly improves the ability of an existing generator to reveal fairness bugs;
    \item provides more reliable information on causality than simple correlation analysis;
    \item is capable of making AI system/model considerably more robust to bias with acceptable extra overhead.
\end{itemize}

In the future, we seek to combine the causal relationship between AI model hyperparameters and fairness with \approach, leading to richer causality-guided fairness testing while extending them for the cases of self-adaptive systems~\cite{DBLP:conf/icse/YeChen25,DBLP:conf/wcre/Chen22}.



